\documentclass[aps,prx,twocolumn,superscriptaddress]{revtex4-2}  
\usepackage{graphicx}  
\usepackage{dcolumn}   
\usepackage{bm}        
\usepackage{amsmath}   
\usepackage{amssymb}   
\usepackage{tikz}
\usetikzlibrary{shapes,arrows}
\usepackage{mathrsfs}
\usepackage{maybemath}
\setcitestyle{super}

\newcommand{\sandw}[3]{\langle #1 | #2 | #3 \rangle}

\newcommand{\bra}[1]{\langle #1|}
\newcommand{\ket}[1]{|#1\rangle}
\newcommand{\Tr}[0]{\textrm{Tr}}
\newcommand{\rarr}[0]{\rightarrow}

\newcommand{\eps}[0]{\varepsilon}
\newcommand{\pphi}[0]{\varphi}

\newcommand{\lp}[0]{\left}
\newcommand{\rp}[0]{\right}

\newcommand{\de}[0]{\partial}
\newcommand{\KS}[0]{\mathrm{KS}}

\newcommand{\leqs}[0]{\leqslant}

\newcommand{\thalf}[0]{\tfrac{1}{2}}
\newcommand{\rr}[0]{\mathbf{r}}

\newcommand{\LR}[0]{\mathrm{L \cdots R}}
\newcommand{\ho}[0]{\mathrm{ho}}
\newcommand{\lu}[0]{\mathrm{lu}}
\newcommand{\CT}[0]{\mathrm{CT}}
\newcommand{\LDA}[0]{\mathrm{LDA}}

\begin{document}

\title{From Kohn-Sham to many-electron energies via step structures in the exchange-correlation potential}

\author{Eli Kraisler}
\thanks{These authors contributed equally}
\email[Author to whom correspondence should be addressed:]{eli.kraisler@mail.huji.ac.il}
\affiliation{Fritz Haber Center for Molecular Dynamics and Institute of Chemistry, The Hebrew University of Jerusalem, 9091401 Jerusalem, Israel}

\author{M.\ J.\ P.\ Hodgson}
\thanks{These authors contributed equally}
\affiliation{Department of Physics, Durham University, South Road, Durham, DH1 3LE, United Kingdom}
\affiliation{Max-Planck-Institut f\"ur Mikrostrukturphysik, Weinberg 2, D-06120 Halle, Germany}

\author{E.\ K.\ U.\  Gross} 
\affiliation{Fritz Haber Center for Molecular Dynamics and Institute of Chemistry, The Hebrew University of Jerusalem, 9091401 Jerusalem, Israel}

\date{\today}

\begin{abstract}

Accurately describing excited states within Kohn-Sham (KS) density functional theory (DFT), particularly those which induce ionization and charge transfer, remains a great challenge. Common exchange-correlation (xc) approximations are unreliable for excited states owing, in part, to the absence of a derivative discontinuity in the xc energy ($\Delta$), which relates a many-electron energy difference to the corresponding KS energy difference. We demonstrate, analytically and numerically, how the relationship between KS and many-electron energies leads to the step structures observed in the exact xc potential in four scenarios: electron addition, molecular dissociation, excitation of a finite system, and charge transfer.  We further show that steps in the potential can be obtained also with common xc approximations, as simple as the LDA, when addressed from the ensemble perspective. The article therefore highlights how capturing the relationship between KS and many-electron energies with advanced xc approximations is crucial for accurately calculating excitations, as well as the ground-state density and energy of systems which consist of distinct subsystems.
\end{abstract}

\maketitle

\section{Introduction}

Describing many-electron excited states at an affordable computational cost remains an important goal within solid state physics, quantum chemistry and materials science~\cite{verma2020status}. In principle, this is possible within density functional theory (DFT)~\cite{DG, PY, Primer, EngelDreizler11, Burke12, Becke14, Jones15} as the ground-state density, $n(\bm{r})$, contains all the information about the many-electron system's ground and excited states according to the first Hohenberg-Kohn (HK) theorem~\cite{HK64}. However, in practice such a description is extremely challenging. The excitation spectrum, the fundamental gap (the difference between the ionization potential (IP), $I$, and the electron affinity (EA), $A$) and charge-transfer energies (the difference between the IP of the donor, $I_\mathrm{d}$, and the EA of the acceptor, $A_\mathrm{a}$) are of particular importance~\cite{pearson2005chemical,Tran07,Tran09,Eisenberg09, HSEsol, ChanCeder10, Tozer03, Maitra05,Toher05, Koentopp06,Ke07, Hofmann12, Nossa13, Fuks16,Kronik_JCTC_review12,KronikKuemmel_PES,Kummel17,Gould18,AschebrockKummel19}. The unreliable performance of standard exchange-correlation (xc) approximations for these quantities is in contrast to the remarkable success of Kohn-Sham (KS) DFT for various applications to ground state properties of materials~\cite{Kaxiras03, Martin, Cramer04, Kohanoff06, ShollSteckel11, Giustino14_materials, DiValentin_TopCurrChem, KronikNeaton16, Kummel17, Maurer19}. In this article we explore the exact relationship between KS excitation energies and the corresponding many-electron quantities with standard and ensemble DFT. We study the consequences of this relationship on the exact KS potential and its importance for the advancement of approximate xc density functionals. 

Unlike other commonly used methods for electronic structure calculations, e.g., many-body perturbation theory~\cite{FetterWalecka,Gross_MB,Reining_MB}, within KS-DFT the relationship between the KS energy levels, $\left \{ \eps_i \right \}$, and the many-electron energies, $\left\{ E_i \right \}$, is not generally straightforward. For example, while for the exact KS potential 
the highest occupied (ho) KS energy level, $\eps^\ho$, equals minus the IP~\cite{PPLB82, PerdewLevy83, LevyPerdewSahni84, NATO85_Perdew, AlmbladhVonBarth85, PerdewLevy97, Harbola99}, $-I$, the fundamental gap, $E_\mathrm{g} = I - A$, does not simply equal the KS gap, $E_\mathrm{g}^\KS = \eps^\lu - \eps^\ho$ (i.e., the difference between the lowest unoccupied (lu) and the ho KS energies), even for the exact KS potential. Instead, the KS gap differs from the fundamental gap by $\Delta$, known as the derivative discontinuity~\cite{PPLB82,ShamSchluter83, PerdewLevy83, Perdew85, ZhangYang00, LeinKummel05, Mundt05, Cohen12, Baerends13, KraislerKronik13, MoriS14, Mosquera14, Mosquera14a, KraislerKronik14, Goerling15, jones2015density}:  $E_\mathrm{g} = I - A = \varepsilon^\lu - \varepsilon^\ho + \Delta$. $\Delta$ manifests in the exact xc potential as a uniform shift when the number of electrons within the system infinitesimally surpasses an integer. It occurs because the xc energy of the system has discontinuities in its derivative as a function of electron number, $N$, at integer values of $N$. 

Similarly, it has been shown recently~\cite{HodgsonKraisler17} that the charge-transfer energy in stretched systems differs from the corresponding KS energy difference by the charge-transfer derivative discontinuity (CTDD), $\Delta_\CT$, which occurs when a fraction of charge is transferred from one subsystem to another within the whole system.  
The CTDD proved to be an important concept for accurately modeling charge transfer within KS theory in practice~\cite{schulz2019description}.

In 1995 Levy proposed that the optical (uncharged) gap, i.e., the energy to excite an electron from the ground to its first excited state ($\hbar \omega_\mathrm{og}$), is related to the corresponding KS gap ($\eps^\mathrm{lu} - \eps^\mathrm{ho} = \hbar \omega^\mathrm{KS}_\mathrm{og}$) via a derivative discontinuity~\cite{PhysRevA.52.R4313}, as such $\hbar \omega_\mathrm{og} = \hbar \omega^\mathrm{KS}_\mathrm{og} + \Delta_\mathrm{og}$.

All the discontinuities mentioned above -- $\Delta$, $\Delta_\CT$ and $\Delta_\mathrm{og}$ -- are important and rather delicate properties of the exact xc functional. Their existence gives rise to step structures in the exact xc potential -- sudden changes in the magnitude of the potential over a short region of space. These steps have a strong nonlocal dependence on the electron density, which partly explains why they are not captured by most existing approximations. 

In Ref.~\onlinecite{HodgsonKraisler17} the relationship between the derivative discontinuity $\Delta$ in the xc energy and the spatial step $S$ that appears in the exact xc potential of stretched diatomics was established.  In this article we further study the step structure of the exact xc potential and relate it to the excitation energies of the interacting many-electron system. Particularly, we show how the steps are crucial in the prediction of the fundamental gap, excitation energies, such as charge transfer, and the correct distribution of charge in stretched systems. 

This article is organized as follows. Section~\ref{sec:theory} gives a detailed introduction to the interatomic step $S$ within a stretched diatomic molecule in its ground state. Then the derivative discontinuity, $\Delta$, which occurs for ground-state systems with a fractional electron number, is discussed. Finally the CTDD, $\Delta_\CT$, is analytically studied for both a stretched diatomic molecule with a fractional $N$ and for a stretched diatomic molecule that experiences charge transfer upon excitation. Section~\ref{sec:NumericalDetails} provides the numerical details of the calculations performed in this work. Section~\ref{sec:relationship} discusses the relationship between $\Delta$ and $S$, numerically addressing finite and stretched systems. Section~\ref{sec:CT} presents the exact KS potential obtained from an excited-state calculation of a one-dimensional (1D) stretched diatomic molecule, which undergoes charge transfer. Then, in Sec.~\ref{sec:excited_atom} an excited atom is analyzed to show that steps and plateaus in the KS potential appear not only for a stretched, but also for a finite system, upon excitation within ensemble DFT. In Sec.~\ref{sec:plateaus.approx} we show that steps can be found not only in the usually unreachable exact KS potential, but also in approximate potentials, as simple as the one that stems from the local density approximation (LDA), by means of numerical inversion of the LDA ensemble density. Finally, in Sec.~\ref{sec:Conclusions} we summarize our work.

\section{Properties of the exact exchange-correlation functional} \label{sec:theory}

\subsection{The spatial step $\maybebm{S}$} \label{sec:theory.S}

In general, sharp spatial steps may occur in the \emph{exact} xc potential~\cite{NATO85_AvB,LeinKummel05,ThieleGrossKummel08} at any point where the electron density decays at a rate which abruptly changes. One scenario is an atom with spatially distinct electron shells (see, e.g., Refs.~\cite{RvL95,Hodgson16}). In this case, approaching the atom inwards from infinity, the decay of the outermost shell is substituted by the decay of the next, inner shell. The potential then experiences a step, which can be revealed~\cite{KRIEGER1990256,PhysRevA.45.101,Grabo00}, particularly when using orbital-dependent, exact-exchange-based approximations, within the optimized effective potential (OEP) method~\cite{PhysRev.90.317,PhysRevA.14.36,PhysRevA.85.052508,PhysRevA.88.046501,PhysRevA.88.046502,KK08,D0FD00069H}; however, this approach has known numerical difficulties which arise from the use of a finite basis set~\cite{hirata2001can,staroverov2006optimized,PhysRevA.85.052508,PhysRevA.88.046501,glushkov2009finite}. Solutions have been proposed to overcome these numerical issues~\cite{PhysRevLett.98.256401,hesselmann2007numerically,KK08,D0FD00069H}, however, the OEP method is yet to be adopted as a mainstream approximation within DFT owing to the numerically challenging nature of the approach.

Another, very important scenario is a complex system, which consists of several spatially distinct subsystems, e.g., atoms within a molecule. For such systems one can introduce the local effective ionization potential (LEIP)~\cite{Hodgson16}, which stems from the decay rate of a given subsystem. Moving from one subsystem to another leads to a change in the LEIP, which causes a sharp spatial step in the xc potential. The height of the step is analytically derived below from the density decay rate, following Refs.~\cite{Hodgson16,HodgsonKraisler17}.

\begin{figure}[b]
\includegraphics[width=1.0\linewidth]{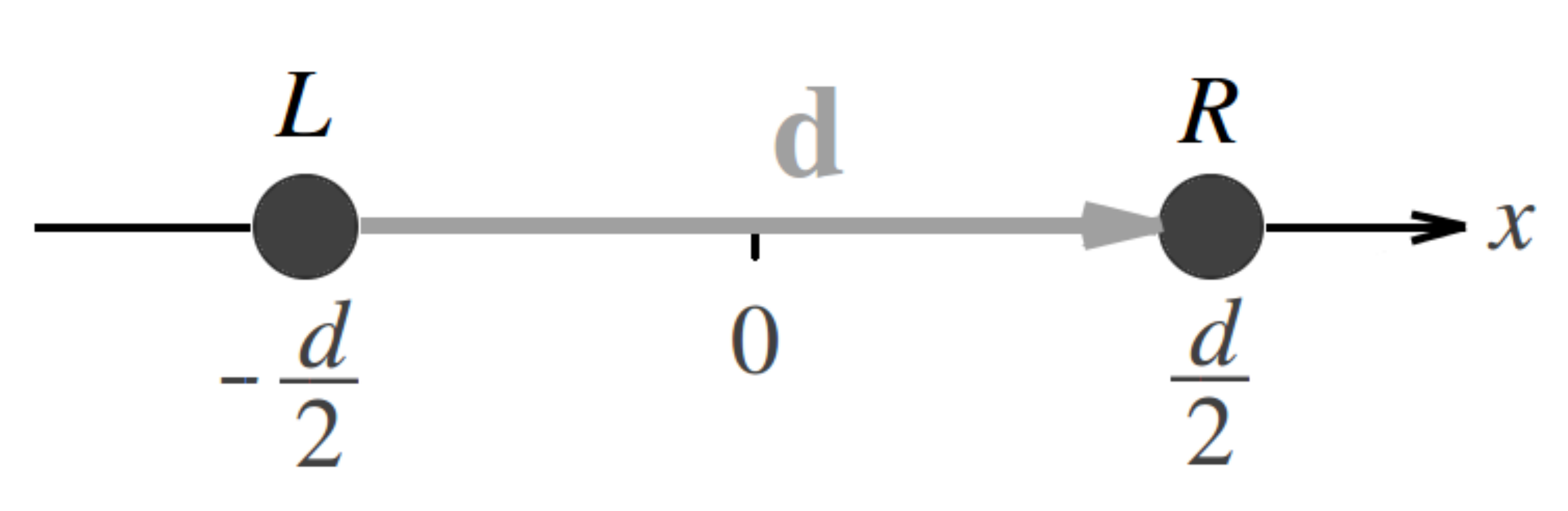} 
\caption{A stretched diatomic molecule, $\LR$, with an interatomic distance $d$.}
\label{fig:diatomicLR}
\end{figure}

A simple and instructive example of a system with a step in the xc potential is a stretched diatomic molecule $\LR$ sketched in Fig.~\ref{fig:diatomicLR}. In this case each atom within the system can be considered a subsystem. Additional, more complicate examples include donor-acceptor pairs, which are important in photovoltaics~\cite{Bredas09,venkataraman2010role,deibel2010role,liu2014strategy,chen2018density,trang2020theoretical} and a molecule between two metallic contacts in a transport experiment~\cite{nitzan2003electron,PhysRevB.69.235411,Hofmann12,stefanucci2015steady,kurth2017transport,Zelovich17,Dunietz18,AschebrockKummel19,Dunietz20}.
Therefore, understanding the steps in the exact KS potential is crucial, as it allows one to accurately describe various scenarios in real materials of high practical importance with KS DFT.

\begin{figure} 
\includegraphics[width=1.25\linewidth,trim={25mm 25mm 0mm 25mm}]{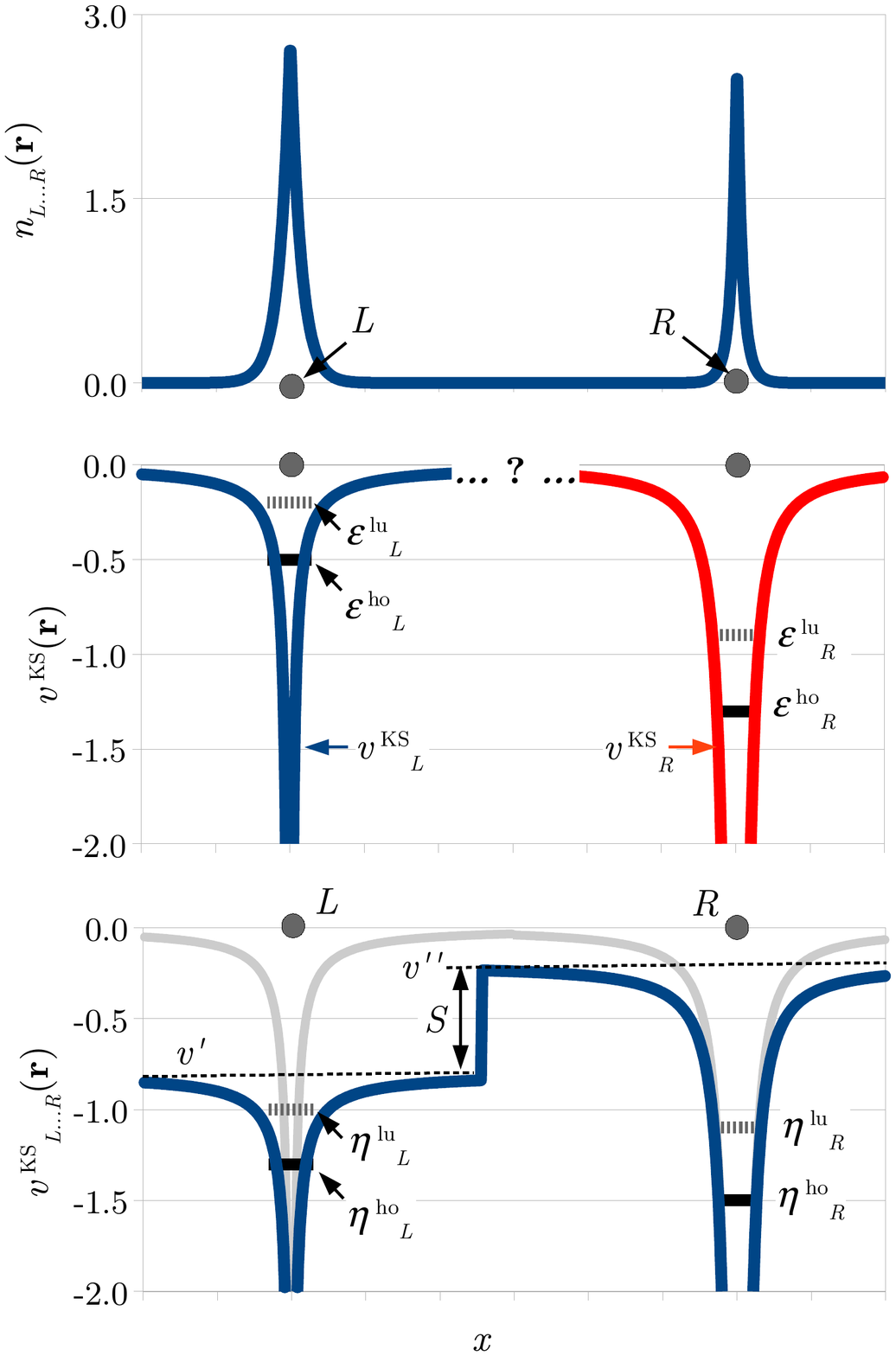}
\caption{Top: A sketch of the density $n_\LR(\bm{r})$ in a stretched diatomic molecule, $\LR$. Middle: The atomic potentials, $v^\KS_\mathrm{L}(\bm{r}+\thalf \bm{d})$ and $v^\KS_\mathrm{R}(\bm{r}-\thalf \bm{d})$, and their ho and lu KS energy levels. The problem of $\eps^\lu_\mathrm{R}$ lying below $\eps^\ho_\mathrm{L}$ is illustrated. Bottom: The molecular KS potential, $v^\KS_\LR(\bm{r})$ (blue), compared to the atomic potentials (gray).  $v^\KS_\LR(\bm{r})$ possesses a step $S$ in between the atoms (as well as a complementary step (down), $-S$ to the right of Atom R, not shown). The molecular ho and lu KS levels are marked. }
 \label{fig:graphs_illustration_LR}
\end{figure}

In the diatomic molecule $\LR$ with interatomic distance $d = |\bm{d}|$, Atom L is located at $x=-\thalf d$ and Atom R at $x= \thalf d$ with $x$ being the interatomic axis (see Fig.~\ref{fig:diatomicLR}). 
In the limit $d \rarr \infty$, the energy of the molecule equals the sum of the energies of the constituent atoms (the subsystems), as such
\begin{equation}\label{eq:E_L...R}
\lim_{d \rarr \infty} E_\LR = E_\mathrm{L} + E_\mathrm{R},
\end{equation}
and the density is the sum of the (shifted) atomic densities:
\begin{equation} \label{eq:n_L...R}
\lim_{d \rarr \infty} n_\LR(\bm{r}) = n_\mathrm{L}(\bm{r} + \thalf \bm{d}) + n_\mathrm{R}(\bm{r} - \thalf \bm{d}), 
\end{equation}
with $N_\mathrm{L}^0$ electrons on Atom L and $N_\mathrm{R}^0$ electrons on R; see Fig.~\ref{fig:graphs_illustration_LR}~(top).   The equilibrium number of electrons in the molecule is thus $N_\LR^0 = N_\mathrm{L}^0 + N_\mathrm{R}^0$.   Equation~\ref{eq:n_L...R} is true for systems that do not experience degeneracy of the ground state in the limit $d \rarr \infty$; those are the systems on which we focus below. 
[Note however that e.g.\ for homonuclear diatomic ions $(\mathrm{A} \cdots \mathrm{A})^+$, any density of the form $n_{\mathrm{A}^{+q}}(\bm{r} + \thalf \bm{d}) + n_{\mathrm{A}^{+(1-q)}}(\bm{r} - \thalf \bm{d})$ is a valid ground-state density in the limit $d \rarr \infty$.]

Now we ask what form the \emph{exact} KS potential of the whole molecule, $v^\KS_\LR(\bm{r})$, takes for large $d$,
\footnote{The criterion for a large separation $d$ is such that in the region between the atoms the L- and R-densities have reached the regime of exponential decay; see below.}
and how it relates to the atomic potentials, $v^\KS_\mathrm{L}(\bm{r})$ and $v^\KS_\mathrm{R}(\bm{r})$. Is it that, similarly to the molecular density, $\lim_{d \rightarrow \infty} v^\KS_\LR(\bm{r})=v^\KS_\mathrm{L}(\bm{r}+\thalf \bm{d}) + v^\KS_\mathrm{R}(\bm{r}-\thalf \bm{d})$? There is reason to think that the limit above holds, at least in the vicinity of each atom, because near, say, Atom L, the molecular potential $v^\KS_\LR(\bm{r})$ has to reproduce the atomic density $n_\mathrm{L}(\bm{r} + \thalf \bm{d})$. From the HK theorem~\cite{HK64} we know that this potential is unique, up to a constant, and equals $v^\KS_\mathrm{L}(\bm{r}+\thalf \bm{d})$~\cite{ProdanKohn05}. The same is, of course, also true for Atom R. However, the simple superposition of the atomic KS potentials can create the following problem (see Fig.~\ref{fig:graphs_illustration_LR}~(middle)): the lu KS energy level of one of the atoms (say, Atom R), $\eps^\lu_\mathrm{R}$, can lie below the ho level of the other atom (Atom L), $\eps^\ho_\mathrm{L}$. Then, from the perspective of the KS system, the electron which should localize on L will spuriously do so on R, resulting in the wrong number of electrons on each atom
\footnote{It is also possible, both in the exact case and within common xc approximations, that such hopping of an electron will not solve the problem: in the system $\mathrm{L}^+ \cdots \mathrm{R}^-$ the lu of $\mathrm{L}^+$ will lie below the ho of $\mathrm{R}^-$, which will require the electron to jump back. In common approximations, such as the LSDA, this results in a spuriously fractional number of electrons on each of the atoms (see Refs.~\cite{PPLB82,KraislerKronik15} and references therein), violating the principle of integer preference~\cite{Perdew90}, and being the manifestation of the \emph{delocalization error}~\cite{MoriS08}}.
In the case in which the atoms within the molecule are bonded, the molecular ho levels of Atoms L and R ought to be aligned; this does not always happen if the two atomic potentials are simply superimposed. 

What must the exact KS potential do to maintain the correct atomic densities in the vicinity of each atom whilst yielding the correct distribution of charge within the molecule? The answer is to raise the level of the potential around one of the atoms, in our case Atom R, forming a \emph{plateau}, which results in a spatially abrupt \emph{step} in the KS potential between the atoms (and a complementary step far to the right of Atom R)~\cite{NATO85_Perdew,NATO85_AvB}. In the vicinity of Atom R the molecular potential equals $v^\KS_\mathrm{R}(\bm{r}-\thalf \bm{d})$, up to a constant, hence no violation of the HK theorem occurs. The density in this vicinity equals $n_\mathrm{R}(\bm{r} - \thalf \bm{d})$, as required. 

Following Ref.~\onlinecite{Hodgson16}, we now show how the height of the step in the KS potential of a stretched diatomic molecule is related, in the general case, to the IPs of the constituent atoms, $I_\mathrm{L}$ and $I_\mathrm{R}$, and the \emph{molecular} orbital energies of the system as a whole (see also Ref.~\onlinecite{HodgsonKraisler17} and references therein). We consider, therefore, a diatomic molecule $\LR$ with a large, but finite separation $d$ and assume that it has been solved within KS DFT, and the molecular KS potential, $v^\KS_\LR(\bm{r})$, as well as the molecular energy levels are known; see Fig.~\ref{fig:graphs_illustration_LR}~(bottom). We denote here the \emph{molecular} KS energy levels by $\{ \eta_i \}$ to clearly distinguish them from the \emph{atomic} KS energy levels, $\{ \eps_i \}$. We also explicitly indicate whether the molecular orbitals localize on one of the atoms by the subscripts L and R. Generally, in the vicinity of Atom L the molecular KS potential $v^\KS_\LR(\bm{r})$ is identical to the atomic potential, $v^\KS_\mathrm{L}(\bm{r}+\thalf \bm{d})$, up to a constant, $v'$, and in the vicinity of R, $v^\KS_\LR(\bm{r})$ is identical to $v^\KS_\mathrm{R}(\bm{r}-\thalf \bm{d})$, up to $v''$. The difference $v''-v'$ is therefore the interatomic step heigh, $S$
~\footnote{For generality, we introduced two constants here, $v'$ and $v''$, to allow both atomic potentials to be vertically shifted. In the case depicted in Fig.~\ref{fig:graphs_illustration_LR} it is actually convenient to set $v'$ to 0, thus far from both atoms $v^\KS_\LR(\bm{r})$ approaches 0.}.
Furthermore, in the vicinity of Atom L the molecular density $n_{\LR}(\bm{r})$, which equals the (shifted) atomic density, $n_\mathrm{L}(\bm{r}+\thalf \bm{d})$ (see Eq.~(\ref{eq:n_L...R})), and decays as $\sim \exp(-2 \sqrt{2I_\mathrm{L}} |\bm{r}+\thalf \bm{d}|)$
~\footnote{Hartree atomic units are used throughout.}
\cite{KatrielDavidson80,LevyPerdewSahni84,PhysRevA.16.1782,AlmbladhVonBarth85,GoriGiorgi18,Kraisler_IJC20}.
From the KS perspective, the decay of the atomic density is governed by the square of the ho KS orbital, which is localized on L, $|\pphi_\mathrm{L}^\ho(\bm{r})|^2$. This orbital decays as
\footnote{We note in passing that the criterion for the separation $d$ to be considered large follows from the exponential decay rate analysis we just performed: $d$ has to be larger than the decay lengths of both Atoms L and R, i.e., $d \gg 1/\sqrt{I_\mathrm{L}}$ and $d \gg 1/\sqrt{I_\mathrm{R}}$.}
\begin{align} \label{eq:phhi_L_ho}
|\pphi_\mathrm{L}^\ho(\bm{r})|^2 \sim \exp \lp(-2 \sqrt{-2(\eta^\ho_\mathrm{L} - v')} |\bm{r}+\thalf \bm{d}| \rp).
\end{align}
As the exact KS density equals the many-electron density, the two decay rates are equal and hence $v' = \eta^\ho_\mathrm{L} + I_\mathrm{L}$. Similar analysis for Atom R yields $v'' = \eta^\ho_\mathrm{R} + I_\mathrm{R}$. Combining these two results, and recalling that $S=v''-v'$, we arrive at an expression for the interatomic step~\cite{Hodgson16}:
\begin{equation} \label{eq:S}
S = I_\mathrm{R} - I_\mathrm{L} + \eta^\ho_\mathrm{R} - \eta^\ho_\mathrm{L}.
\end{equation}
Importantly, the constraint that the multiplicative KS potential must yield a single-particle density which exactly equals the many-electron density leads to the step $S$ in the potential~\cite{PhysRevB.99.045129}. The step is generally nonzero, because the KS energy differences do not equal the many-electron energy differences, as mentioned in the Introduction. In the particular case here, the many-electron energy difference, $ I_\mathrm{R} - I_\mathrm{L}$ does not equal the KS energy difference, $\eta^\ho_\mathrm{L} - \eta^\ho_\mathrm{R}$. The step forms at the point in the electron density where the decay from the left meets the decay from the right, and the LEIP abruptly changes. 

We wish to emphasize that the right-hand side of Eq.~(\ref{eq:S}) includes the \emph{molecular} energy levels, $\left \{ \eta_i \right \}$, and not the \emph{atomic} levels, $\left \{ \eps_i \right \}$. Therefore, in general, Eq.~(\ref{eq:S}) does not allow one to directly obtain the step height in the molecular potential, $S$, relying only on atomic calculations. This equation rather shows the relationship between $S$, the molecular KS energies and the many-electron energies, $I_\mathrm{L}$ and $I_\mathrm{R}$, associated with each atom.

Equation~(\ref{eq:S}) refers to the general case, where L and R can be any atoms, and therefore the energies $\eta^\ho_\mathrm{L}$ and $\eta^\ho_\mathrm{R}$ need not be assumed equal. The latter is true when L and/or R are closed-shell atoms. In the particular case that L and R are bonded, the ho KS orbital stretches over both atoms and therefore it follows that, in the notation adopted here, $\eta^\ho_\mathrm{R} = \eta^\ho_\mathrm{L}$. As a result, Eq.~(\ref{eq:S}) reduces to the famous result $S = I_\mathrm{R} - I_\mathrm{L}$ by Almbladh and von Barth~\cite{NATO85_AvB}$_,$\footnote{In the bonded case both the bonding and the anti-bonding molecular orbitals delocalize over both atoms, and in the infinite limit the bonding orbital can equally be described by two half-filled orbitals of the same energy, one localized on L, and one on R.}.

Depending on the atoms L and R, either $I_\mathrm{L}$ or $I_\mathrm{R}$ is the \emph{overall} IP of the molecule; in the case depicted in Fig.~\ref{fig:graphs_illustration_LR} it is $I_\mathrm{L}$. Thus, the overall highest occupied molecular orbital (HOMO) energy is $\eta^\ho_\mathrm{L}$ and is equal to the atomic orbital $\eps^\ho_\mathrm{L}$ when $v'=0$. Furthermore, due to the IP theorem in DFT~\cite{PPLB82,LevyPerdewSahni84,PerdewLevy97, Harbola98,Harbola99,Yang12}, which we discuss in detail below, $\eps^\ho_\mathrm{L} = -I_\mathrm{L}$. It then follows that Eq.~(\ref{eq:S}) reduces to $S = I_\mathrm{R} + \eta^\ho_\mathrm{R}$. It does not necessarily follow, however, that $S$ vanishes. A generally nonzero $S$ stems from the inclusion of the molecular energy, $\eta^\ho_\mathrm{R}$, opposed to the atomic energy, $\eps^\ho_\mathrm{R}$, in  Eq.~(\ref{eq:S}). The atomic energy $\eps^\ho_\mathrm{R}$ equals $-I_\mathrm{R}$, whereas the molecular energy $\eta^\ho_\mathrm{R}$ does not, as it is elevated relative to the atomic energy by the step height $S$: $\eta^\ho_\mathrm{R} = \eps^\ho_\mathrm{R} + S$

Our decomposition of this molecule into fragments is reminiscent of Partition DFT (PDFT) \cite{cohen2007foundations} in which the exact KS potential is separated into the KS potential for each individual subsystem plus the `partition potential'. In the limit that the subsystems are completely separated -- in our case the two atoms -- the partition potential consists of the interatomic step described above \cite{oueis2018exact}. The partition potential is a functional of the density of each fragment of the system \cite{elliott2010partition} and hence is nonlocal in character \cite{gomez2019partition}. In addition, the exact partition potential is known to contain derivative discontinuities \cite{nafziger2011molecular}. The perspective allowed by PDFT offers an approach to developing approximations which capture these discontinuous features, yield accurate binding energies of disassociated diatomics \cite{nafziger2011molecular,nafziger2015fragment,jiang2018constructing} or a reliable description of charge transfer \cite{cohen2009charge,nafziger2014density}. The partition potential has also been shown to be a chemically significant reactivity potential \cite{gomez2017partition,chavez2020towards}. 

\subsection{The uniform jump $\maybebm{\Delta}$} \label{sec:theory.Delta}

The uniform jump $\Delta$ occurs in the KS potential when the number of electrons, $N$, varies continuously, and infinitesimally surpasses an integer value. A fractional number of electrons in our systems of interest may be considered as a time average of the number of electrons in an \textit{open} system, namely in a system which is free to exchange electrons with its surroundings (see, e.g., Ref.~\onlinecite[\S 14]{LL3}). The ground state of such a system can no longer be described by a \textit{pure} quantum-mechanical state. Instead, it is a statistical mixture, or \emph{ensemble}, of pure (integer-electron) states~\cite{PPLB82}. 

In the following we consider three types of many-electron systems. First, in this section, we describe in detail a finite system that is connected to an electron reservoir, which allows $N$ to change continuously. Second, in Sec.~\ref{sec:theory.CTDD} we consider a stretched diatomic molecule $\LR$, whose total number of electrons can vary continuously, and for which any additional charge localizes on Atom R, whereas any charge deficiency results in decrease of charge around Atom L. Third, in Sec.~\ref{sec:theory.CTDD} we consider a stretched diatomic molecule $\LR$, whose total number of electrons is fixed at a given integer value, but the number of electrons on each atom can become fractional by transferring charge between the atoms.  

We start with a finite system, like an atom or a molecule, with $N = N_0 + \alpha$ electrons, where $N_0$ is an integer number, and $0 \leqs \alpha \leqs 1$. As mentioned above, the ground state of such a system is an ensemble, which combines states each with a different integer number of electrons. For systems with Coulomb interaction at zero temperature, this ensemble consists only of states for $N_0$ and $N_0+1$ electrons, $\ket{\Psi_{N_0}}$ and $\ket{\Psi_{N_0+1}}$:
\begin{equation}\label{eq:Lambda}
\hat{\Lambda} = (1-\alpha) \ket{\Psi_{N_0}}\bra{\Psi_{N_0}} + \alpha \ket{\Psi_{N_0+1}}\bra{\Psi_{N_0+1}},
\end{equation}
with the statistical weights of $(1-\alpha)$ and $\alpha$, respectively~\cite{DG,Lieb,RvL_PhD,RvL_adv, PPLB82}. As a direct consequence of Eq.~(\ref{eq:Lambda}), the expectation value of any operator $\hat O$ in the ensemble state is $O = \Tr \{ \hat{\Lambda} \hat{O} \} = (1-\alpha) \sandw{\Psi_{N_0}}{\hat O}{\Psi_{N_0}} + \alpha \sandw{\Psi_{N_0+1}}{\hat O}{\Psi_{N_0+1}}$~\cite{PPLB82}. In particular, the average density of a system with $N$ electrons is
\begin{equation} \label{eq:n_piecewise}
 n(\bm{r};N) = (1-\alpha) n(\bm{r};N_0) + \alpha n(\bm{r};N_0+1),
\end{equation}
where $n(\bm{r};N_0)$ is the ground-state density for the $N_0$-electron system and $n(\bm{r};N_0+1)$ is the ground-state density for the $(N_0+1)$-electron system. Furthermore, the total energy as a function of $N$ equals 
\begin{equation} \label{eq:E_piecewise}
E(N) = (1-\alpha) E(N_0) + \alpha E(N_0+1).
\end{equation}

\begin{figure} 
\includegraphics[width=0.95\linewidth]{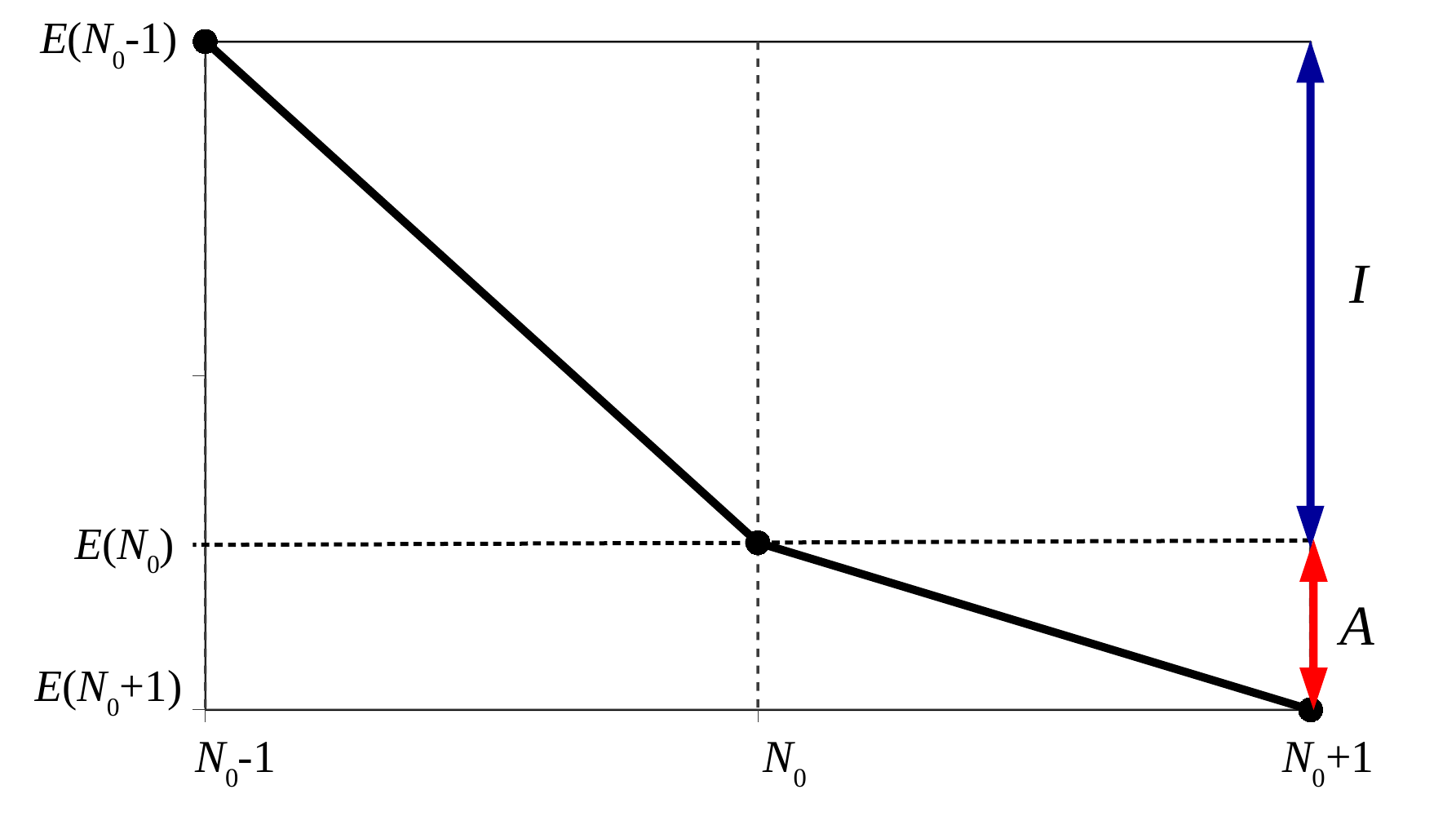} \\
\vspace{3mm}
\includegraphics[width=0.95\linewidth]{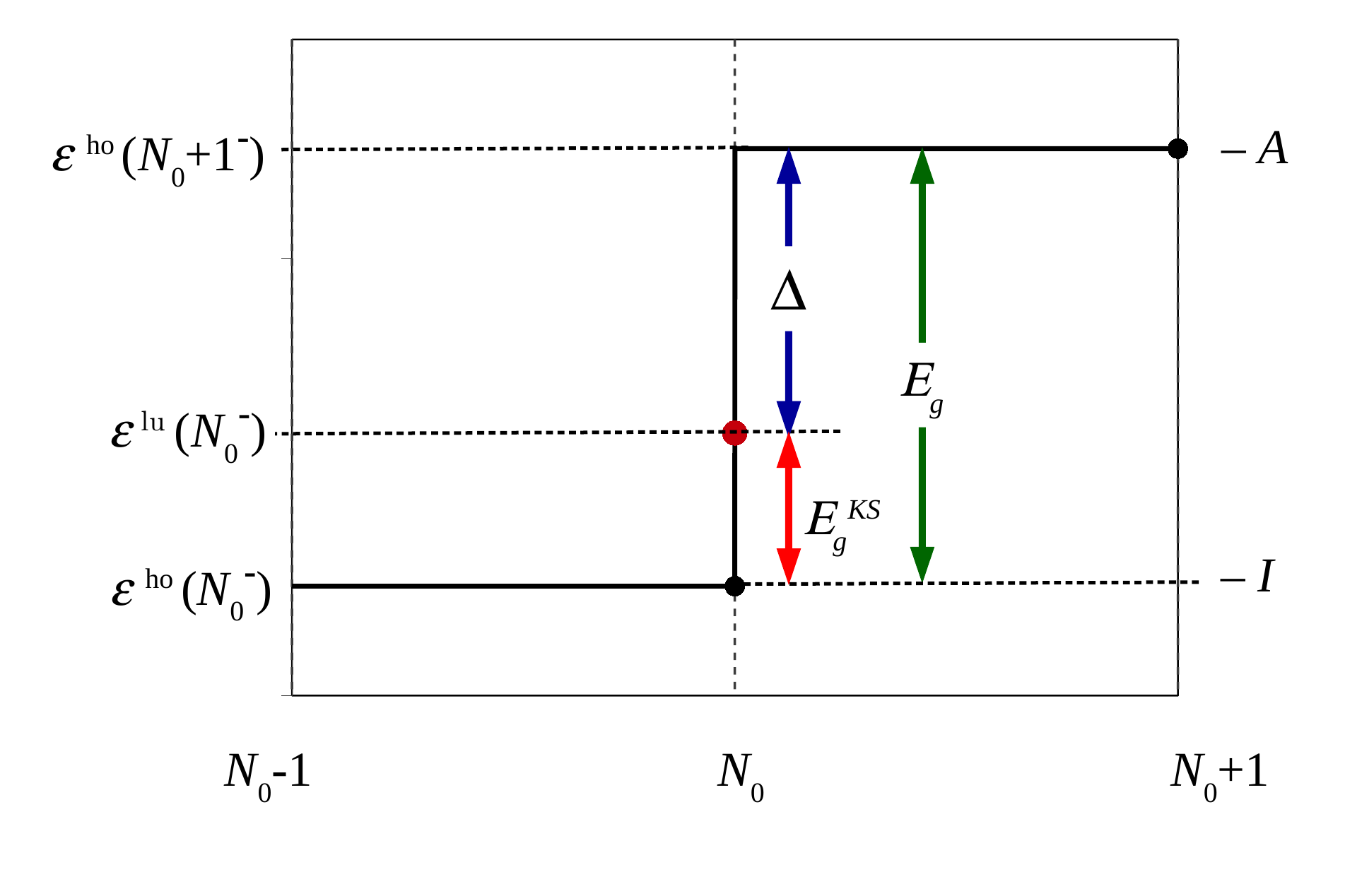} \\
\vspace{3mm}
\includegraphics[width=0.95\linewidth]{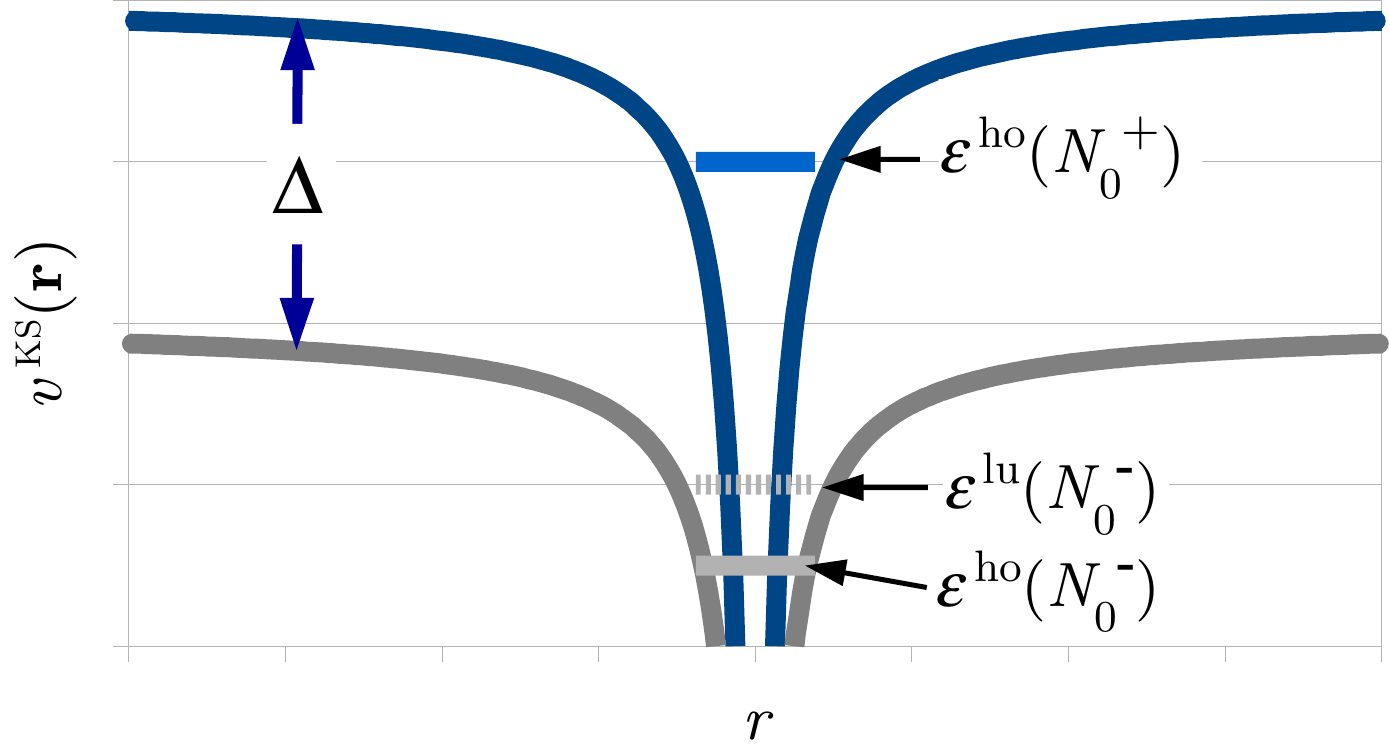}
\caption{Top: dependence of the energy $E(N)$ on the number of electrons, $N$, for a finite system. The IP and EA are marked on the graph. Middle: dependence of the chemical potential, $\mu(N)$ -- which equals the ho energy level -- on $N$. The fundamental gap, $E_\mathrm{g}$, the KS gap, $E_\mathrm{g}^\KS$ and the uniform jump $\Delta$ are marked on the graph. Bottom: the KS potential, $v^\KS(\bm{r})$ at $N=N_0^-$ (gray), compared to KS potential at $N=N_0^+$ (blue). The ho and lu energy levels at $N=N_0^-$, the ho level at $N=N_0^+$ as well as the uniform jump $\Delta$ are marked. }
\label{fig:graphs_finite_system}
\end{figure}

As can be seen in Fig.~\ref{fig:graphs_finite_system}(top), $E(N)$ is piecewise-linear in $N$: for any fractional $N$, the energy is linear, but it can change its slope when $N$ passes an integer. Consequently, the chemical potential, $\mu = \de E/ \de N$, is a stair-step function of $N$. For example, in the ground state:
\begin{equation}\label{eq:mu_of_N}
    \mu(N) = \left\{
              \begin{array}{ccrcl}
                - I       & : & N_0-1 & < N \leqs & N_0   \\
                - A       & : & N_0   & < N \leqs & N_0+1 \\
              \end{array}
           \right.,
\end{equation}
where $I = E(N_0-1) - E(N_0)$ is the IP and $A = E(N_0) - E(N_0+1)$ is the EA of the system. Clearly, the chemical potential is generally discontinuous at integer $N$; the height of this discontinuity equals the fundamental gap of the system, $E_\mathrm{g} = I-A$. 

Furthermore, from a combination of the piecewise-linearity of the energy and Janak's theorem~\cite{Janak78}, which states that the $i^\mathrm{th}$ KS eigenenergy, $\eps_i = \de E / \de f_i$ -- the derivative of the total energy with respect to the occupation of the $i^\mathrm{th}$ level, $f_i$ -- we find that the ho KS energy level, $\eps^\ho(N)$, equals the chemical potential, $\mu(N)$, and is also discontinuous at integer $N$ (see Fig.~\ref{fig:graphs_finite_system}(middle)). This is the content of the IP theorem in DFT~\cite{PPLB82,LevyPerdewSahni84,PerdewLevy97, Harbola98,Harbola99,Yang12}: for the exact xc potential, infinitesimally below an integer, $\eps^\ho(N_0^-) = - I$ and infinitesimally above $\eps^\ho(N_0^+) = -A$. The IP theorem in KS DFT is an exact result, for the exact xc potential.

Satisfying the aforementioned IP theorem creates a challenge for the exact xc potential, $v_\mathrm{xc}(\bm{r})$. From the perspective of the KS system, increasing $N$ above an integer means occupying the next KS level, $\eps^\lu(N_0^-)$. As $\eps^\lu(N_0^-)$ does not necessarily equal $-A$, \textit{even for the exact KS potential} (see Fig.~\ref{fig:graphs_finite_system}(middle)), the only thing the exact potential can do in order to satisfy the IP theorem is to \emph{discontinuously} change as $N$ infinitesimally surpasses an integer. However, due to the continuity of the density with $N$ (see Eq.~(\ref{eq:n_piecewise})) and the HK theorem, the discontinuity of the KS potential can change only by a \emph{spatially uniform constant} (see Fig.~\ref{fig:graphs_finite_system}(bottom)), which is usually denoted $\Delta$. This discontinuity in the KS potential, $v_\KS(\bm{r})$, can only come from $v_\mathrm{xc}(\bm{r})$, because the Hartree potential is continuous and the external potential is $N$-independent. Therefore,
\begin{equation} \label{eq:Delta.1}
\Delta = \lim_{\alpha \rarr 0^+} v_\mathrm{xc}(\bm{r};N_0 + \alpha) - v_\mathrm{xc}(\bm{r};N_0 - \alpha).
\end{equation} 
The value of $\Delta$ is easy to deduce from the arguments above: it is the difference between the value that 
$\eps^\ho(N_0^+)$ ought to have, namely $-A$, and the value it has in absence of discontinuity, $ \eps^\lu(N_0^-)$: $\Delta = -A - \eps^\lu(N_0^-)$. Together with $\eps^\ho(N_0^-) + I = 0$, and dropping here the argument $N_0^-$ for brevity, we arrive at the following familiar form for $\Delta$:
\begin{equation} \label{eq:Delta.2}
\Delta = E_\mathrm{g} - E_\mathrm{g}^\KS = I - A - (\eps^\lu - \eps^\ho),
\end{equation}
where $\Delta$ is expressed as the difference between the fundamental gap of the system, $E_\mathrm{g} = I-A$, and the KS gap, $E_\mathrm{g}^\KS = \eps^\lu - \eps^\ho$. The derivative discontinuity is a topic of great importance and has received much attention over the years~\cite{PPLB82,ShamSchluter83, PerdewLevy83, Perdew85, ZhangYang00, Yang00, LeinKummel05, Mundt05, Cohen12, Yang12, Baerends13, MoriS14, Mosquera14, Mosquera14a, KraislerKronik14, Goerling15, doi:10.1002/qua.26190}. Yet, many common approximate xc functionals lack this important feature; advanced approximations are being developed to reconstruct it (see, e.g., \cite{Tran07, Tran09, Sai11, Zheng11, Refaely11, Kronik_JCTC_review12, Yang12, Yang13, Atalla13, ArmientoKummel13, Refaely13, KraislerKronik13, KraislerKronik14,  Dabo14, Borghi14, Mosquera14, Mosquera14a, Borghi15, Refaely15, KraislerKronik15, KraislerSchmidt15, Nguyen15, Goerling15, LiZhengYang15, Yang16, Atalla16, Tran16, KronikNeaton16, NguyenColonna18, SenjeanFromager18, Gould18, KronikKummel18_review, Gould19, DeurFromager19, AschebrockKummel19, WingKronik19}).

\subsection{Charge-transfer derivative discontinuity} \label{sec:theory.CTDD}

Let us now consider a stretched diatomic molecule $\LR$, where the separation between the atoms is large enough for the energy and density of the molecule to satisfy Eqs.~(\ref{eq:E_L...R}) and~(\ref{eq:n_L...R}). At first, the molecule possesses $N^0_\mathrm{L}$ electrons on Atom L and $N^0_\mathrm{R}$ electrons on Atom R, so the total number of electrons equals $N_\LR^0 = N_\mathrm{L}^0 + N_\mathrm{R}^0$. Next, we allow the total number of electrons to vary continuously: $N_\LR = N_\LR^0 + \alpha$ ($-1 \leqs \alpha \leqs 1$). We consider the specific case for which any additional charge localizes on Atom R, whereas any charge deficiency results in decrease of charge around Atom L. As we show by a direct charge transfer calculation in Sec.~\ref{sec:CT} below, this case is indeed specific but not esoteric -- it is the prototype case for a donor-acceptor pair.

\begin{figure} 
\includegraphics[width=1.0\linewidth]{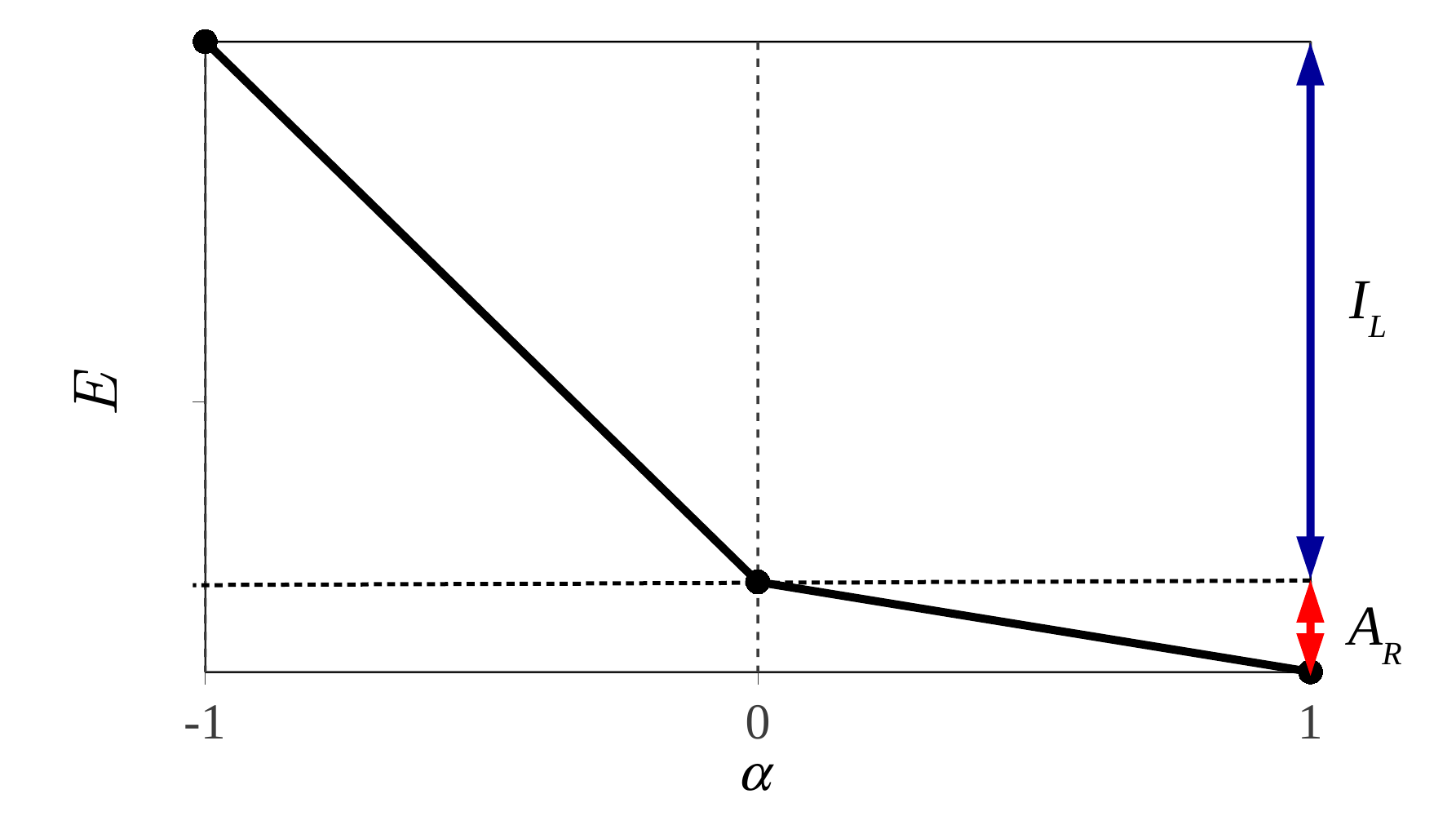} 
\caption{Dependence of the total energy of a stretched diatomic molecule $\LR$ on $\alpha$ -- the deviation of the total number of electrons from its integer value, $N_\LR^0$. The slopes of the graph are associated with the IP of the left atom and the EA of the right atom.} \label{fig:E_LR_alpha}
\end{figure}

Combining Eqs.~(\ref{eq:E_L...R}) and~(\ref{eq:E_piecewise}) we can conclude that the total energy of the molecule is piecewise-linear with the number of electrons (see Fig.~\ref{fig:E_LR_alpha}): 
\begin{equation}\label{eq:E_L...R_of_N}
    E_\LR(\alpha) = \left\{
              \begin{array}{ccc}
                E_\LR^0 - I_\mathrm{L} \cdot \alpha  & : & -1 \leqs \alpha  \leqs 0   \\
                E_\LR^0 - A_\mathrm{R} \cdot \alpha  & : &  0 \leqs \alpha  \leqs 1.   \\
              \end{array}
           \right.
\end{equation}

The chemical potential of the molecule as a whole, being the derivative of its energy with respect to $N_\LR$, or equivalently to $\alpha$, is a stair-step function discontinuous at integers, qualitatively similar to the chemical potential depicted in Fig.~\ref{fig:graphs_finite_system}(middle):
\begin{equation}\label{eq:mu_L...R_of_N}
    \mu_\LR(\alpha) = \left\{
              \begin{array}{ccc}
                - I_\mathrm{L}       & : & -1 < \alpha  \leqs 0   \\
                - A_\mathrm{R}       & : &  0 < \alpha  \leqs 1.   \\
              \end{array}
           \right.
\end{equation}
Notably, here the height of the discontinuity in $\mu_\LR$ is the left-to-right charge-transfer energy, $E^\CT_\mathrm{L \rarr R} = I_\mathrm{L}-A_\mathrm{R}$, namely the energy required to remove one electron from Atom L minus the energy gained by adding an electron to an infinitely distant Atom R. As for the finite system discussed above, the stretched molecule $\LR$ also obeys the IP theorem. Namely, the overall HOMO energy, $\eta^\ho(N_\LR)$, has to equal $\mu_\LR(N_\LR)$. For $N_\LR$ slightly below $N_\LR^0$ the overall ho energy equals $\eta^\ho_\mathrm{L}(N_\LR^{0-})$, which in our case, as explained in Sec.~\ref{sec:theory.S}, equals $-I_\mathrm{L}$. As the overall number of electrons increases above $N_\LR^0$, the overall ho level is localized around Atom R and has to equal $-A_\mathrm{R}$. As a result, the molecular potential $v^\KS_\LR(\bm{r})$ jumps by the constant
\begin{equation} \label{eq:CTDD}
\Delta^\CT_\mathrm{L \rarr R} = I_\mathrm{L} - A_\mathrm{R} - (\eta^\lu_\mathrm{R} - \eta^\ho_\mathrm{L})
\end{equation}
(cf.~Eq.~(\ref{eq:Delta.2})). This quantity was first introduced in Ref.~\onlinecite{HodgsonKraisler17}, where it has been termed \emph{charge-transfer derivative discontinuity}.
$\Delta^\CT_\mathrm{L \rarr R}$ is the difference between the charge-transfer energy, $E^\CT_\mathrm{L \rarr R} = I_\mathrm{L} - A_\mathrm{R}$ and the corresponding quantity in the KS system $(\eta^\lu_\mathrm{R} - \eta^\ho_\mathrm{L})$ (cf.~Eq.~(\ref{eq:Delta.2})).

\begin{figure} 
\includegraphics[width=1.0\linewidth]{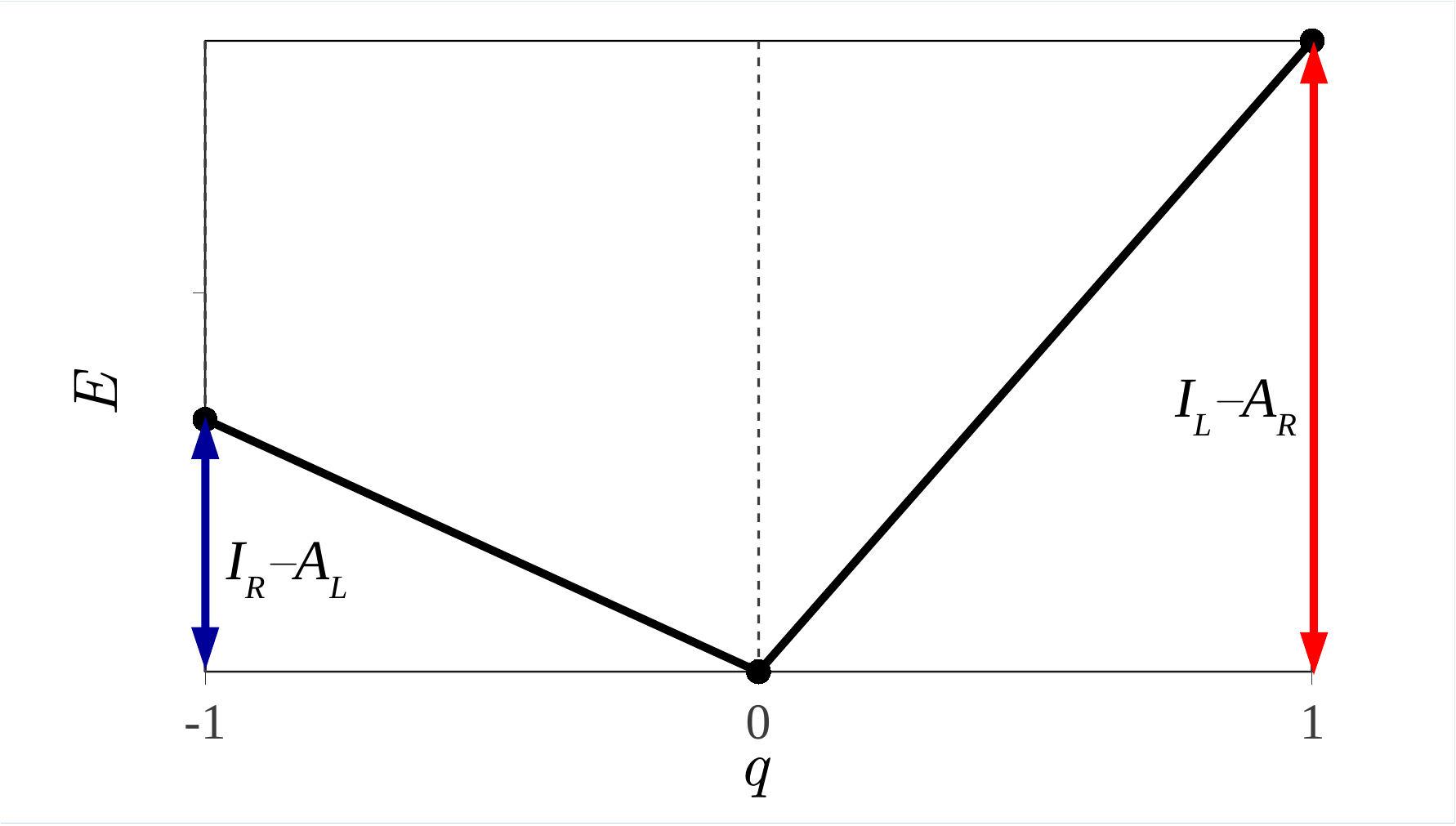} 
\caption{Dependence of the total energy of a stretched diatomic molecule $\LR$ on $q$ -- a fraction of an electron transferred from Atom L to Atom R. The slopes of the graph are associated with the IPs and the EAs of the constituent atoms.} \label{fig:E_LR_q}
\end{figure}

Finally, we consider a stretched but finite diatomic molecule in which the atomic separation is large enough to define individual atoms within the molecule but in which the electrons localized on the left atom experience the Coulomb repulsion of the electron localized on the right atom and vice versa. The total number of electrons within the molecule, $N_\LR^0$, is \textit{constant} and integer. When the molecule is excited a fraction of $q$ electrons is transferred from Atom L to Atom R. We define an ensemble consisting of the ground state, $\left | \Psi_0\right >$, of the molecule and the first excited state $\left | \Psi_1\right >$ where the latter has charge-transfer character, i.e.\ the nature of $\left | \Psi_1\right >$ is such that compared to the ground state, one electron is transferred from Atom L to Atom R. The statistical operator describing this ensemble is give by
\begin{equation} \label{eq:EGSCTES}
\hat{\Gamma} = (1-q) \left | \Psi_0\right > \left < \Psi_0 \right | + q \left | \Psi_1\right > \left < \Psi_1 \right |.
\end{equation}
Both states, $\left | \Psi_0\right >$ and  $\left | \Psi_1\right >$ have fixed (integer) particle number $N_0$. The ensemble expectation value of any operator $\hat{O}$, by virtue of Eq.~(\ref{eq:EGSCTES}), $O = \Tr \{\hat{\Gamma} \hat{O} \}  = (1-q) \left < \Psi_0 \right | \hat{O} \left | \Psi_0\right > + q \left < \Psi_1 \right | \hat{O} \left | \Psi_1\right >$. In particular, the ensemble density is given by 
 \begin{equation}  \label{eq:excited_den_frac}
n(\bm{r};q) = (1-q) \cdot n_0(\bm{r}) + q \cdot n_1(\bm{r}),
\end{equation}
where $n_0(\bm{r})$ and $n_1(\bm{r})$ are the densities of the ground state and the first excited state, respectively. Likewise the total ensemble energy as a function of $q$ equals
\begin{equation} \label{eq:EEGSCTS}
E_\LR (q) = (1-q) E_0 + q E_1 = E_0 + q \left( E_1 - E_0 \right),
\end{equation}
where the subscript 0 corresponds to the ground state, whereas the subscript 1 corresponds to the first excited state. Therefore, $E_1 = E_0 + E_\mathrm{CT}$; for this system with a large but \textit{finite} atomic separation, $E_\mathrm{CT} = \tilde{I}_\mathrm{L}-\tilde{A}_\mathrm{R}$ for $q>0$ and $E_\mathrm{CT} = \tilde{I}_\mathrm{R}-\tilde{A}_\mathrm{L}$ for $q<0$, where $\tilde{I}_\mathrm{L}$ is the ionization energy of the whole molecule which corresponds to an electron localized to the left atom while $\tilde{A}_\mathrm{R}$ is the molecule's affinity and corresponds to the addition of an electron to the right atom \textit{once the electron on the left atom has been ionized} -- this is the nature of a charge-transfer excitation. Consequently, both $\tilde{I}_\mathrm{L}$ and $\tilde{A}_\mathrm{R}$ are influenced by the Coulomb interaction between the left and right atoms; this effect has previously been emitted because the atoms were assumed to be infinitely separated. $\lim_{d \rightarrow \infty}  \tilde{I}_\mathrm{L}-\tilde{A}_\mathrm{R} =  I_\mathrm{L}-A_\mathrm{R}$ (as defined above). By modeling the system with a finite separation we more closely model a real donor-acceptor pair for short- to medium-range charge transfer. The difference between $\tilde{I}_\mathrm{L}-\tilde{A}_\mathrm{R}$ and $I_\mathrm{L}-A_\mathrm{R}$ is the electron-hole electrostatic interaction. For large separation between the donor and acceptor, it is usually approximated as $-1/d$ \cite{PPLB82,dreuw2003long,gritsenko2004asymptotic}.

Plugging this definition for $E_\mathrm{CT}$ in this system into Eq.~(\ref{eq:EEGSCTS}), we obtain
\begin{equation}
E_\LR (q) = E_0 + q \left( \tilde{I}_\mathrm{L} - \tilde{A}_\mathrm{R} \right) \mathrm{for} \ q>0.
\end{equation}
Analogously, for a charge transfer from R to L, we obtain 
\begin{equation}
E_\LR (q) = E_0 - q \left( \tilde{I}_\mathrm{R} - \tilde{A}_\mathrm{L} \right) \mathrm{for} \ q<0.
\end{equation}
Hence the total energy is piecewise-linear with respect to $q$ (see Fig.~\ref{fig:E_LR_q}). Therefore, its derivative, $m(q) = \de E_\LR / \de q$, which is the change in energy as a result of transfer of charge, is a stair-step function:
 \begin{equation}
    m(q) = \left\{
              \begin{array}{ccrcl}
                -\left(\tilde{I}_\mathrm{R} - \tilde{A}_\mathrm{L}\right)  & : & q < 0   \\ [7pt]
                  \tilde{I}_\mathrm{L} - \tilde{A}_\mathrm{R}  & : & q > 0.    \\
              \end{array}
           \right.
\end{equation}
From the Gross-Oliveira-Kohn (GOK) theorem~\cite{GOK1,GOK2,GOK3}, we can express the charge-transfer energy as such
\begin{equation}\label{eq:CT_en_def}
E_\mathrm{CT} = \tilde{I}_\mathrm{L} - \tilde{A}_\mathrm{R} = \lim_{q \rightarrow 0^+} \eta_{N_0+1}^q - \eta_{N_0}^q + \left . \frac{\partial E^q_\mathrm{xc}[n]}{\partial q} \right |_{n = n_q},
\end{equation}
where $\eta_i^q$ is the $i^\mathrm{th}$ KS energy of the ensemble system. As $q \rightarrow 0^+ \ \eta_{N_0+1}^q - \eta_{N_0}^q = \eta^\lu_\mathrm{R} - \eta^\ho_\mathrm{L}$. Therefore, recalling that in the limit of infinite atomic separation Eq.~(\ref{eq:CT_en_def}) is equivalent to Eq.~(\ref{eq:CTDD}), we arrive at an expression for the CTDD for the ensemble system, defined in terms of the derivative of the ensemble xc energy:
\begin{equation}\label{eq:CTDD_en_def}
\Delta^\CT_\mathrm{L \rarr R} = \lim_{q \rightarrow 0^+} \left . \frac{\partial E^q_\mathrm{xc}[n]}{\partial q} \right |_{n = n_q}.
\end{equation}
This expression allows one to calculate the CTDD from any explicit $q$-dependent xc functional~\cite{PhysRevB.95.035120,loos2020weight,fromager2020individual}. In Ref.~\onlinecite{veldman2009energy} $\Delta^\CT_\mathrm{L \rarr R}$ -- as it is defined by Eq.~(\ref{eq:CTDD_en_def}) -- was evaluated experimentally for donor-acceptor pairs.

Note that in the limit that Atom L and Atom R become infinitely separated, $m(q)$ equals the difference between the chemical potentials of the constituent atoms
\begin{equation}
m(q) = \mu_\mathrm{R}(N^0_R+q) - \mu_\mathrm{L}(N^0_L-q),
\end{equation}
with the atomic chemical potentials given by Eq.~(\ref{eq:mu_of_N}).
The discontinuity in $m(q)$ around 0, denoted here $D = \lim_{q \rarr 0^+} m(q) - m(-q)$, equals
\begin{equation}\label{eq:D1}
D = I_\mathrm{L} - A_\mathrm{R} + I_\mathrm{R} - A_\mathrm{L} = E^\CT_\mathrm{L \rarr R} + E^\CT_\mathrm{R \rarr L},
\end{equation}
being the sum of the left-to-right and the right-to-left charge-transfer energies. It can also be expressed as the sum of the atomic fundamental gaps: $D = E_{\mathrm{g},\mathrm{L}} + E_{\mathrm{g},\mathrm{R}}$.
Using Eq.~(\ref{eq:CTDD}), $D$ can be also expressed in terms of the KS quantities:
\begin{equation}\label{eq:D2}
D = (\eta^\lu_\mathrm{R} - \eta^\ho_\mathrm{L}) + (\eta^\lu_\mathrm{L} - \eta^\ho_\mathrm{R}) + \Delta^\CT_\mathrm{L \rarr R} + \Delta^\CT_\mathrm{R \rarr L},
\end{equation}
in direct analogy with results presented above.
$D$ may also be expressed solely in terms of the KS gaps and $\Delta$'s of the constituent atoms using Eq.~(\ref{eq:Delta.2}):
\begin{equation}\label{eq:D3}
D = E^\KS_{\mathrm{g},\mathrm{L}} + E^\KS_{\mathrm{g},\mathrm{R}} + \Delta_\mathrm{L} + \Delta_\mathrm{R}.
\end{equation}
Hence, for this stretched system the derivative discontinuity, $D$, can equally be expressed in the KS system in terms of the derivative discontinuities of the \textit{individual} atoms and also in terms of the charge-transfer derivative discontinuities of the system \textit{as a whole}. We shall see below in Sec.~\ref{sec:CT} that the interatomic step, $S$, derived in Sec.~\ref{sec:theory.S} is related to both the derivative discontinuity of the individual atoms and to the CTDDs.

Finally, we emphasize two additional results. From Eqs.~(\ref{eq:D2}) and~(\ref{eq:D3}) we arrive at the following relation for the CTDDs,
\begin{align} \label{eq:DCT_DL_DR}
\Delta^\CT_\mathrm{L \rarr R} + \Delta^\CT_\mathrm{R \rarr L} = \Delta_\mathrm{L} + \Delta_\mathrm{R},
\end{align}
which shows the close relationship between them to the atomic $\Delta$'s.  
Furthermore, we wish to draw attention to the following relation, which emerges from Eq.~(\ref{eq:D1}):
\begin{align} \label{eq:ECT_Eg}
E^\CT_\mathrm{L \rarr R} + E^\CT_\mathrm{R \rarr L} = E_{\mathrm{g},\mathrm{L}} + E_{\mathrm{g},\mathrm{R}},
\end{align}
meaning that the sum of the left-to-right and the right-to-left charge-transfer energies, between any two distant subsystems, equals the sum of the fundamental gaps of these subsystems.
Details about the implications of the CTDD to the xc potential are provided below in Sections~\ref{sec:relationship}, \ref{sec:CT}, \ref{sec:excited_atom} and~\ref{sec:plateaus.approx}.

\section{Numerical details} \label{sec:NumericalDetails}

We use a 1D model to investigate the structure of the exact KS potential. Our 1D models -- in Secs.~\ref{sec:relationship.Delta}, \ref{sec:CT} and \ref{sec:excited_atom} -- employ the \texttt{iDEA} code~\cite{Hodgson13} in which the exact, fully-correlated many-electron wavefunction may be calculated for an arbitrary external potential. In addition to the ground state, the many-electron excited states are calculated by solving the many-electron Schr\"odinger equation~\cite{PhysRevB.99.161102}. As a result we have access to the exact many-electron ground-state and excited-state electron densities, from which the exact corresponding KS potentials can be calculated by a numerical inversion of the KS equations. Our inversion algorithm to calculate the KS potential is that of Ref.~\cite{Hodgson13}. It can be summarized as follows. Given a target density $n_\textrm{tar}(\rr)$ and an initial guess for the KS potential, $v_\KS^{(0)}(\rr)$, the following iterative procedure is performed: For the $k$-th iteration, a DFT calculation with $v_\KS^{(k)}(\rr)$ is made and the density $n^{(k)}(\rr)$ is obtained. Then, the KS potential for the next iteration is updated, as follows: $v_\KS^{(k+1)}(\rr) = v_\KS^{(k)}(\rr) + \mu [(n^{(k)}(\rr))^p - (n_\textrm{tar}(\rr))^p]$, where $\lambda$ and $p$ are parameters (typically, $\lambda = 0.1$ and $p = 0.05$). The procedure continues up to numerical convergence, which in our case happens when the mean absolute error between the many-electron and KS densities is $<10^{-9}$ Bohr$^{-3}$. More details for this algorithm can be found in Ref.~\onlinecite{Hodgson13}. 

Results for Sec.~\ref{sec:plateaus.approx} were obtained using the \texttt{ORCHID} program~\cite{KraislerMakovKelson10}, version 3.1, on a natural logarithmic radial grid, $r \in [e^c/Z,L]$, with $c=-13$, $L=35$ Bohr and $Z$ being the atomic number. The total energy and the eigenvalues are converged below $10^{-6}$ Hartree. The inversion procedure~\cite{Hodgson13} used the parameters $p=0.1$ and $\mu = 0.72$. The convergence criterion for the inversion procedure is $\ln{\lp(n(r)/n_\textrm{target}(r)\rp)} < 10^{-4}$, enforced for $r \in [e^c/Z,L']$, with $L'=30$ Bohr. Finally, the parameters $a$ and $b$ required for the alignment of the KS potentials, which show the asymptotic behavior of $\sim a/r + b$ (see details in Sec.~\ref{sec:plateaus.approx} and the Supplemental material), have been obtained by a linear fit of the potential vs.\ $1/r$ at 20 and 30 Bohr.

\section{The relationship between $\maybebm{S}$ and $\maybebm{\Delta}$} \label{sec:relationship}

The properties $S$ and $\Delta$ of the exact xc potential discussed in Secs.~\ref{sec:theory.S} and~\ref{sec:theory.Delta}, respectively, have been known for a long time~\cite{PPLB82,Perdew85,NATO85_Perdew,NATO85_AvB,RvL95,Gritsenko96,zhang2000perspective,PhysRevB.62.16364}, but whether these two are completely independent or related properties, remained elusive until recently~\cite{HodgsonKraisler17}. Indeed, $S$ and $\Delta$ are not one and the same: first, they can be derived from two different perspectives, as performed in Sec.~\ref{sec:theory}. Second, the EA and the lu energy, which contribute to $\Delta$ (Eq.~(\ref{eq:Delta.2})), are absent from the expression for $S$ (Eq.~(\ref{eq:S})). Finally, the shift $\Delta$ occurs when varying the charge of the system, whereas $S$ occurs at a fixed, integer number of electrons. However, it was realized early on that both $S$ and $\Delta$ occur for a finite system when the decay rate of the electron density abruptly changes~\cite{NATO85_AvB,NATO85_Perdew}. This suggests a close relationship between the two properties. In the following we characterize this relationship in detail, by formulating and subsequently resolving two paradoxes that arise from the combination of the concepts presented in Secs.~\ref{sec:theory.S} and~\ref{sec:theory.Delta}.

\subsection{Uniform jump paradox} \label{sec:relationship.Delta}

\textit{Paradox 1 -- The spatial uniformity of the jump in the KS potential implies $\Delta=0$.}

In Sec.~\ref{sec:theory.Delta} we described a finite system with a varying number of electrons $N$ and concluded that as $N$ passes an integer the KS potential jumps by a spatially uniform constant $\Delta$. Here we address a finite system again, like in Sec.~\ref{sec:theory.Delta}, but now we are applying the approach from Sec.~\ref{sec:theory.S}. In other words, we find $\Delta$ by examining the exponential decay of the density. 

If the number of electrons in the system equals an integer $N_0$ or a little bit less, the density decay is determined by the IP of the system, i.e., $n(\bm{r};N_0) \propto \exp \lp(-2 \sqrt{2I} |\bm{r}| \rp)$ (denoted $I$-decay). From the KS perspective, the density decay is governed by the ho orbital squared, $|\pphi^\ho(\bm{r})|^2 \propto \exp \lp(-2 \sqrt{-2 \eps^\ho(N_0^-)} |\bm{r}| \rp)$. As the exact KS density equals the many-electron density, $\eps^\ho(N_0^-) = -I$. If the number of electrons is now slightly increased above $N_0$ by a small fraction of an electron, $\alpha$, the density becomes a linear combination of $n(\bm{r};N_0)$ and $n(\bm{r};N_0+1)$, as in Eq.~(\ref{eq:n_piecewise}). The term $n(\bm{r};N_0+1)$ decays $\propto \exp \lp(-2 \sqrt{2A} |\bm{r}| \rp)$ ($A$-decay) which is slower than the decay of $n(\bm{r};N_0)$ because $I>A$ for all known systems (known as the convexity conjecture~\cite{PPLB82,DG,Lieb,Cohen12}). Therefore the $A$-decay asymptotically dominates the density decay. From the KS perspective, the decay of the density is dominated by the now highest, partially occupied orbital (the former lu orbital). The problem arises when taking Fig.~\ref{fig:graphs_finite_system}(bottom) at face value, namely assuming that the KS potential indeed jumps by a completely uniform constant $\Delta$. Then, one may think that the decay of the highest, partially occupied orbital is $\propto \exp \lp(-2 \sqrt{-2 (\eps^\ho(N_0^+)-\Delta)} |\bm{r}| \rp)$, i.e., the decay rate is governed by the ho energy, $\eps^\ho(N_0^+)$, \emph{relative} to the overall potential shift, $\Delta$ (cf.\ Eq.~(\ref{eq:phhi_L_ho})). Recalling that $\eps^\ho(N_0^+) = \eps^\lu(N_0^-) + \Delta$, one may further infer that the density decays $\propto \exp \lp(-2 \sqrt{-2 \eps^\lu(N_0^-)} |\bm{r}| \rp)$. This leads to the paradoxical conclusion that $\eps^\lu(N_0^-) = -A$ and hence $\Delta = 0$. In other words, if the jump $\Delta$ is uniform, its height is zero.

\begin{figure}
  \centering
  \includegraphics[width=1.0\linewidth]{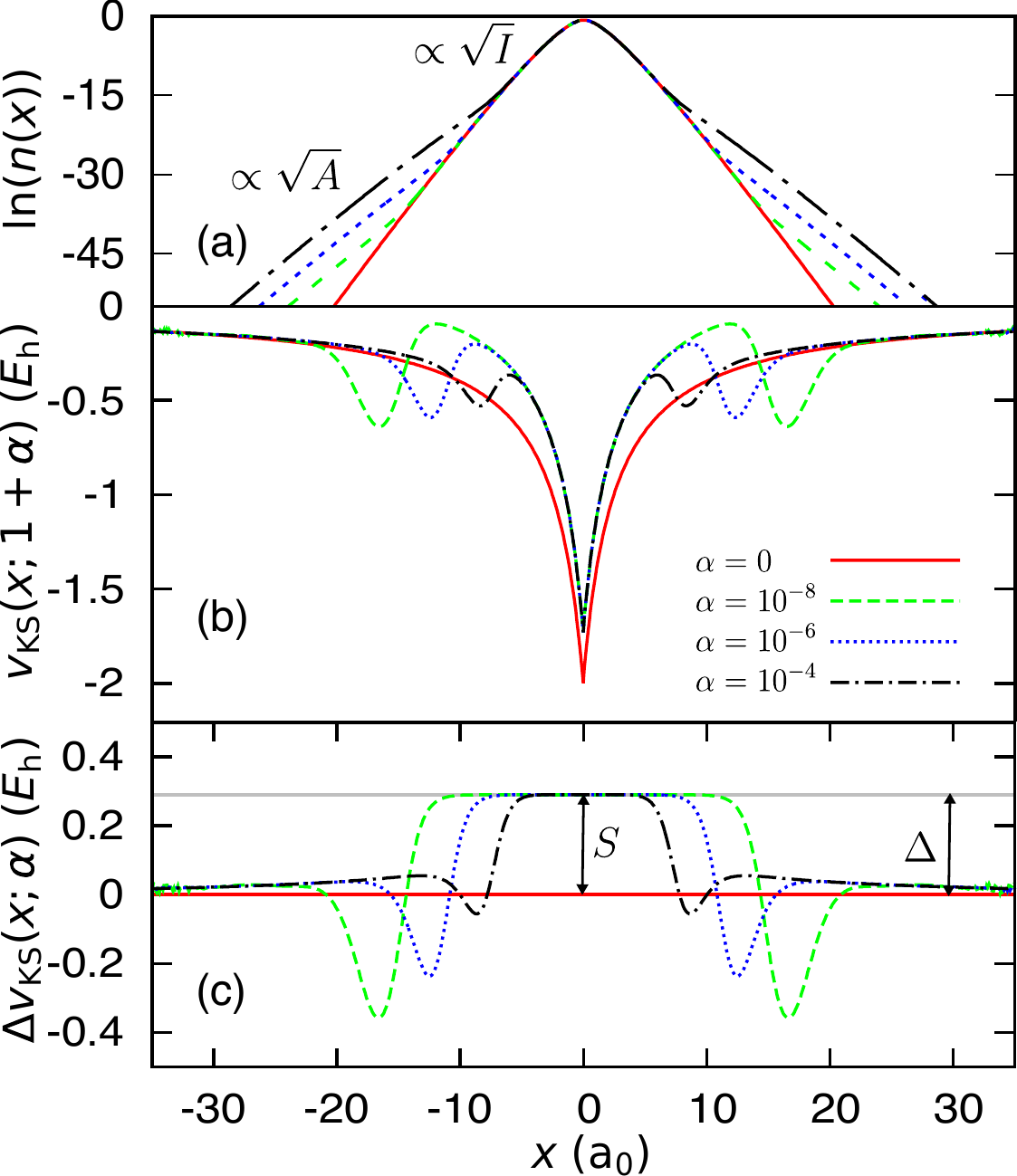}
\caption{
(a) The natural logarithm of the electron densities for an atom consisting of $1+\alpha$ same-spin electrons, for varying values of $\alpha$ (see legend on panel (b) below). For $\alpha > 0$ there are two regions of exponential decay: the $I$- and the $A$-decay regions. The smaller the value of $\alpha$, the further from the atom the change in decay.
(b) The corresponding KS potential for various $\alpha$ (see legend). For $\alpha > 0$, the potential has a plateau comprised of two spatial steps that occur at the points in the density where the decay changes. The plateau elevates the potential around the nucleus by the amount $S$.
(c) The difference between the different KS potentials presented on panel (b) and the KS potential for $\alpha = 0$ (solid red line on (b)). The height of the plateau, $S$, equals the derivative discontinuity $\Delta$ (solid gray line) obtained separately.}
\label{fig:1D_atom}
\end{figure}

To resolve this paradox we look more closely at Eq.~(\ref{eq:n_piecewise}), keeping in mind that in our case $\alpha \rarr 0^+$. Although $n(\bm{r};N_0+1)$ decays slower and is thus the asymptotically dominant term, it is multiplied by the small coefficient, $\alpha$. As a result, we have a competition between the two decay rates: when we reduce $\alpha$ to 0 while looking at a fixed and large $r$, the region in which the $A$-decay is dominant moves away from the nucleus as the term $\alpha n(\bm{r}; N_0+1)$ vanishes and the term $(1-\alpha) n(\bm{r}; N_0)$ prevails. The process is illustrated in Fig.~\ref{fig:1D_atom} for an exactly solved 1D model of an atom with $v_\textrm{ext}(x) = -2.0/(0.4 \cdot \left | x \right |+1)$, with $1+\alpha$ same-spin electrons~\footnote{The spin of the electron affects the gap in a quantitate but not qualitative way~\cite{capelle2010spin}.}. It is useful to look at the natural logarithm of the density in order to clearly see the decay rates, as such a region of an exponential decay appears as a linear line of negative slope. Indeed, in Fig.~\ref{fig:1D_atom}(a) we clearly observe the $I$- and $A$-regions of exponential decay. As $\alpha$ decreases, the $A$-decay region appears further away from the nucleus. Next, recalling our conclusion from Sec.~\ref{sec:theory.S} that a change in the decay rate of the density (no matter what the reason) leads to a step in the KS potential, we indeed find in Fig.~\ref{fig:1D_atom}(b) that for all positive $\alpha$ the KS potential is elevated near the origin, comparing to the $(\alpha=0)$-case, and presents steps far from the origin, at the point where the decay rate changes and hence where the LEIP changes. In Fig.~\ref{fig:1D_atom}(c), subtracting the ($\alpha=0$)-potential from all the potentials of Fig.~\ref{fig:1D_atom}(b), we clearly see a plateau around the origin, in agreement with previous studies (see, e.g., Refs.~\cite{NATO85_AvB,NATO85_Perdew,LeinKummel05,GouldToulouse14,HodgsonKraisler17}).
As $\alpha$ vanishes, the width of the plateau increases, approaching infinity. However, at any finite $\alpha$ the plateau width is finite and asymptotically the KS potential approaches the value of 0 (and not $\Delta$), i.e., the shift for finite $\alpha$ is \textit{not} uniform. This resolves our paradox: the correct decay rate of the density in the region of $A$-decay is $\propto \exp \lp(-2 \sqrt{-2 \eps^\ho(N_0^+)} |\bm{r}| \rp)$, which leads to the conclusion that $\eps^\ho(N_0^+) = \eps^\lu(N_0^-) + \Delta = -A$, as required; whereas in the region of $I$-decay the potential is elevated by $\Delta$. As a result, steps form in the potential as shown in Fig.~\ref{fig:1D_atom}. Thus in this case, for a finite system with varying $N$, the quantities $\Delta$ and $S$ have the following relationship: $\lim_{\alpha \rarr 0^+} S = \Delta$. For the system presented in Fig.~\ref{fig:1D_atom} this has been numerically verified as $\Delta$ was obtained also from total-energy differences.

Finally, we wish to add several comments on plateaus in finite systems. First, the shape of the steps observed includes characteristic dips clearly seen in Fig.~\ref{fig:1D_atom}(c) (cf.~Refs.~\onlinecite{Tempel09,HelbigTokatlyRubio09,GouldToulouse14,HodgsonKraisler17,Aschebrock17b}).
These features are numerically robust, meaning that their magnitude is significantly higher than the numerical error in the inverted potential; their presence in the potential is required to yield the exact KS density. 
Second, the value of the KS potential of a finite system far from its center is an example for an \emph{order-of-limits} problem, namely $\lim_{|\bm{r}| \rarr \infty} \lim_{\alpha \rarr 0^+} v_\KS(\bm{r},N_0+\alpha) = \Delta$, whereas $\lim_{\alpha \rarr 0^+} \lim_{|\bm{r}| \rarr \infty} v_\KS(\bm{r},N_0+\alpha) = 0$. In words, if we examine the value of the KS potential at some finite point $|\bm{r}|$ while continuously decreasing $\alpha$ to zero, for a certain $\alpha$ the plateau will be wide enough to reach $|\bm{r}|$ and elevate the potential there. Taking then $|\bm{r}|$ to infinity will result with the height $\Delta$ for the KS potential. Conversely, taking $|\bm{r}|$ to infinity first while keeping $\alpha$ finite, ensures that for any finite $\alpha$, no matter how small, we will reach the edge of the plateau and the potential value will drop to 0. 

\subsection{Charge transfer paradox} \label{sec:relationship.S}

\textit{Paradox 2 -- The transfer of charge in a diatomic molecule results in a plateau, $\Delta$, around the acceptor atom. Yet, the overall interatomic step height must remain $S$.}

To further explore the relationship between $\Delta$ and $S$ we study the stretched diatomic molecule presented in Sec.~\ref{sec:theory.S}, but now taking into account also the results of Sec.~\ref{sec:theory.Delta}. We consider two scenarios that model charge transfer (cf.~Sec.~\ref{sec:theory.CTDD}): (i) The overall number of electrons in the stretched molecule is increased; the additional charge localizes on one of the atoms, say, Atom R. (ii) When we increase the number of electrons on Atom R, we decrease the number of electrons on Atom L by means of charge-transfer excitation of the molecule so that the overall number of electrons is constant. From the results shown in Fig.~\ref{fig:1D_atom}(c), we would expect a plateau of height $\Delta_\mathrm{R}$ to emerge around the acceptor atom, in our case Atom R (with no significant change around L). But this is contrary to the results of Sec.~\ref{sec:theory.S}: there exists a plateau of height $S$ around Atom R, irrespective of any \textit{infinitesimal} transfer of charge, to ensure the correct distribution of charge in the ground-state KS system\footnote{For a stretched diatomic molecule the transfer of an infinitesimal amount of charge cannot yield a change in the height of the interatomic step because, as shown in Sec.~\ref{sec:relationship.Delta} and \ref{sec:excited_atom}, when the additional or excited charge is infinitesimal the xc potential can only change by an overall constant in the vicinity of the atoms, i.e., the only change to the potential is at the periphery of the system.}. As $S \neq \Delta_\mathrm{R}$, and (thinking of the complimentary scenario of right-to-left charge transfer) $S \neq \Delta_\mathrm{L}$ either, there appears to be a contradiction.

\begin{figure}
  \centering
  \includegraphics[width=1.0\linewidth]{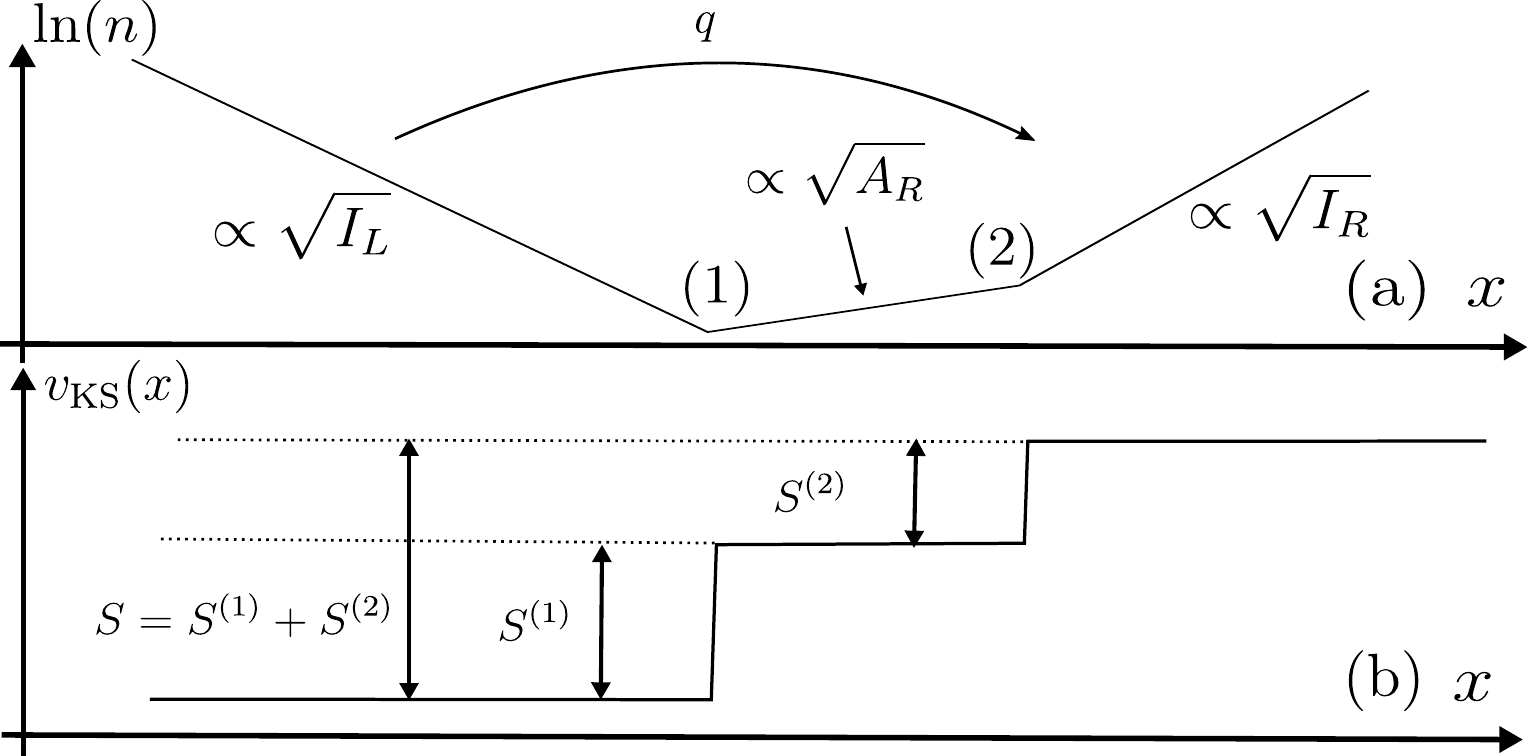}
\caption{(a) A diagram of $\ln{(n)}$ far from, and in between, the atoms of a molecule $\LR$. Three regions of density decay are present: $I_\mathrm{R}$-, $A_\mathrm{R}$- and $I_\mathrm{L}$-regions. Transition from the $I_\mathrm{R}$- to the $A_\mathrm{R}$-region occurs at point (2) and from the $A_\mathrm{R}$- to the $I_\mathrm{L}$-region at point (1). The changes in the density give rise to two steps in the KS potential (b).}
\label{fig:sketch_ln_n}
\end{figure}

To resolve this paradox, we refer again to the density of the system. For both Cases (i) and (ii) the natural logarithm of the density in between the two atoms is sketched in Fig.~\ref{fig:sketch_ln_n}(a). We expect \emph{three} regions of exponential decay between the atoms: going from right to left, the density decay is first governed by $I_\mathrm{R}$ and then by $A_\mathrm{R}$ (changing at point (2); cf.~Fig.~\ref{fig:1D_atom}(a)), due to the extra charge on Atom R. Then, the $A_\mathrm{R}$-decay meets the $I_\mathrm{L}$-decay at point (1), simply due to the fact that the two atoms form one molecule. As a result, we expect not one, but \emph{two} steps in the KS potential between the atoms in this diatomic molecule (Fig.~\ref{fig:sketch_ln_n}(b)). The height of the steps can be deduced analytically~\cite{HodgsonKraisler17}, similarly to the derivation of Eq.~(\ref{eq:S}): the step $S^{(2)}$, which depends solely on quantities related to Atom R, equals $\Delta_\mathrm{R}$, whereas the step $S^{(1)}$ equals $-\Delta^\CT_\mathrm{L \rarr R}$. Importantly, the steps $S^{(1)}$ and $S^{(2)}$ combine to yield the overall step $S$ of Eq.~(\ref{eq:S}). 
This resolves the paradox raised above: indeed, a plateau of height $\Delta_\mathrm{R}$ is expected to form on the receiving Atom R upon charge transfer or addition. But in conjunction, in the region of Atom L, the KS potential shifts when the `local electron number' \textit{decreases} below an integer. The combination of these two plateaus yields an overall interatomic step of height $S$.

The internal structure of the step $S$ in Case (i) has been illustrated and extensively discussed in Ref.~\onlinecite{HodgsonKraisler17}. The two steps, $S^{(1)}$ and $S^{(2)}$, have been identified both in a 1D model of a stretched diatomic molecule and in a 3D (Li $\cdots$ Be)$^{3+}$ ion. Case (ii) is numerically illustrated in Sec.~\ref{sec:CT} below for a charge transfer in a stretched 1D diatomic molecule induced by exciting the system.  

\section{Charge transfer in a diatomic molecule} \label{sec:CT}

Simulation of a charge transfer process, and particularly obtaining the \emph{exact} KS potential that describes the process is by no means a trivial task~\cite{gould2018charge}. To this end it is necessary to exactly obtain not only the ground state of the system, but also its first excited state that corresponds to a charge transfer. 

\begin{figure}[htbp]
  \centering
  \includegraphics[width=1.0\linewidth]{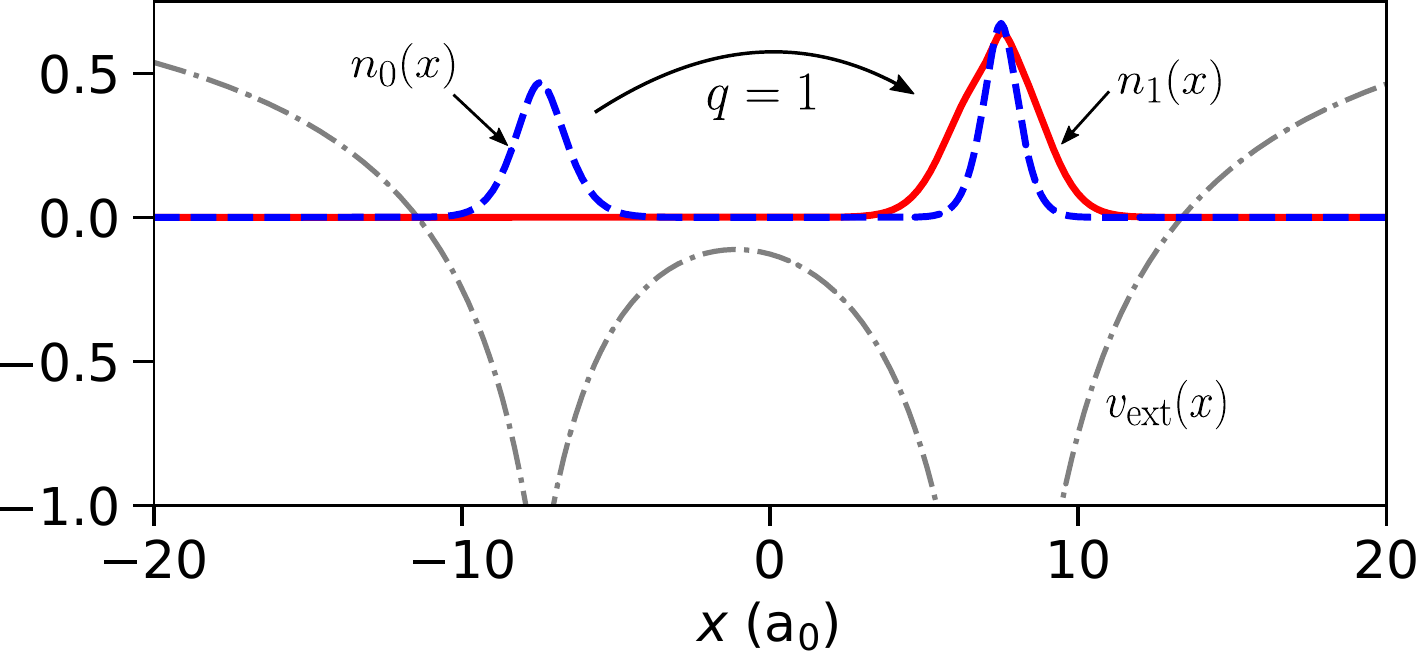}
\caption{Charge-transfer in a $2$-electron system: The external potential, $v_\mathrm{ext}(x)$ consists of two separated atom-like wells (dashed-dotted gray). The exact ground-state density $n_0(x)$ (dashed blue) with one electron on each atom. The exact first excited-state density $n_1(x)$ (solid red) with both electrons localized on Atom R. } 
\label{fig:CT_MB}
\end{figure}

In this section we present a prototypical 1D stretched diatomic molecule $\LR$, which we excite in order to transfer charge from Atom L to Atom R. Our system consists of an \textit{integer} number of \textit{same-spin} electrons, in this case $N^0_\LR = 2$. Figure~\ref{fig:CT_MB} illustrates the charge-transfer process: the external potential, $v_\mathrm{ext}(x) = -4/(0.6 \cdot |x-7.5|+1) -2/(0.4 \cdot |x+7.5|+1)$ is asymmetric, chosen such that the ground-state electron density corresponds to a system with one electron localized on Atom L and one electron on Atom R, whereas in the first excited state both electrons are localized on Atom R. Hence, by exciting this system we can initiate a transfer of charge from L to R. We first find the exact many-electron ground-state density $n_0(x)$ and the first excited-state density $n_1(x)$. Then we construct an ensemble electron density, which corresponds to a transfer of a fraction of $q$ electrons from left to right by a linear combination of the ground-state and excited-state densities, given by Eq.~(\ref{eq:excited_den_frac}) where $0 \leqs q \leqs 1$~\cite{gould2018charge}. We emphasize that \emph{all} the densities present in Eq.~(\ref{eq:excited_den_frac}) integrate to an integer number of electrons.

The GOK theorem ensures a one-to-one mapping between the density and the local potential for this excited system, provided that $0 \leqs q \leqslant 0.5$. Hence there exists a KS system, which exactly reproduces the electron density of Eq.~(\ref{eq:excited_den_frac}), and thus we can obtain this KS potential from the density $n(x;q)$ by numerical inversion (Sec.~\ref{sec:NumericalDetails}). In our case, where $N^0_\LR = 2$, the density is given in terms of the KS orbitals by
~\footnote{Note our system consists of same-spin electrons and hence each electron occupies a distinct KS orbital.}
$n(x;q) = \left | \phi_0(x) \right |^2 + (1-q) \cdot \left | \phi_1(x) \right |^2 + q \cdot \left | \phi_2(x) \right |^2$. When the system is excited, a fraction of the electron ($q$) initially occupying the first excited KS orbital is transferred into the second excited KS orbital localized in our case on Atom R, while the overall number of electrons stays constant and integer; in this sense this type of excitation is uncharged (the number of electrons within the overall system is unchanged) but in the vicinity of each atom, this excitation corresponds to a charged excitation (the number of electrons changes locally). This observation may explain why approximate KS theories, such as linear response time-dependent DFT (TDDFT), struggle to accurately describe charge transfer~\cite{Hofmann12,Fuks14,Maitra_2017}.

The exact ensemble xc potential for charge transfer was first studied by Pribram-Jones \textit{et al.}~\cite{pribram2014excitations}. The authors modelled a spin singlet which in its ground state consisted of two electrons localized to one potential well; the first excited state corresponded to an electron localized each to a distinct potential well. The authors found an interatomic step in the exact xc potential upon charge transfer, the overall height of which acted to align the chemical potentials of the two wells~\cite{pribram2014excitations}. However, Pribram-Jones \textit{et al.} did not observe a plateau which corresponds to $\Delta$ localized to the acceptor as in their ground state both electrons were localized to the donor and hence initially no electrons were localized to the acceptor. Our model charge-transfer system consists of one electron localized to the donor and another (same-spin) electron localized to the acceptor in the ground state. Therefore our donor-acceptor is more general in character and hence upon excitation we expect to observe the double step structure, one which corresponds to $\Delta$ for the acceptor atom and one to the CTDD, as described in Sec.~\ref{sec:relationship.S}. 

\begin{figure}[htbp]
  \centering
  \includegraphics[width=1.0\linewidth]{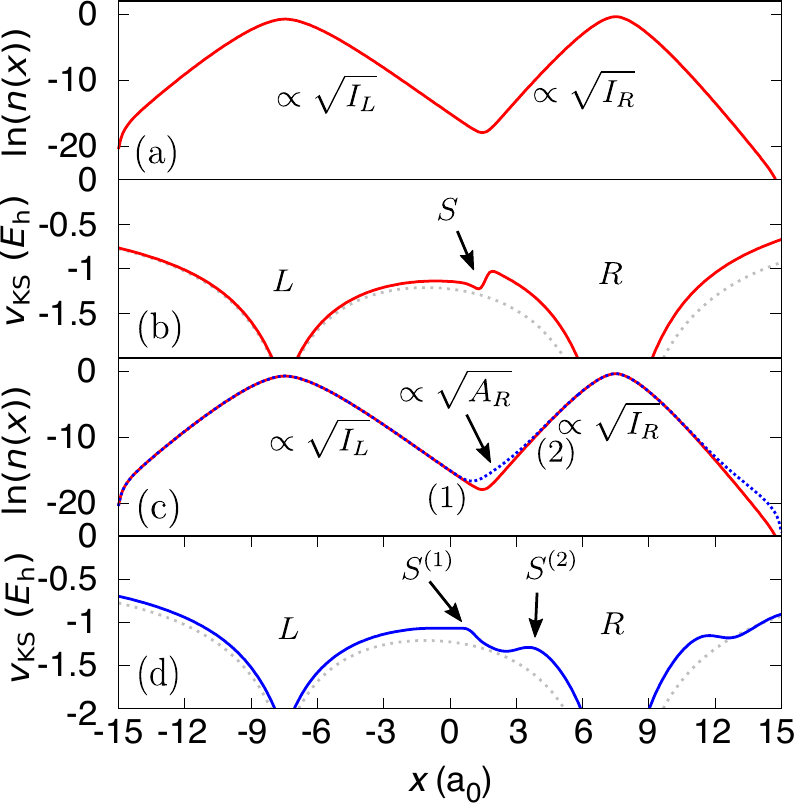}
\caption{
(a) The natural logarithm of the ground-state density (solid red). 
(b) The exact KS potential corresponding to (a) (solid red). The step in the potential occurs at the point where the decay of the density changes. The external potential is shown (dotted gray) -- also in (d).
(c) The natural logarithm of the partially excited density corresponding to $q = 5 \times 10^{-4}$ (dashed blue) and the natural logarithm of the ground-state density for reference (solid red). Three regions of decay of the excited density: $I_\mathrm{L}$-, $A_\mathrm{R}$- and $I_\mathrm{R}$-decay regions are apparent. At the interface between these regions of the decay the density decay rate changes suddenly (points (1) and (2)). 
(d) The exact KS potential corresponding to the excited density (solid blue). Two plateaus are present: one corresponding to the derivative discontinuity of the right atom, $\Delta_\mathrm{R}$, the other corresponds to the CTDD, $\Delta^\CT_\mathrm{L \rarr R}$. These steps combine to give an overall step whose height is given by Eq.~(\ref{eq:S}).
}
\label{fig:CT}
\end{figure}

Figure~\ref{fig:CT}(a) shows the natural logarithm of the exact \textit{ground-state} electron density, $\ln{(n_0(x))}$, for our diatomic molecule: each electron occupies its own potential well, and far from the well the density decays exponentially. There are two regions of decay between the atoms -- the $I_\mathrm{L}$- and the $I_\mathrm{R}$-decay -- and hence one step at the point where the decay of the density changes yielding a change in the LEIP; see Sec.~\ref{sec:theory.S}. The height of this step is given by Eq.~(\ref{eq:S}). Figure~\ref{fig:CT}(b) shows the KS potential corresponding to this ground-state density. The potential has an interatomic step which acts to localize one electron on each atom in the KS system, as required. Another step of height $-S$ is expected far to the right of Atom R, when the $I_\mathrm{L}$-decay will prevail over the $I_\mathrm{R}$-decay (not shown on the figure). Both steps together form a plateau of height $S$ around Atom R. Figure~\ref{fig:CT}(c) shows $\ln{(n(x;q))}$, the natural logarithm of the exact excited many-electron density, given by Eq.~(\ref{eq:excited_den_frac}), with $q = 5 \times 10^{-4}$. For reference, $\ln{(n_0(x))}$ is also shown. There are now three regions of exponential decay in the density $n(x;q)$: the $I_\mathrm{L}$-, $A_\mathrm{R}$- and the $I_\mathrm{R}$-decay, as we expected (cf.\ Sec.~\ref{sec:relationship.S}). These three regions of decay give rise to two steps in the corresponding exact KS potential, at the points in the density where the decay rate changes. Figure~\ref{fig:CT}(d) shows the corresponding exact KS potential of our \textit{excited} system with the two steps apparent, $S^{(1)}$ and $S^{(2)}$ (arrows). The right (acceptor) atom experiences the jump in the KS potential characteristic of the derivative discontinuity, i.e., $S^{(2)} = \Delta_\mathrm{R}$, owing to the local number of electrons of Atom R surpassing an integer by a small amount ($q$). Simultaneously a plateau forms in the vicinity of the left donor atom. The height of the plateau is $\Delta^\CT_\mathrm{L \rightarrow R}$, i.e., the CTDD associated with transferring an electron from left to right atom (Eq.~(\ref{eq:CTDD})). $S^{(1)}$ is therefore equal to $-\Delta^\CT_\mathrm{L \rightarrow R}$ (the minus sign describes the fact that $S^{(1)}$ is a step \emph{down} between the atoms; whereas $\Delta_\mathrm{R}$ is a step up). The sum of the two steps equals the overall step of Eq.~(\ref{eq:S}).

Figure~\ref{fig:CT} is notably similar to Fig.~2 in Ref.~\onlinecite{HodgsonKraisler17}, where the same 1D diatomic molecule is modeled but for a system with a \textit{fractional} number of electrons $N_\LR = 2.0005$ in the \textit{ground state}. This means that the approach chosen in Ref.~\onlinecite{HodgsonKraisler17} to reveal the internal structure of the interatomic step $S$ and find the CTDD, employing calculation which are much cheaper numerically, is appropriate. Therefore, there is reason to assume that modeling of full charge transfer for 3D systems, as the one analyzed in Ref.~\onlinecite{HodgsonKraisler17} and others mentioned in Sec.~\ref{sec:relationship.S}, will also yield extremely similar results to those already obtained by varying the total number of electrons.

To summarize, simulation of a charge transfer by means of excitation of a 1D diatomic molecule showed that the interatomic step 
\begin{equation} \label{eq:internal_str_S}
S = \Delta_R-\Delta^\CT_\mathrm{L \rightarrow R},
\end{equation}
hence it has an internal structure, as expected: it consists of the $\Delta$ of the acceptor atom, in our case Atom R, and the (negative of the) relevant CTDD, $\Delta^\CT_\mathrm{L \rarr R}$. If charge is transferred from right to left a similar picture is expected: the overall step will split as $S = -\Delta_\mathrm{L} + \Delta^\CT_\mathrm{R \rarr L}$ (cf.\ Eq.~(\ref{eq:DCT_DL_DR})). Therefore, also in the case of a stretched diatomic molecule the relationship between the interatomic step $S$ and the $\Delta$'s of the constituent atoms is established (Eq.~(\ref{eq:internal_str_S})) via the CTDD (Eq.~(\ref{eq:CTDD})).

\section{Discontinuities in excited finite systems with integer electron number} \label{sec:excited_atom}

In Sec.~\ref{sec:CT} we demonstrated that derivative discontinuities arise upon excitation of a stretched system, which induces charge transfer. But what happens to a \emph{finite} (and not stretched) system, upon excitation from its ground to first excited state, not necessarily related to a transfer of charge? Shall we expect steps in the potential also in this case? To explore this question we model a single atom with an \emph{integer} $N$ in its ground and excited states, to find whether its KS potential forms any plateaus upon excitation. This concept was first proposed by Levy~\cite{PhysRevA.52.R4313} and was analyzed numerically by Yang \textit{et al.}~\cite{YangZH14}. Below we study how the change in the exact KS potential of the excited ensemble state varies with the ensemble weight, $\beta$, which allows us to compare this scenario with those studied above. 

We model a single atom in 1D with the external potential $v_\mathrm{ext}(x) = -2.0/(0.4\left |x \right |+1)$ with $N_0 = 2$ (again, same-spin electrons). We calculate the exact ground-state and the first excited-state density. We then find the ensemble electron density employing the 1D version of Eq.~(\ref{eq:excited_den_frac}), where $q = \beta$ in this case, for $\beta = 10^{-4}, 10^{-3}$ and $10^{-2}$, and invert the KS equations to find the corresponding exact KS potential associated with each density. 

\begin{figure}[htbp]
  \centering
  \includegraphics[width=1.0\linewidth]{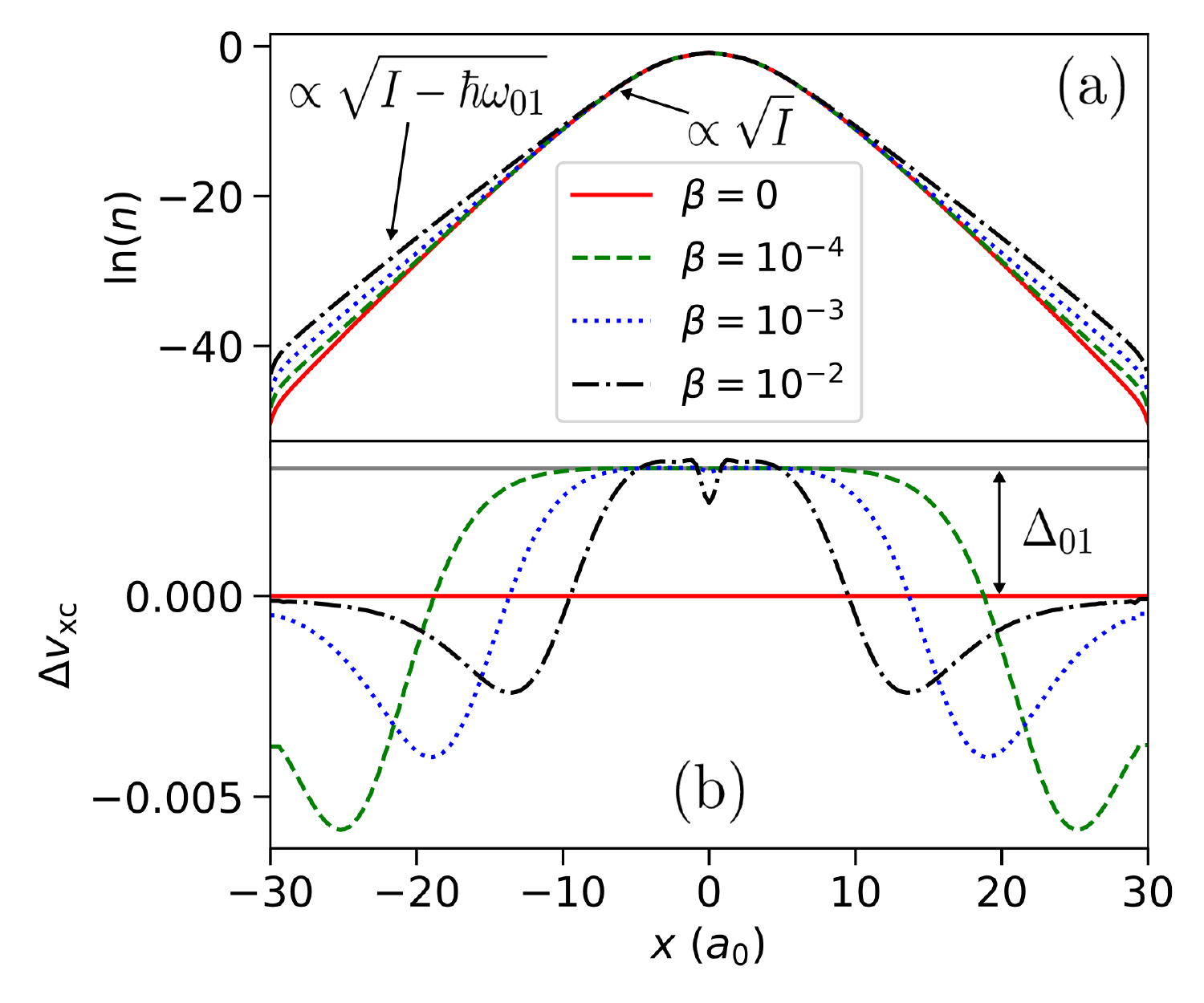}
\caption{(a) The natural logarithm of the excited density for varying values of $\beta$ and the natural logarithm of the ground-state density (for reference). The excited density has two regions of exponential decay: $I$-decay and ($I-\hbar \omega_{01}$)-decay. (b) The difference between the exact excited state xc potential and the exact ground-state xc potential; $\Delta v_\mathrm{xc} =  v^{01}_\mathrm{xc}(x)-v_\mathrm{xc}(x)$. At the point where the exponential decay rate in the density changes , sharp steps form in the exact KS potential. The resulting plateau raises the level of the KS potential in the central region by $\Delta_{01}$ (see text).
}
\label{fig:excited_atom}
\end{figure}

Figure~\ref{fig:excited_atom}(a) shows the natural logarithm of the electron density for the ground state ($\beta = 0$) and for the ensemble system with $\beta = 10^{-4}, 10^{-3}$ and $10^{-2}$. The excited density has two regions of decay in each case: closer to the origin the $I$-decay region is present, but then the decay rate changes and the density decays slower. The rate of decay of this excited density is determined by $I - \hbar \omega_{0 1}$, where $\hbar \omega_{0 1}$ is the energy required to excite the many-electron system from the ground to the first excited state. Due to this change in the density decay rate we expect steps in the potential of the corresponding KS system. 

The steps are clearly seen in Fig.~\ref{fig:excited_atom}(b), which shows $\Delta v_\mathrm{xc}$ -- the difference between the xc potential of the excited system and the ground-state xc potential. In the central region of the system the excited KS potential is elevated by a plateau of height $\Delta_{01}$. The height of the plateau can be analytically deduced, as before: $\Delta_{01} =  I -\left( I-\hbar \omega_{01} \right) - (\eps^\beta_{N_0+1} - \eps^\beta_{N_0})$, where $\eps^\beta_i$ is the $i^\mathrm{th}$ KS energy of the ensemble system. As $\beta \rightarrow 0^+$, $\eps^\beta_{N_0+1} - \eps^\beta_{N_0} = \eps^\lu - \eps^\ho = \hbar \omega^\mathrm{KS}_\mathrm{og}$, is the energy required to excite a KS electron from the ho to the lu KS orbital. $\hbar \omega_{01} = \hbar \omega_\mathrm{og}$, the many-electron optical gap. Thus, $\Delta_{0 1} = \Delta_\mathrm{og}$ and 
\begin{equation}
\hbar \omega_\mathrm{og} = \hbar \omega^\mathrm{KS}_\mathrm{og} + \Delta_\mathrm{og}.
\end{equation}
This equation is the same result found by Levy~\cite{PhysRevA.52.R4313} and Yang \textit{et al.}~\cite{YangZH14}. It is a time-independent way of calculating exact excitation energies~\cite{PhysRevB.95.035120}, similar to the calculation of the fundamental gap (discussed above)~\cite{senjean2018unified}. 

We find that $\Delta_\mathrm{og}$ is always relatively small, below 0.03 Hartree ($<1$ eV), for different 1D atoms with a slightly less or more confining external potential, e.g., $v_\mathrm{ext}(x) = -8.0/(\left |x \right |+1)$, with $N=2$ -- this implies that exciting one KS electron for this system is indeed a good model for the many-electron excitation of the two-electron system. Hence, for this system, $\hbar \omega^\mathrm{KS}_\mathrm{og} \approx \hbar \omega_\mathrm{og}$ which implies that as long as $\Delta_\mathrm{og}$ is small, the ground-state KS energy levels are reasonably good approximations to the many-electron excitation energies in their own right, i.e., neglecting the contribution of the Hartree-xc (Hxc) kernel within TDDFT, which has been observed by others~\cite{PhysRevA.54.3912,umrigar1998unoccupied,Baerends13,PhysRevB.99.161102}. For more strongly correlated systems, or indeed the charge-transfer system above, this is not the case, and the role of the Hxc kernel or the corresponding $\Delta$ becomes crucial \cite{Hellgren12_PRA,hellgren2013discontinuous,PhysRevB.101.115109}.

From the analysis in the sections above, we conclude that any electron donor experiences a discontinuous shift in its xc potential despite the local number of electrons \textit{decreasing} below an integer. This discontinuity emerges because a truly isolated system with a fractional number of electrons cannot exist in reality; there must be a source of electrons, e.g., an electron reservoir (the donor), with which a finite system, like an atom or molecule, can exchange electrons (the acceptor). Imagine that the chemical potential of the reservoir is adjusted such that an infinitesimal amount of charge is transferred to the finite system. The xc potential of the system as a whole (reservoir plus the finite system) experiences a uniform shift of height $\Delta_\mathrm{CT}$, which is the CTDD associated with transferring an electron from the reservoir to the finite system; see Sec.~\ref{sec:theory.CTDD}. This shift in the potential is truly uniform as it manifests as a result of an excitation experienced by the whole system, like the atom in this section 

As the amount of charge transferred from the reservoir is steadily increased, a plateau localizes in the vicinity of the acceptor which is associated with the derivative discontinuity of that finite system, $\Delta$. In conjunction, the shift in the xc potential associated with the CTDD localizes to the donor. This occurs for the diatomic molecule of Fig.~\ref{fig:CT}; in this case the donor atom acts as the electron reservoir. The charge-transfer derivative discontinuity, $\Delta^\CT_\mathrm{L \rightarrow R}$, manifests as a uniform shift in the xc potential of the donor-acceptor when the transferred (excited) charge is infinitesimal. As the amount of charge is increased a plateau of height $\Delta_\mathrm{R}$ localizes to the acceptor atom which in the vicinity of just the acceptor looks to be uniform -- $\left | S^{(2)} \right | = \Delta_\mathrm{R}$ in Fig.~\ref{fig:CT}. In conjunction, a complementary plateau forms around the donor atom of height $\Delta^\CT_\mathrm{L \rightarrow R}$ because the donor and acceptor form \textit{one system} -- $\left | S^{(1)} \right | = \Delta^\CT_\mathrm{L \rightarrow R}$. Consequently, the shift to the xc potential associated with the derivative discontinuity of the finite system when the local number of electron increases above an integer, $\Delta$, can never be truly uniform.

\section{Plateaus in approximate xc potentials}  \label{sec:plateaus.approx}

So far we have addressed \emph{exact} many-electron densities and the corresponding exact KS potentials obtained from the densities by means of numerical inversion. But what happens when working within one of the common \emph{approximations} to the xc functional, like the local density approximation (LDA) or a generalized gradient approximation (GGA)? Does the resultant approximate KS potential possess any steps or form any plateaus in the various scenarios discussed above?

The immediate answer to this question is negative. It is well-known that if one addresses a finite system with a varying number of electrons, $N = N_0 + \alpha$, with, e.g., the LDA in its standard implementation (i.e.\ constructing the density for fractional $N$ by occupying the last KS level with $\alpha$ electrons), one obtains a gradually changing xc potential, without any plateau of the sort presented in Fig.~\ref{fig:1D_atom}(c). 

However, in the spirit of the present work, it is possible to obtain the KS potential for fractional $N$, relying on LDA densities, also in a different way: First, one solves the system self-consistently for $N_0$ and separately for $N_0+1$ electrons, within a given xc approximation. Second, one creates the ensemble density, $n(\bm{r};N)$, using Eq.~(\ref{eq:n_piecewise}), thus assuring piecewise-linearity of the density. Third, one obtains the KS potential, up to a constant, via numerical inversion of the ensemble density. 

\begin{figure}
  \centering
  \includegraphics[width=1.0\linewidth,trim={0mm 7mm 0mm 0mm}]{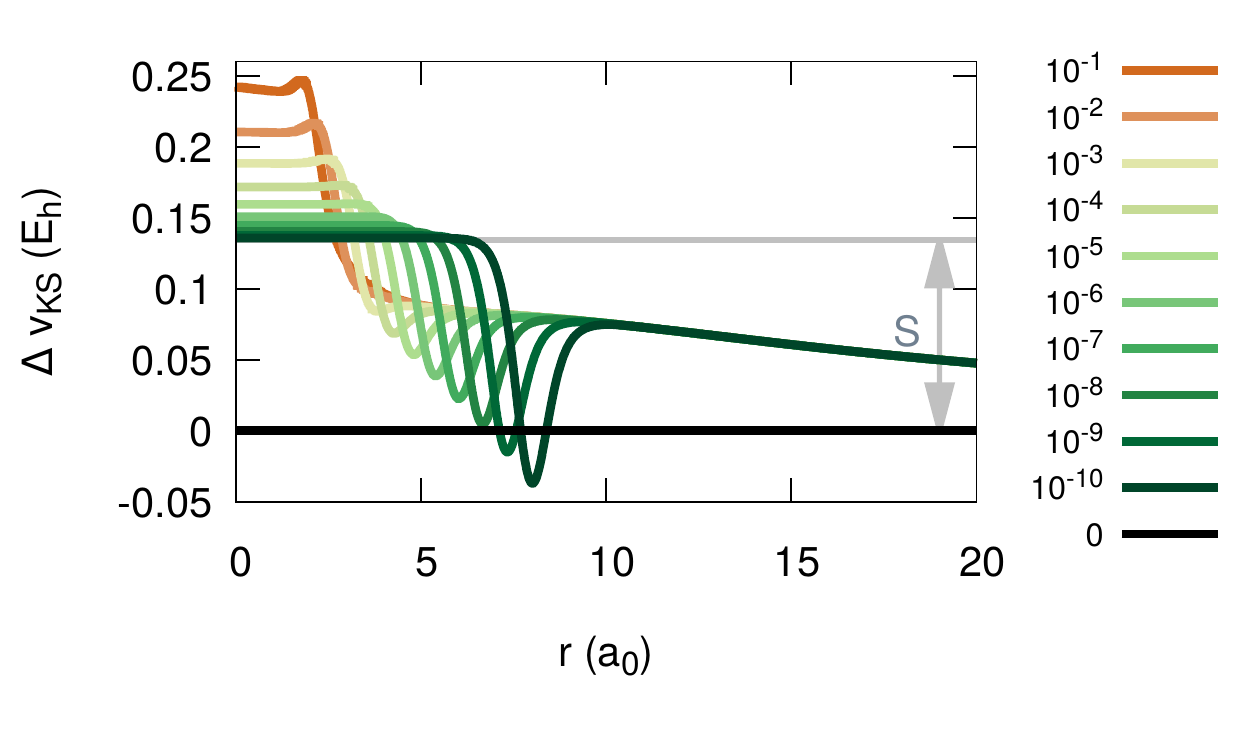}
\caption{Difference between the inverted LDA (invLDA) KS potential for Li with $2+\alpha$ electrons and the KS potential for 2 electrons, for various values of $\alpha$ (see legend). As $\alpha \rarr 0^+$, a plateau of height $S$ is formed around the origin.}
\label{fig:plateau_Li_LDA}
\end{figure}

We obtained this `inverted LDA' (invLDA)  potential for the Li ion ($N_0 = 2$) for varying $\alpha$. Remarkably, the potentials show a clear asymptotic behavior of $\sim a/r + b$ far from the nucleus (with $a$, $b$ being $\alpha$-dependent parameters), rather than the exponential decay of the standard LDA. This allows us to align each potential such that it decays to 0 (and not to some finite constant, $b$) and subsequently subtract from it the KS potential for $N=2$. The resultant differences are shown in Fig.~\ref{fig:plateau_Li_LDA}. 
We can clearly see that the invLDA KS potential \emph{does form a plateau} of height $S_\LDA=0.134$ Hartree in the vicinity of the nucleus. As $\alpha \rarr 0^+$, the height of the plateau converges and its width logarithmically approaches infinity (cf.~\cite{NATO85_Perdew_p284-286,HodgsonKraisler17})

\begin{figure}
  \centering
  \includegraphics[width=1.0\linewidth]{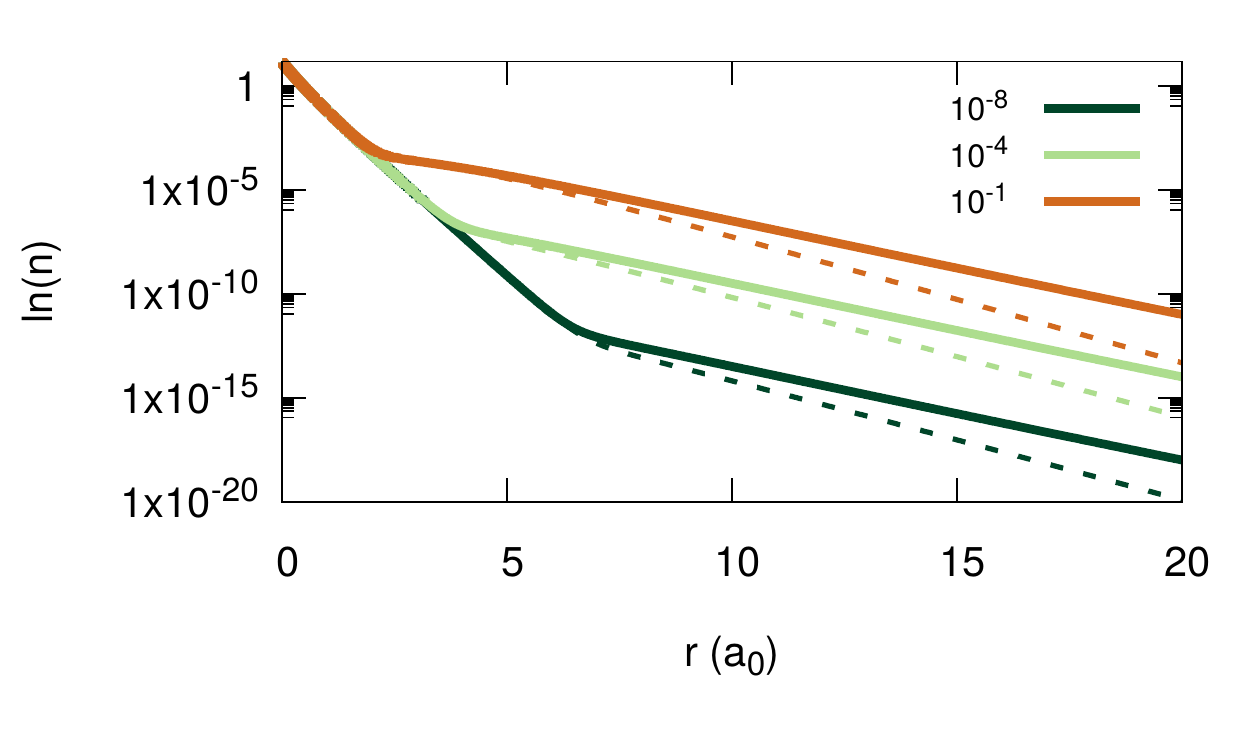}
\caption{Natural natural logarithm of the density for Li with $2+\alpha$ electrons, for various values of $\alpha$ (see legend). Dashed lines correspond to densities obtained within the LDA in its standard implementation, whereas solid lines correspond to densities obtained as an ensemble linear combination, using Eq.~(\ref{eq:n_piecewise}), relying on LDA densities for 2 and 3 electrons.}
\label{fig:n_Li_LDA}
\end{figure}

A qualitative understanding of the emergence of plateaus in the invLDA can be gained by looking at the density decay rates, presented in Fig.~\ref{fig:n_Li_LDA}. Surely, the decay rate of the ensemble densities obtained via Eq.~(\ref{eq:n_piecewise}) is slower than the decay rate of the density obtained from a standard LDA calculation with fractional occupations. Then, clearly, whereas the change in the decay rates of the latter yields a plateau of height zero, a density with a slower decay will yield a non-zero plateau.

Next, we establish the quantitative relationship between $S_\LDA$ found with invLDA and $\Delta_\LDA$ for Li$^+$ obtained from KS-LDA quantities. For Li$^+$ with LDA, $I_\LDA = 2.8712$ Hartree and $A_\LDA = 0.1924$ Hartree (calculated from total energies of Li, Li$^+$ and Li$^{++}$), $\eps^\ho_\LDA = -2.1899$ Hartree and $\eps^\lu_\LDA = -0.2399$ Hartree. Hence, according to Eq.~(\ref{eq:Delta.2}), $\Delta_\LDA = 0.7288$ Hartree. Alternatively, $\Delta'_\LDA = -A_\LDA -\eps^\lu_\LDA = 0.0475$ Hartree. For the exact xc functional, $\Delta'=\Delta$, but for an approximate one, like the LDA, the above equality is not necessarily true, because the IP theorem is not obeyed. In any case, neither $\Delta_\LDA$ nor $\Delta'_\LDA$ seem to equal $S_\LDA$.

\begin{figure}
  \centering
  \includegraphics[width=1.0\linewidth,trim={0mm 7mm 0mm 0mm}]{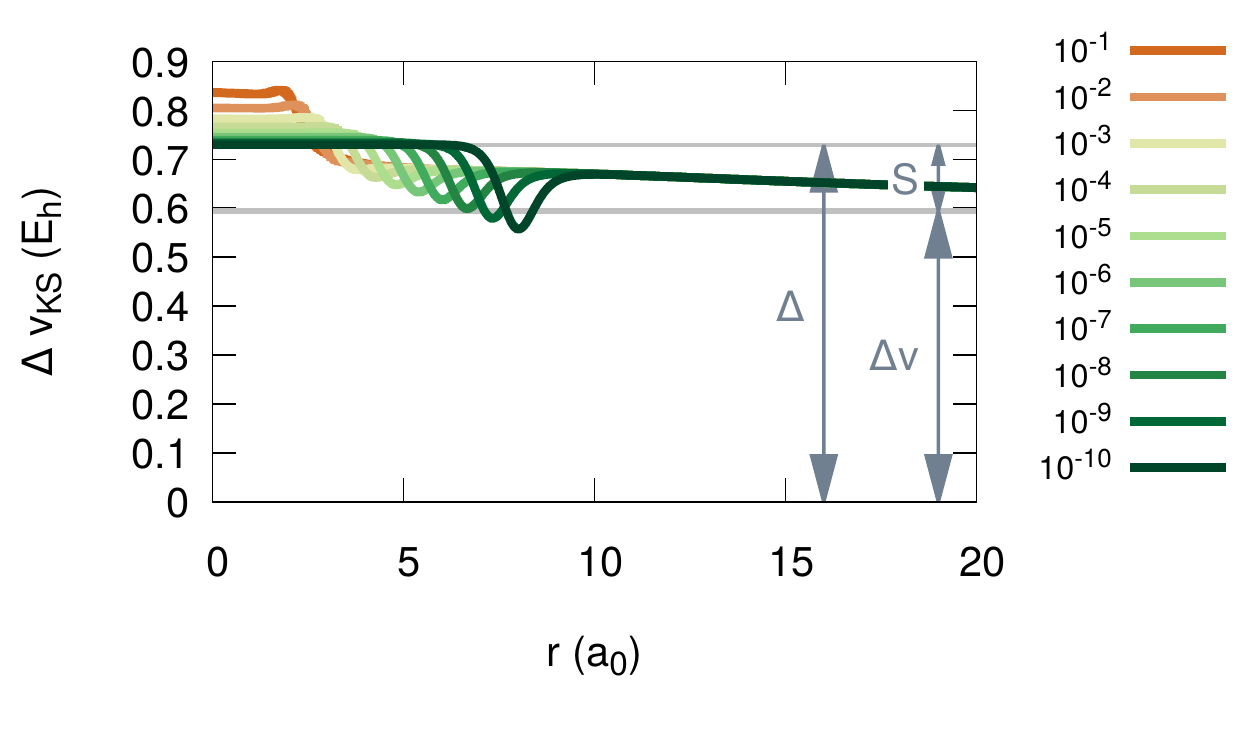}
\caption{Difference between the inverted LDA (invLDA) KS potential for Li with $2+\alpha$ electrons and the KS potential for 2 electrons, for various values of $\alpha$ (see legend), aligned to satisfy the IP theorem. Relationship between the plateau height, $S$, the alignment potential difference, $\Delta v$ and the discontinuity $\Delta$ (see text for definitions) is illustrated.}
\label{fig:plateau_Dv_Li_LDA}
\end{figure}

We resolve the above conundrum by \emph{realignment} of the KS potentials to satisfy the IP theorem
~\footnote{Obviously, now the KS potential does not approach zero as $|\bm{r}| \rarr \infty$. While this should happen for the exact KS potential, this does not happen for the LDA (or invLDA).}.
This means that for each $\alpha$ the KS potential is shifted by the amount required for the ho level to equal the IP. For $N_0=2$, this shift is $v_0 = -I_\LDA - \eps^\ho_\LDA(N_0^-) = -0.6812$ Hartree. Notably, for all $\alpha >0$, the \emph{same} shift of $v_1 = -A_\LDA-\eps^\ho_\LDA(N_0^-+\alpha) = -0.0868$ Hartree is required. We denote $\Delta v = v_1-v_0 = -0.5945$ Hartree and recall that $\eps^\ho(N_0^+) = \eps^\lu + \lim_{\alpha \rarr 0^+} S$, to find that 
\begin{align}
\Delta_\LDA = \lim_{\alpha \rarr 0^+} S_\LDA + \Delta v.
\end{align}
For the exact potential $\Delta v = 0$, and we return to the basic relationship between $\Delta$ and $S$ derived in Sec.~\ref{sec:relationship.S}. This result is presented graphically in Fig.~\ref{fig:plateau_Dv_Li_LDA}.
Results presented in this section are for the LDA. Calculations with the local spin-density approximation (LSDA) and with the Perdew-Burke-Ernzerhof (PBE) GGA yield similar results and are detailed in the Supplemental Material.

To summarize, within approximate KS DFT calculations for finite systems with a fractional $N$ there are two, equally legitimate approaches to obtain the KS potential. They lead to two qualitatively different results: the standard approach yields a smoothly varying potential, without steps, which exponentially decays at infinity. The invLDA approach yields steps in the KS potential, and the asymptotic decay is $\sim a/r$. We relate these improvements to the piecewise-linearity in the density, which is enforced in the invLDA approach.  This internal inconsistency within semi-local xc approximations closely relates to another inconsistency: the IP of finite systems, like atoms and small molecules can be obtained with common xc approximations from total-energy differences with high accuracy of a few percent, whereas obtaining the same quantity directly from the ho energy level results in discrepancies of $\sim 50\%$ (see, e.g., Refs.~\onlinecite{KraislerMakovKelson10,KraislerSchmidt15} and references therein); whereas when the associated $\Delta$ is added to the KS energy difference the exact many-electron energy difference is obtained for the exact xc potential (as shown above).

\section{Conclusions} \label{sec:Conclusions}

In this article we studied the relationship between the Kohn-Sham energies and the many-electron energies of various systems, such as atoms and diatomic molecules, and related them to the step structures that appear in the exchange-correlation (xc) potential.

Steps can occur in the exact potential in different scenarios. In this article we address four: (i) a finite system (an atom) in the ground state with a varying number of electrons (Sec.~\ref{sec:theory.Delta} and \ref{sec:relationship.Delta}); (ii) a finite, excited system with a constant number of electrons (Sec.~\ref{sec:excited_atom}); (iii) a system comprised of subsystems (stretched diatomic molecule) in the ground state with a varying overall number of electrons (Sec.~\ref{sec:theory.S} and \ref{sec:CT}); (iv) a system comprised of subsystems that experiences a charge transfer upon excitation (Sec.~\ref{sec:theory.CTDD} and \ref{sec:CT}). By these examples we address the processes of ionisation, excitation, dissociation and charge transfer. 

As a general rule, steps in the potential occur at points where the exponential decay rate of the density changes, and hence changes the `local effective ionization potential' (LEIP)~\cite{Hodgson16}. This rule is true irrespectively of the specific physical or chemical process the system undergoes, be it adding a small fraction of an electron to the system, exciting the system, inducing transfer of charge or even bringing two subsystems together. In a sense, the complex step structure of the potential is the price one pays for the decision to describe an interacting many-electron system via a non-interacting system with a multiplicative potential~\cite{PhysRevB.99.045129}. An expression for the height of the step in the exact KS potential can be derived from the changes in the LEIP. 

By analyzing the exact KS potential, we show the general relationship between the step structures in the potential and derivative discontinuities in the xc energy: in the cases discussed here, the many-electron energy difference equals the corresponding KS energy difference plus the associated derivative discontinuity. 

The well-known derivative discontinuity of the xc energy ($\Delta$) of a system with a varying number of electrons relates the fundamental gap and the Kohn-Sham (KS) gap: $E_\mathrm{g} = E^\mathrm{KS}_\mathrm{g} + \Delta$. This relationship manifests in the potential as a uniform shift as the system's electron number infinitesimally surpasses an integer value. For a small finite fraction of an additional electron, spatial step structures form in the exact xc potential on the periphery of the system in order to elevate the level of the potential in the center by $\Delta$; as this additional amount of electron tends to zero the plateau created by the steps becomes the uniform shift. 

The relationship between a particular step structure in the xc potential and derivative discontinuities is not always straightforward. The infamous interatomic step, $S$, which forms in a stretched diatomic molecule in order to correctly distribute the electron density throughout the system has usually been regarded as unrelated to the derivative discontinuity because the system typically consists of a fixed number of electrons and the height of the step is seemingly unrelated to the $\Delta$'s of any of the constituent atoms. We demonstrate that upon the transfer of charge from one atom to another within the diatomic molecule, the acceptor atom experiences a shift which corresponds to $\Delta$ of that atom owing to the `local number of electrons' on that atom surpasses an integer, $\Delta_\mathrm{a}$. Simultaneously the donor atom experiences a shift which corresponds to the charge-transfer derivative discontinuity (CTDD)~\cite{HodgsonKraisler17}, $\Delta_\mathrm{d \rightarrow a}^\CT$.

We demonstrate that this discontinuity occurs within the exact KS potential within ensemble DFT of a system which undergoes charge transfer when excited. Analysis of this potential can offer valuable insight for the development of advanced approximations to the xc energy within ensemble DFT. In this case, we show that $S = \Delta_\mathrm{a} - \Delta_\mathrm{d \rightarrow a}^\CT$ and hence the interatomic step is comprised of two derivative discontinuities, which are revealed when charge transfer occurs. In addition, this derivative discontinuity occurs when a fraction of an electron is added to the overall system while the additional charge localizes on one of the atoms. In both cases $\Delta_\CT$ is related to the discontinuity of the derivative of the xc energy of the stretched molecule. 

We also show that the many-electron excitation energy from the ground to the first excited state is related to the KS energy difference plus the associated derivative discontinuity~\cite{PhysRevA.52.R4313}. We demonstrate this numerically for a single atom and show that this excitation is well approximated by the ground-state KS energy differences for this system alone, i.e., in this case the $\Delta$ is small. This implies that the Hartree-xc kernel plays a small role in yielding accurate spectra for our single atom. This is not the case for the charge-transfer system, however, as we typically find the CTDD to be large. Hence in this case the Hxc kernel must have important features which, at least in part, correspond to the CTDD in the potential. Capturing these features in approximations to the ground-state and excited xc potential of DFT and ensemble DFT respectively, as well as the xc kernel of time-dependent DFT, is crucial for accurately obtaining many-electron excitation energies from KS theory.

Finally, we demonstrate that step structures are obtainable also from approximate xc functionals, as simple as the LDA. With the `inverted LDA' (invLDA) approach introduced here, we construct an ensemble of LDA densities with integer number of electrons for each. Upon `reverse-engineering' these densities we find that the corresponding potential possesses step structures, which resemble those present in the exact potential. Ensuring that our invLDA potentials obey the IP theorem, we establish the relationship between the step height and the derivative discontinuity in approximate xc functionals.   

\acknowledgments{We acknowledge Rex Godby for providing us with computational resources and Neepa Maitra and Axel Schild for fruitful discussions. EKUG acknowledges financial support from the European Research
Council Advanced Grant FACT (ERC-2017-AdG-788890).} 

\section*{Description of Supplemental Material}
This paper is accompanied by a supplemental material pdf file, which provides technical details and peripheral information as to the inversion procedure of densities obtained for atoms and ions within common exchange-correlation (xc) approximations, as presented in Section~\ref{sec:plateaus.approx}

\bibliographystyle{achemso}
\bibliography{bib_EK_2020_02_20,refs_from_Matt}

\providecommand{\latin}[1]{#1}
\makeatletter
\providecommand{\doi}
  {\begingroup\let\do\@makeother\dospecials
  \catcode`\{=1 \catcode`\}=2 \doi@aux}
\providecommand{\doi@aux}[1]{\endgroup\texttt{#1}}
\makeatother
\providecommand*\mcitethebibliography{\thebibliography}
\csname @ifundefined\endcsname{endmcitethebibliography}
  {\let\endmcitethebibliography\endthebibliography}{}
\begin{mcitethebibliography}{190}
\providecommand*\natexlab[1]{#1}
\providecommand*\mciteSetBstSublistMode[1]{}
\providecommand*\mciteSetBstMaxWidthForm[2]{}
\providecommand*\mciteBstWouldAddEndPuncttrue
  {\def\EndOfBibitem{\unskip.}}
\providecommand*\mciteBstWouldAddEndPunctfalse
  {\let\EndOfBibitem\relax}
\providecommand*\mciteSetBstMidEndSepPunct[3]{}
\providecommand*\mciteSetBstSublistLabelBeginEnd[3]{}
\providecommand*\EndOfBibitem{}
\mciteSetBstSublistMode{f}
\mciteSetBstMaxWidthForm{subitem}{(\alph{mcitesubitemcount})}
\mciteSetBstSublistLabelBeginEnd
  {\mcitemaxwidthsubitemform\space}
  {\relax}
  {\relax}

\bibitem[Verma and Truhlar(2020)Verma, and Truhlar]{verma2020status}
Verma,~P.; Truhlar,~D.~G. Status and Challenges of Density Functional Theory.
  \emph{Trends in Chemistry} \textbf{2020}, \emph{2}, 302--318\relax
\mciteBstWouldAddEndPuncttrue
\mciteSetBstMidEndSepPunct{\mcitedefaultmidpunct}
{\mcitedefaultendpunct}{\mcitedefaultseppunct}\relax
\EndOfBibitem
\bibitem[{R.M. Dreizler} and {E.K.U. Gross}(1990){R.M. Dreizler}, and {E.K.U.
  Gross}]{DG}
{R.M. Dreizler},; {E.K.U. Gross}, \emph{{D}ensity {F}unctional {T}heory};
  Springer Verlag, 1990\relax
\mciteBstWouldAddEndPuncttrue
\mciteSetBstMidEndSepPunct{\mcitedefaultmidpunct}
{\mcitedefaultendpunct}{\mcitedefaultseppunct}\relax
\EndOfBibitem
\bibitem[Parr and Yang(1989)Parr, and Yang]{PY}
Parr,~R.~G.; Yang,~W. \emph{Density-Functional Theory of Atoms and Molecules};
  Oxford University Press, 1989\relax
\mciteBstWouldAddEndPuncttrue
\mciteSetBstMidEndSepPunct{\mcitedefaultmidpunct}
{\mcitedefaultendpunct}{\mcitedefaultseppunct}\relax
\EndOfBibitem
\bibitem[Fiolhais \latin{et~al.}(2003)Fiolhais, Nogueira, and Marques]{Primer}
Fiolhais,~C., Nogueira,~F., Marques,~M. A.~L., Eds. \emph{A {P}rimer in
  {D}ensity {F}unctional {T}heory}; Springer, 2003\relax
\mciteBstWouldAddEndPuncttrue
\mciteSetBstMidEndSepPunct{\mcitedefaultmidpunct}
{\mcitedefaultendpunct}{\mcitedefaultseppunct}\relax
\EndOfBibitem
\bibitem[Engel and Dreizler(2011)Engel, and Dreizler]{EngelDreizler11}
Engel,~E.; Dreizler,~R. \emph{Density Functional Theory: An Advanced Course};
  Springer, 2011\relax
\mciteBstWouldAddEndPuncttrue
\mciteSetBstMidEndSepPunct{\mcitedefaultmidpunct}
{\mcitedefaultendpunct}{\mcitedefaultseppunct}\relax
\EndOfBibitem
\bibitem[Burke(2012)]{Burke12}
Burke,~K. {Perspective on density functional theory}. \emph{J. Chem. Phys.}
  \textbf{2012}, \emph{136}, 150901\relax
\mciteBstWouldAddEndPuncttrue
\mciteSetBstMidEndSepPunct{\mcitedefaultmidpunct}
{\mcitedefaultendpunct}{\mcitedefaultseppunct}\relax
\EndOfBibitem
\bibitem[Becke(2014)]{Becke14}
Becke,~A.~D. {Perspective: Fifty years of density-functional theory in chemical
  physics.} \emph{J. Chem. Phys.} \textbf{2014}, \emph{140}, 18A301\relax
\mciteBstWouldAddEndPuncttrue
\mciteSetBstMidEndSepPunct{\mcitedefaultmidpunct}
{\mcitedefaultendpunct}{\mcitedefaultseppunct}\relax
\EndOfBibitem
\bibitem[Jones(2015)]{Jones15}
Jones,~R.~O. Density functional theory: Its origins, rise to prominence, and
  future. \emph{Rev. Mod. Phys.} \textbf{2015}, \emph{87}, 897--923\relax
\mciteBstWouldAddEndPuncttrue
\mciteSetBstMidEndSepPunct{\mcitedefaultmidpunct}
{\mcitedefaultendpunct}{\mcitedefaultseppunct}\relax
\EndOfBibitem
\bibitem[Hohenberg and Kohn(1964)Hohenberg, and Kohn]{HK64}
Hohenberg,~P.; Kohn,~W. Inhomogeneous Electron Gas. \emph{Phys. Rev.}
  \textbf{1964}, \emph{136}, B864--B871\relax
\mciteBstWouldAddEndPuncttrue
\mciteSetBstMidEndSepPunct{\mcitedefaultmidpunct}
{\mcitedefaultendpunct}{\mcitedefaultseppunct}\relax
\EndOfBibitem
\bibitem[Pearson and Pearson(2005)Pearson, and Pearson]{pearson2005chemical}
Pearson,~R.~G.; Pearson,~R.~G. Chemical hardness and density functional theory.
  \emph{J. Chem. Sci.} \textbf{2005}, \emph{117}\relax
\mciteBstWouldAddEndPuncttrue
\mciteSetBstMidEndSepPunct{\mcitedefaultmidpunct}
{\mcitedefaultendpunct}{\mcitedefaultseppunct}\relax
\EndOfBibitem
\bibitem[Tran \latin{et~al.}(2007)Tran, Blaha, and Schwarz]{Tran07}
Tran,~F.; Blaha,~P.; Schwarz,~K. {Band gap calculations with Becke-Johnson
  exchange potential}. \emph{J. Phys.: Condens. Matter} \textbf{2007},
  \emph{19}, 196208\relax
\mciteBstWouldAddEndPuncttrue
\mciteSetBstMidEndSepPunct{\mcitedefaultmidpunct}
{\mcitedefaultendpunct}{\mcitedefaultseppunct}\relax
\EndOfBibitem
\bibitem[Tran and Blaha(2009)Tran, and Blaha]{Tran09}
Tran,~F.; Blaha,~P. {Accurate Band Gaps of Semiconductors and Insulators with a
  Semilocal Exchange-Correlation Potential}. \emph{Phys. Rev. Lett.}
  \textbf{2009}, \emph{102}, 226401\relax
\mciteBstWouldAddEndPuncttrue
\mciteSetBstMidEndSepPunct{\mcitedefaultmidpunct}
{\mcitedefaultendpunct}{\mcitedefaultseppunct}\relax
\EndOfBibitem
\bibitem[Eisenberg and Baer(2009)Eisenberg, and Baer]{Eisenberg09}
Eisenberg,~H.~R.; Baer,~R. {A new generalized Kohn-Sham method for fundamental
  band-gaps in solids}. \emph{Phys. Chem. Chem. Phys.} \textbf{2009},
  \emph{11}, 4674\relax
\mciteBstWouldAddEndPuncttrue
\mciteSetBstMidEndSepPunct{\mcitedefaultmidpunct}
{\mcitedefaultendpunct}{\mcitedefaultseppunct}\relax
\EndOfBibitem
\bibitem[Schimka \latin{et~al.}(2011)Schimka, Harl, and Kresse]{HSEsol}
Schimka,~L.; Harl,~J.; Kresse,~G. {Improved hybrid functional for solids: the
  HSEsol functional.} \emph{J. Chem. Phys.} \textbf{2011}, \emph{134},
  024116\relax
\mciteBstWouldAddEndPuncttrue
\mciteSetBstMidEndSepPunct{\mcitedefaultmidpunct}
{\mcitedefaultendpunct}{\mcitedefaultseppunct}\relax
\EndOfBibitem
\bibitem[Chan and Ceder(2010)Chan, and Ceder]{ChanCeder10}
Chan,~M. K.~Y.; Ceder,~G. Efficient Band Gap Prediction for Solids. \emph{Phys.
  Rev. Lett.} \textbf{2010}, \emph{105}, 196403\relax
\mciteBstWouldAddEndPuncttrue
\mciteSetBstMidEndSepPunct{\mcitedefaultmidpunct}
{\mcitedefaultendpunct}{\mcitedefaultseppunct}\relax
\EndOfBibitem
\bibitem[Tozer(2003)]{Tozer03}
Tozer,~D.~J. {Relationship between long-range charge-transfer excitation energy
  error and integer discontinuity in Kohn-Sham theory}. \emph{J. Chem. Phys.}
  \textbf{2003}, \emph{119}, 12697--12699\relax
\mciteBstWouldAddEndPuncttrue
\mciteSetBstMidEndSepPunct{\mcitedefaultmidpunct}
{\mcitedefaultendpunct}{\mcitedefaultseppunct}\relax
\EndOfBibitem
\bibitem[Maitra(2005)]{Maitra05}
Maitra,~N.~T. {Undoing static correlation: long-range charge transfer in
  time-dependent density-functional theory.} \emph{J. Chem. Phys.}
  \textbf{2005}, \emph{122}, 234104\relax
\mciteBstWouldAddEndPuncttrue
\mciteSetBstMidEndSepPunct{\mcitedefaultmidpunct}
{\mcitedefaultendpunct}{\mcitedefaultseppunct}\relax
\EndOfBibitem
\bibitem[Toher \latin{et~al.}(2005)Toher, Filippetti, Sanvito, and
  Burke]{Toher05}
Toher,~C.; Filippetti,~A.; Sanvito,~S.; Burke,~K. {Self-Interaction Errors in
  Density-Functional Calculations of Electronic Transport}. \emph{Phys. Rev.
  Lett.} \textbf{2005}, \emph{95}, 146402\relax
\mciteBstWouldAddEndPuncttrue
\mciteSetBstMidEndSepPunct{\mcitedefaultmidpunct}
{\mcitedefaultendpunct}{\mcitedefaultseppunct}\relax
\EndOfBibitem
\bibitem[Koentopp \latin{et~al.}(2006)Koentopp, Burke, and Evers]{Koentopp06}
Koentopp,~M.; Burke,~K.; Evers,~F. {Zero-bias molecular electronics:
  Exchange-correlation corrections to Landauer's formula}. \emph{Phys. Rev. B}
  \textbf{2006}, \emph{73}, 121403\relax
\mciteBstWouldAddEndPuncttrue
\mciteSetBstMidEndSepPunct{\mcitedefaultmidpunct}
{\mcitedefaultendpunct}{\mcitedefaultseppunct}\relax
\EndOfBibitem
\bibitem[Ke \latin{et~al.}(2007)Ke, Baranger, and Yang]{Ke07}
Ke,~S.-H.; Baranger,~H.~U.; Yang,~W. {Role of the exchange-correlation
  potential in ab initio electron transport calculations.} \emph{J. Chem.
  Phys.} \textbf{2007}, \emph{126}, 201102\relax
\mciteBstWouldAddEndPuncttrue
\mciteSetBstMidEndSepPunct{\mcitedefaultmidpunct}
{\mcitedefaultendpunct}{\mcitedefaultseppunct}\relax
\EndOfBibitem
\bibitem[Hofmann and K\"{u}mmel(2012)Hofmann, and K\"{u}mmel]{Hofmann12}
Hofmann,~D.; K\"{u}mmel,~S. {Integer particle preference during charge transfer
  in Kohn-Sham theory}. \emph{Phys. Rev. B} \textbf{2012}, \emph{86},
  201109\relax
\mciteBstWouldAddEndPuncttrue
\mciteSetBstMidEndSepPunct{\mcitedefaultmidpunct}
{\mcitedefaultendpunct}{\mcitedefaultseppunct}\relax
\EndOfBibitem
\bibitem[Nossa \latin{et~al.}(2013)Nossa, Islam, Canali, and Pederson]{Nossa13}
Nossa,~J.~F.; Islam,~M.~F.; Canali,~C.~M.; Pederson,~M.~R. Electric control of
  a $\{ \mathrm{Fe}_4 \}$ single-molecule magnet in a single-electron
  transistor. \emph{Phys. Rev. B} \textbf{2013}, \emph{88}, 224423\relax
\mciteBstWouldAddEndPuncttrue
\mciteSetBstMidEndSepPunct{\mcitedefaultmidpunct}
{\mcitedefaultendpunct}{\mcitedefaultseppunct}\relax
\EndOfBibitem
\bibitem[Fuks(2016)]{Fuks16}
Fuks,~J.~I. {Time-dependent density functional theory for charge-transfer
  dynamics: review of the causes of failure and success}. \emph{Eur. Phys. J.
  B} \textbf{2016}, \emph{89}, 236\relax
\mciteBstWouldAddEndPuncttrue
\mciteSetBstMidEndSepPunct{\mcitedefaultmidpunct}
{\mcitedefaultendpunct}{\mcitedefaultseppunct}\relax
\EndOfBibitem
\bibitem[Kronik \latin{et~al.}(2012)Kronik, Stein, Refaely-Abramson, and
  Baer]{Kronik_JCTC_review12}
Kronik,~L.; Stein,~T.; Refaely-Abramson,~S.; Baer,~R. {Excitation Gaps of
  Finite-Sized Systems from Optimally Tuned Range-Separated Hybrid
  Functionals}. \emph{J. Chem. Theory Comp.} \textbf{2012}, \emph{8},
  1515--1531\relax
\mciteBstWouldAddEndPuncttrue
\mciteSetBstMidEndSepPunct{\mcitedefaultmidpunct}
{\mcitedefaultendpunct}{\mcitedefaultseppunct}\relax
\EndOfBibitem
\bibitem[Kronik and K\"{u}mmel(2014)Kronik, and K\"{u}mmel]{KronikKuemmel_PES}
Kronik,~L.; K\"{u}mmel,~S. In \emph{Topics of Current Chemistry: First
  Principles Approaches to Spectroscopic Properties of Complex Materials};
  di~Valentin,~C., Botti,~S., Coccoccioni,~M., Eds.; Springer: Berlin, 2014;
  Vol. 347; pp 137--192\relax
\mciteBstWouldAddEndPuncttrue
\mciteSetBstMidEndSepPunct{\mcitedefaultmidpunct}
{\mcitedefaultendpunct}{\mcitedefaultseppunct}\relax
\EndOfBibitem
\bibitem[K\"{u}mmel(2017)]{Kummel17}
K\"{u}mmel,~S. Charge-Transfer Excitations: A Challenge for Time-Dependent
  Density Functional Theory That Has Been Met. \emph{Adv. Energy Mater.}
  \textbf{2017}, \emph{7}, 1700440\relax
\mciteBstWouldAddEndPuncttrue
\mciteSetBstMidEndSepPunct{\mcitedefaultmidpunct}
{\mcitedefaultendpunct}{\mcitedefaultseppunct}\relax
\EndOfBibitem
\bibitem[Gould \latin{et~al.}(2018)Gould, Kronik, and Pittalis]{Gould18}
Gould,~T.; Kronik,~L.; Pittalis,~S. {Charge transfer excitations from exact and
  approximate ensemble Kohn-Sham theory}. \emph{J. Chem. Phys.} \textbf{2018},
  \emph{148}, 174101\relax
\mciteBstWouldAddEndPuncttrue
\mciteSetBstMidEndSepPunct{\mcitedefaultmidpunct}
{\mcitedefaultendpunct}{\mcitedefaultseppunct}\relax
\EndOfBibitem
\bibitem[Aschebrock and K\"ummel(2019)Aschebrock, and
  K\"ummel]{AschebrockKummel19}
Aschebrock,~T.; K\"ummel,~S. Ultranonlocality and accurate band gaps from a
  meta-generalized gradient approximation. \emph{Phys. Rev. Research}
  \textbf{2019}, \emph{1}, 033082\relax
\mciteBstWouldAddEndPuncttrue
\mciteSetBstMidEndSepPunct{\mcitedefaultmidpunct}
{\mcitedefaultendpunct}{\mcitedefaultseppunct}\relax
\EndOfBibitem
\bibitem[Kaxiras(2003)]{Kaxiras03}
Kaxiras,~E. \emph{{Atomic and Electronic Structure of Solids}}; Cambridge
  University Press, 2003\relax
\mciteBstWouldAddEndPuncttrue
\mciteSetBstMidEndSepPunct{\mcitedefaultmidpunct}
{\mcitedefaultendpunct}{\mcitedefaultseppunct}\relax
\EndOfBibitem
\bibitem[Martin(2004)]{Martin}
Martin,~R.~M. \emph{Electronic Structure}; Cambridge Unviersity Press,
  2004\relax
\mciteBstWouldAddEndPuncttrue
\mciteSetBstMidEndSepPunct{\mcitedefaultmidpunct}
{\mcitedefaultendpunct}{\mcitedefaultseppunct}\relax
\EndOfBibitem
\bibitem[Cramer(2004)]{Cramer04}
Cramer,~C. \emph{Essentials Of Computational Chemistry: Theories And Models};
  Wiley, 2004\relax
\mciteBstWouldAddEndPuncttrue
\mciteSetBstMidEndSepPunct{\mcitedefaultmidpunct}
{\mcitedefaultendpunct}{\mcitedefaultseppunct}\relax
\EndOfBibitem
\bibitem[Kohanoff(2006)]{Kohanoff06}
Kohanoff,~J. \emph{Electronic Structure Calculations For Solids And Molecules:
  Theory And Computational Methods}; Cambridge University Press, 2006\relax
\mciteBstWouldAddEndPuncttrue
\mciteSetBstMidEndSepPunct{\mcitedefaultmidpunct}
{\mcitedefaultendpunct}{\mcitedefaultseppunct}\relax
\EndOfBibitem
\bibitem[Sholl and Steckel(2011)Sholl, and Steckel]{ShollSteckel11}
Sholl,~D.; Steckel,~J. \emph{Density Functional Theory: A Practical
  Introduction}; Wiley, 2011\relax
\mciteBstWouldAddEndPuncttrue
\mciteSetBstMidEndSepPunct{\mcitedefaultmidpunct}
{\mcitedefaultendpunct}{\mcitedefaultseppunct}\relax
\EndOfBibitem
\bibitem[Giustino(2014)]{Giustino14_materials}
Giustino,~F. \emph{Materials modelling using density functional theory:
  properties and predictions}; Oxford University Press, 2014\relax
\mciteBstWouldAddEndPuncttrue
\mciteSetBstMidEndSepPunct{\mcitedefaultmidpunct}
{\mcitedefaultendpunct}{\mcitedefaultseppunct}\relax
\EndOfBibitem
\bibitem[DiValentin \latin{et~al.}(2014)DiValentin, Botti, and
  Cococcioni]{DiValentin_TopCurrChem}
DiValentin,~C., Botti,~S., Cococcioni,~M., Eds. \emph{First Principle
  Approaches to Spectroscopic Properties of Complex Materials}; Topics in
  Current Chemistry; Springer, 2014; Vol. 347\relax
\mciteBstWouldAddEndPuncttrue
\mciteSetBstMidEndSepPunct{\mcitedefaultmidpunct}
{\mcitedefaultendpunct}{\mcitedefaultseppunct}\relax
\EndOfBibitem
\bibitem[Kronik and Neaton(2016)Kronik, and Neaton]{KronikNeaton16}
Kronik,~L.; Neaton,~J.~B. {Excited-State Properties of Molecular Solids from
  First Principles}. \emph{Annu. Rev. Phys. Chem.} \textbf{2016}, \emph{67},
  587--616\relax
\mciteBstWouldAddEndPuncttrue
\mciteSetBstMidEndSepPunct{\mcitedefaultmidpunct}
{\mcitedefaultendpunct}{\mcitedefaultseppunct}\relax
\EndOfBibitem
\bibitem[Maurer \latin{et~al.}(2019)Maurer, Freysoldt, Reilly, Brandenburg,
  Hofmann, Bj{\"{o}}rkman, Leb{\`{e}}gue, and Tkatchenko]{Maurer19}
Maurer,~R.~J.; Freysoldt,~C.; Reilly,~A.~M.; Brandenburg,~J.~G.;
  Hofmann,~O.~T.; Bj{\"{o}}rkman,~T.; Leb{\`{e}}gue,~S.; Tkatchenko,~A.
  {Advances in Density-Functional Calculations for Materials Modeling}.
  \emph{Annu. Rev. Mater. Res.} \textbf{2019}, \emph{49}, 3.1--3.30\relax
\mciteBstWouldAddEndPuncttrue
\mciteSetBstMidEndSepPunct{\mcitedefaultmidpunct}
{\mcitedefaultendpunct}{\mcitedefaultseppunct}\relax
\EndOfBibitem
\bibitem[Fetter and Walecka(1971)Fetter, and Walecka]{FetterWalecka}
Fetter,~A.~L.; Walecka,~J.~D. \emph{{Quantum Theory of Many-Particle Systems}};
  MacGraw-Hill: New York, 1971\relax
\mciteBstWouldAddEndPuncttrue
\mciteSetBstMidEndSepPunct{\mcitedefaultmidpunct}
{\mcitedefaultendpunct}{\mcitedefaultseppunct}\relax
\EndOfBibitem
\bibitem[{E.~K.~U. Gross} \latin{et~al.}(1991){E.~K.~U. Gross}, {E. Runge}, and
  {O. Heinonen}]{Gross_MB}
{E.~K.~U. Gross},; {E. Runge},; {O. Heinonen}, \emph{{M}any-{P}article
  {T}heory}; Adam Hilger, 1991\relax
\mciteBstWouldAddEndPuncttrue
\mciteSetBstMidEndSepPunct{\mcitedefaultmidpunct}
{\mcitedefaultendpunct}{\mcitedefaultseppunct}\relax
\EndOfBibitem
\bibitem[Martin \latin{et~al.}(2016)Martin, Reining, and Ceperley]{Reining_MB}
Martin,~R.~M.; Reining,~L.; Ceperley,~D.~M. \emph{{I}nteracting {E}lectrons:
  {T}heory and {C}omputational {A}pproaches}; Cambridge University Press,
  2016\relax
\mciteBstWouldAddEndPuncttrue
\mciteSetBstMidEndSepPunct{\mcitedefaultmidpunct}
{\mcitedefaultendpunct}{\mcitedefaultseppunct}\relax
\EndOfBibitem
\bibitem[Perdew \latin{et~al.}(1982)Perdew, Parr, Levy, and Balduz]{PPLB82}
Perdew,~J.~P.; Parr,~R.~G.; Levy,~M.; Balduz,~J.~L. Density-Functional Theory
  for Fractional Particle Number: Derivative Discontinuities of the Energy.
  \emph{Phys. Rev. Lett.} \textbf{1982}, \emph{49}, 1691--1694\relax
\mciteBstWouldAddEndPuncttrue
\mciteSetBstMidEndSepPunct{\mcitedefaultmidpunct}
{\mcitedefaultendpunct}{\mcitedefaultseppunct}\relax
\EndOfBibitem
\bibitem[Perdew and Levy(1983)Perdew, and Levy]{PerdewLevy83}
Perdew,~J.~P.; Levy,~M. {Physical content of the exact Kohn-Sham orbital
  energies: Band gaps and derivative discontinuities}. \emph{Phys. Rev. Lett.}
  \textbf{1983}, \emph{51}, 1884--1887\relax
\mciteBstWouldAddEndPuncttrue
\mciteSetBstMidEndSepPunct{\mcitedefaultmidpunct}
{\mcitedefaultendpunct}{\mcitedefaultseppunct}\relax
\EndOfBibitem
\bibitem[Levy \latin{et~al.}(1984)Levy, Perdew, and Sahni]{LevyPerdewSahni84}
Levy,~M.; Perdew,~J.~P.; Sahni,~V. {Exact differential equation for the density
  and ionization energy of a many-particle system}. \emph{Phys. Rev. A}
  \textbf{1984}, \emph{30}, 2745--2748\relax
\mciteBstWouldAddEndPuncttrue
\mciteSetBstMidEndSepPunct{\mcitedefaultmidpunct}
{\mcitedefaultendpunct}{\mcitedefaultseppunct}\relax
\EndOfBibitem
\bibitem[Perdew(1985)]{NATO85_Perdew}
Perdew,~J.~P. In \emph{Density Functional Methods in Physics}; Dreizler,~R.~M.,
  da~Provid\^{e}ncia,~J., Eds.; NATO ASI Series; Plenum Press, 1985; Vol. 123;
  pp 265--308\relax
\mciteBstWouldAddEndPuncttrue
\mciteSetBstMidEndSepPunct{\mcitedefaultmidpunct}
{\mcitedefaultendpunct}{\mcitedefaultseppunct}\relax
\EndOfBibitem
\bibitem[Almbladh and {von Barth}(1985)Almbladh, and {von
  Barth}]{AlmbladhVonBarth85}
Almbladh,~C.~O.; {von Barth},~U. Exact results for the charge and spin
  densities, exchange-correlation potentials, and density-functional
  eigenvalues. \emph{Phys. Rev. B} \textbf{1985}, \emph{31}, 3231--3244\relax
\mciteBstWouldAddEndPuncttrue
\mciteSetBstMidEndSepPunct{\mcitedefaultmidpunct}
{\mcitedefaultendpunct}{\mcitedefaultseppunct}\relax
\EndOfBibitem
\bibitem[Perdew and Levy(1997)Perdew, and Levy]{PerdewLevy97}
Perdew,~J.~P.; Levy,~M. {Comment on ``Significance of the highest occupied
  Kohn-Sham eigenvalue''}. \emph{Phys. Rev. B} \textbf{1997}, \emph{56},
  16021--16028\relax
\mciteBstWouldAddEndPuncttrue
\mciteSetBstMidEndSepPunct{\mcitedefaultmidpunct}
{\mcitedefaultendpunct}{\mcitedefaultseppunct}\relax
\EndOfBibitem
\bibitem[Harbola(1999)]{Harbola99}
Harbola,~M.~K. Relationship between the highest occupied Kohn-Sham orbital
  eigenvalue and ionization energy. \emph{Phys. Rev. B} \textbf{1999},
  \emph{60}, 4545\relax
\mciteBstWouldAddEndPuncttrue
\mciteSetBstMidEndSepPunct{\mcitedefaultmidpunct}
{\mcitedefaultendpunct}{\mcitedefaultseppunct}\relax
\EndOfBibitem
\bibitem[Sham and Schl\"{u}ter(1983)Sham, and Schl\"{u}ter]{ShamSchluter83}
Sham,~L.~J.; Schl\"{u}ter,~M. {Density-functional theory of the energy gap}.
  \emph{Phys. Rev. Lett.} \textbf{1983}, \emph{51}, 1888--1891\relax
\mciteBstWouldAddEndPuncttrue
\mciteSetBstMidEndSepPunct{\mcitedefaultmidpunct}
{\mcitedefaultendpunct}{\mcitedefaultseppunct}\relax
\EndOfBibitem
\bibitem[Perdew(1985)]{Perdew85}
Perdew,~J.~P. Density functional theory and the band gap problem. \emph{Int J.
  Quantum Chem.} \textbf{1985}, \emph{28}, 497--523\relax
\mciteBstWouldAddEndPuncttrue
\mciteSetBstMidEndSepPunct{\mcitedefaultmidpunct}
{\mcitedefaultendpunct}{\mcitedefaultseppunct}\relax
\EndOfBibitem
\bibitem[Zhang and Yang(2000)Zhang, and Yang]{ZhangYang00}
Zhang,~Y.; Yang,~W. Perspective on ``Density-functional theory for fractional
  particle number: derivative discontinuities of the energy''. \emph{Theo.
  Chem. Acc.} \textbf{2000}, \emph{103}, 346--348\relax
\mciteBstWouldAddEndPuncttrue
\mciteSetBstMidEndSepPunct{\mcitedefaultmidpunct}
{\mcitedefaultendpunct}{\mcitedefaultseppunct}\relax
\EndOfBibitem
\bibitem[Lein and K\"ummel(2005)Lein, and K\"ummel]{LeinKummel05}
Lein,~M.; K\"ummel,~S. Exact Time-Dependent Exchange-Correlation Potentials for
  Strong-Field Electron Dynamics. \emph{Phys. Rev. Lett.} \textbf{2005},
  \emph{94}, 143003\relax
\mciteBstWouldAddEndPuncttrue
\mciteSetBstMidEndSepPunct{\mcitedefaultmidpunct}
{\mcitedefaultendpunct}{\mcitedefaultseppunct}\relax
\EndOfBibitem
\bibitem[Mundt and K\"{u}mmel(2005)Mundt, and K\"{u}mmel]{Mundt05}
Mundt,~M.; K\"{u}mmel,~S. {Derivative Discontinuities in Time-Dependent
  Density-Functional Theory}. \emph{Phys. Rev. Lett.} \textbf{2005}, \emph{95},
  203004\relax
\mciteBstWouldAddEndPuncttrue
\mciteSetBstMidEndSepPunct{\mcitedefaultmidpunct}
{\mcitedefaultendpunct}{\mcitedefaultseppunct}\relax
\EndOfBibitem
\bibitem[Cohen \latin{et~al.}(2012)Cohen, Mori-S\'{a}nchez, and Yang]{Cohen12}
Cohen,~A.~J.; Mori-S\'{a}nchez,~P.; Yang,~W. Challenges for Density Functional
  Theory. \emph{Chem. Rev.} \textbf{2012}, \emph{112}, 289--320\relax
\mciteBstWouldAddEndPuncttrue
\mciteSetBstMidEndSepPunct{\mcitedefaultmidpunct}
{\mcitedefaultendpunct}{\mcitedefaultseppunct}\relax
\EndOfBibitem
\bibitem[Baerends \latin{et~al.}(2013)Baerends, Gritsenko, and van
  Meer]{Baerends13}
Baerends,~E.; Gritsenko,~O.; van Meer,~R. {The Kohn-Sham gap, the fundamental
  gap and the optical gap: the physical meaning of occupied and virtual
  Kohn-Sham orbital energies}. \emph{Phys. Chem. Chem. Phys.} \textbf{2013},
  \emph{15}, 16408--16425\relax
\mciteBstWouldAddEndPuncttrue
\mciteSetBstMidEndSepPunct{\mcitedefaultmidpunct}
{\mcitedefaultendpunct}{\mcitedefaultseppunct}\relax
\EndOfBibitem
\bibitem[Kraisler and Kronik(2013)Kraisler, and Kronik]{KraislerKronik13}
Kraisler,~E.; Kronik,~L. {Piecewise Linearity of Approximate Density
  Functionals Revisited: Implications for Frontier Orbital Energies}.
  \emph{Phys. Rev. Lett.} \textbf{2013}, \emph{110}, 126403\relax
\mciteBstWouldAddEndPuncttrue
\mciteSetBstMidEndSepPunct{\mcitedefaultmidpunct}
{\mcitedefaultendpunct}{\mcitedefaultseppunct}\relax
\EndOfBibitem
\bibitem[Mori-S{\'{a}}nchez and Cohen(2014)Mori-S{\'{a}}nchez, and
  Cohen]{MoriS14}
Mori-S{\'{a}}nchez,~P.; Cohen,~A.~J. {The derivative discontinuity of the
  exchange-correlation functional.} \emph{Phys. Chem. Chem. Phys.}
  \textbf{2014}, \emph{16}, 14378--14387\relax
\mciteBstWouldAddEndPuncttrue
\mciteSetBstMidEndSepPunct{\mcitedefaultmidpunct}
{\mcitedefaultendpunct}{\mcitedefaultseppunct}\relax
\EndOfBibitem
\bibitem[Mosquera and Wasserman(2014)Mosquera, and Wasserman]{Mosquera14}
Mosquera,~M.~A.; Wasserman,~A. {Integer discontinuity of density functional
  theory}. \emph{Phys. Rev. A} \textbf{2014}, \emph{89}, 052506\relax
\mciteBstWouldAddEndPuncttrue
\mciteSetBstMidEndSepPunct{\mcitedefaultmidpunct}
{\mcitedefaultendpunct}{\mcitedefaultseppunct}\relax
\EndOfBibitem
\bibitem[Mosquera and Wasserman(2014)Mosquera, and Wasserman]{Mosquera14a}
Mosquera,~M.~A.; Wasserman,~A. {Derivative discontinuities in density
  functional theory}. \emph{Mol. Phys.} \textbf{2014}, \emph{112},
  2997--3013\relax
\mciteBstWouldAddEndPuncttrue
\mciteSetBstMidEndSepPunct{\mcitedefaultmidpunct}
{\mcitedefaultendpunct}{\mcitedefaultseppunct}\relax
\EndOfBibitem
\bibitem[Kraisler and Kronik(2014)Kraisler, and Kronik]{KraislerKronik14}
Kraisler,~E.; Kronik,~L. Fundamental gaps with approximate density functionals:
  The derivative discontinuity revealed from ensemble considerations. \emph{J.
  Chem. Phys.} \textbf{2014}, \emph{140}, 18A540\relax
\mciteBstWouldAddEndPuncttrue
\mciteSetBstMidEndSepPunct{\mcitedefaultmidpunct}
{\mcitedefaultendpunct}{\mcitedefaultseppunct}\relax
\EndOfBibitem
\bibitem[G\"orling(2015)]{Goerling15}
G\"orling,~A. Exchange-correlation potentials with proper discontinuities for
  physically meaningful Kohn-Sham eigenvalues and band structures. \emph{Phys.
  Rev. B} \textbf{2015}, \emph{91}, 245120\relax
\mciteBstWouldAddEndPuncttrue
\mciteSetBstMidEndSepPunct{\mcitedefaultmidpunct}
{\mcitedefaultendpunct}{\mcitedefaultseppunct}\relax
\EndOfBibitem
\bibitem[Jones(2015)]{jones2015density}
Jones,~R.~O. Density functional theory: Its origins, rise to prominence, and
  future. \emph{Rev. Mod. Phys.} \textbf{2015}, \emph{87}, 897\relax
\mciteBstWouldAddEndPuncttrue
\mciteSetBstMidEndSepPunct{\mcitedefaultmidpunct}
{\mcitedefaultendpunct}{\mcitedefaultseppunct}\relax
\EndOfBibitem
\bibitem[Hodgson \latin{et~al.}(2017)Hodgson, Kraisler, Schild, and
  Gross]{HodgsonKraisler17}
Hodgson,~M. J.~P.; Kraisler,~E.; Schild,~A.; Gross,~E.~K.~U. {How interatomic
  steps in the exact Kohn-Sham Potential relate to derivative discontinuities
  of the energy}. \emph{J. Phys. Chem. Lett.} \textbf{2017}, \emph{8},
  5974\relax
\mciteBstWouldAddEndPuncttrue
\mciteSetBstMidEndSepPunct{\mcitedefaultmidpunct}
{\mcitedefaultendpunct}{\mcitedefaultseppunct}\relax
\EndOfBibitem
\bibitem[Schulz and Jacob(2019)Schulz, and Jacob]{schulz2019description}
Schulz,~A.; Jacob,~C.~R. Description of intermolecular charge transfer with
  subsystem density-functional theory. \emph{J. Chem. Phys.} \textbf{2019},
  \emph{151}, 131103\relax
\mciteBstWouldAddEndPuncttrue
\mciteSetBstMidEndSepPunct{\mcitedefaultmidpunct}
{\mcitedefaultendpunct}{\mcitedefaultseppunct}\relax
\EndOfBibitem
\bibitem[Levy(1995)]{PhysRevA.52.R4313}
Levy,~M. Excitation energies from density-functional orbital energies.
  \emph{Phys. Rev. A} \textbf{1995}, \emph{52}, R4313--R4315\relax
\mciteBstWouldAddEndPuncttrue
\mciteSetBstMidEndSepPunct{\mcitedefaultmidpunct}
{\mcitedefaultendpunct}{\mcitedefaultseppunct}\relax
\EndOfBibitem
\bibitem[Almbladh and {von Barth}(1985)Almbladh, and {von Barth}]{NATO85_AvB}
Almbladh,~C.~O.; {von Barth},~U. In \emph{Density Functional Methods in
  Physics}; Dreizler,~R.~M., da~Provid\^{e}ncia,~J., Eds.; NATO ASI Series;
  Plenum Press, 1985; Vol. 123; pp 209--231\relax
\mciteBstWouldAddEndPuncttrue
\mciteSetBstMidEndSepPunct{\mcitedefaultmidpunct}
{\mcitedefaultendpunct}{\mcitedefaultseppunct}\relax
\EndOfBibitem
\bibitem[Thiele \latin{et~al.}(2008)Thiele, Gross, and
  K\"ummel]{ThieleGrossKummel08}
Thiele,~M.; Gross,~E. K.~U.; K\"ummel,~S. Adiabatic Approximation in
  Nonperturbative Time-Dependent Density-Functional Theory. \emph{Phys. Rev.
  Lett.} \textbf{2008}, \emph{100}, 153004\relax
\mciteBstWouldAddEndPuncttrue
\mciteSetBstMidEndSepPunct{\mcitedefaultmidpunct}
{\mcitedefaultendpunct}{\mcitedefaultseppunct}\relax
\EndOfBibitem
\bibitem[van Leeuwen \latin{et~al.}(1995)van Leeuwen, Gritsenko, and
  Baerends]{RvL95}
van Leeuwen,~R.; Gritsenko,~O.; Baerends,~E.~J. Step structure in the atomic
  Kohn-Sham potential. \emph{Z. Phys. D} \textbf{1995}, \emph{33},
  229--238\relax
\mciteBstWouldAddEndPuncttrue
\mciteSetBstMidEndSepPunct{\mcitedefaultmidpunct}
{\mcitedefaultendpunct}{\mcitedefaultseppunct}\relax
\EndOfBibitem
\bibitem[Hodgson \latin{et~al.}(2016)Hodgson, Ramsden, and Godby]{Hodgson16}
Hodgson,~M. J.~P.; Ramsden,~J.~D.; Godby,~R.~W. {Origin of static and dynamic
  steps in exact Kohn-Sham potentials}. \emph{Phys. Rev. B} \textbf{2016},
  \emph{93}, 155146\relax
\mciteBstWouldAddEndPuncttrue
\mciteSetBstMidEndSepPunct{\mcitedefaultmidpunct}
{\mcitedefaultendpunct}{\mcitedefaultseppunct}\relax
\EndOfBibitem
\bibitem[Krieger \latin{et~al.}(1990)Krieger, Li, and Iafrate]{KRIEGER1990256}
Krieger,~J.~B.; Li,~Y.; Iafrate,~G.~J. Derivation and application of an
  accurate Kohn-Sham potential with integer discontinuity. \emph{Phys. Lett. A}
  \textbf{1990}, \emph{146}, 256 -- 260\relax
\mciteBstWouldAddEndPuncttrue
\mciteSetBstMidEndSepPunct{\mcitedefaultmidpunct}
{\mcitedefaultendpunct}{\mcitedefaultseppunct}\relax
\EndOfBibitem
\bibitem[Krieger \latin{et~al.}(1992)Krieger, Li, and Iafrate]{PhysRevA.45.101}
Krieger,~J.~B.; Li,~Y.; Iafrate,~G.~J. Construction and application of an
  accurate local spin-polarized Kohn-Sham potential with integer discontinuity:
  Exchange-only theory. \emph{Phys. Rev. A} \textbf{1992}, \emph{45},
  101--126\relax
\mciteBstWouldAddEndPuncttrue
\mciteSetBstMidEndSepPunct{\mcitedefaultmidpunct}
{\mcitedefaultendpunct}{\mcitedefaultseppunct}\relax
\EndOfBibitem
\bibitem[Grabo \latin{et~al.}(2000)Grabo, Kreibich, Kurth, and Gross]{Grabo00}
Grabo,~T.; Kreibich,~T.; Kurth,~S.; Gross,~E.~K.~U. In \emph{Strong Coulomb
  Correlations in Electronic Structure Calculations: Beyond Local Density
  Approximations}; Avisimov,~V.~I., Ed.; CRC Press, 2000; pp 203--211\relax
\mciteBstWouldAddEndPuncttrue
\mciteSetBstMidEndSepPunct{\mcitedefaultmidpunct}
{\mcitedefaultendpunct}{\mcitedefaultseppunct}\relax
\EndOfBibitem
\bibitem[Sharp and Horton(1953)Sharp, and Horton]{PhysRev.90.317}
Sharp,~R.~T.; Horton,~G.~K. A Variational Approach to the Unipotential
  Many-Electron Problem. \emph{Phys. Rev.} \textbf{1953}, \emph{90},
  317--317\relax
\mciteBstWouldAddEndPuncttrue
\mciteSetBstMidEndSepPunct{\mcitedefaultmidpunct}
{\mcitedefaultendpunct}{\mcitedefaultseppunct}\relax
\EndOfBibitem
\bibitem[Talman and Shadwick(1976)Talman, and Shadwick]{PhysRevA.14.36}
Talman,~J.~D.; Shadwick,~W.~F. Optimized effective atomic central potential.
  \emph{Phys. Rev. A} \textbf{1976}, \emph{14}, 36--40\relax
\mciteBstWouldAddEndPuncttrue
\mciteSetBstMidEndSepPunct{\mcitedefaultmidpunct}
{\mcitedefaultendpunct}{\mcitedefaultseppunct}\relax
\EndOfBibitem
\bibitem[Gidopoulos and Lathiotakis(2012)Gidopoulos, and
  Lathiotakis]{PhysRevA.85.052508}
Gidopoulos,~N.~I.; Lathiotakis,~N.~N. Nonanalyticity of the optimized effective
  potential with finite basis sets. \emph{Phys. Rev. A} \textbf{2012},
  \emph{85}, 052508\relax
\mciteBstWouldAddEndPuncttrue
\mciteSetBstMidEndSepPunct{\mcitedefaultmidpunct}
{\mcitedefaultendpunct}{\mcitedefaultseppunct}\relax
\EndOfBibitem
\bibitem[Friedrich \latin{et~al.}(2013)Friedrich, Betzinger, and
  Bl\"ugel]{PhysRevA.88.046501}
Friedrich,~C.; Betzinger,~M.; Bl\"ugel,~S. Comment on ``Nonanalyticity of the
  optimized effective potential with finite basis sets''. \emph{Phys. Rev. A}
  \textbf{2013}, \emph{88}, 046501\relax
\mciteBstWouldAddEndPuncttrue
\mciteSetBstMidEndSepPunct{\mcitedefaultmidpunct}
{\mcitedefaultendpunct}{\mcitedefaultseppunct}\relax
\EndOfBibitem
\bibitem[Gidopoulos and Lathiotakis(2013)Gidopoulos, and
  Lathiotakis]{PhysRevA.88.046502}
Gidopoulos,~N.~I.; Lathiotakis,~N.~N. Reply to ``Comment on `Nonanalyticity of
  the optimized effective potential with finite basis sets' ''. \emph{Phys.
  Rev. A} \textbf{2013}, \emph{88}, 046502\relax
\mciteBstWouldAddEndPuncttrue
\mciteSetBstMidEndSepPunct{\mcitedefaultmidpunct}
{\mcitedefaultendpunct}{\mcitedefaultseppunct}\relax
\EndOfBibitem
\bibitem[K\"ummel and Kronik(2008)K\"ummel, and Kronik]{KK08}
K\"ummel,~S.; Kronik,~L. Orbital-dependent density functionals: Theory and
  applications. \emph{Rev. Mod. Phys.} \textbf{2008}, \emph{80}, 3--60\relax
\mciteBstWouldAddEndPuncttrue
\mciteSetBstMidEndSepPunct{\mcitedefaultmidpunct}
{\mcitedefaultendpunct}{\mcitedefaultseppunct}\relax
\EndOfBibitem
\bibitem[Callow \latin{et~al.}(2020)Callow, Pearce, Pitts, Lathiotakis,
  Hodgson, and Gidopoulos]{D0FD00069H}
Callow,~T.; Pearce,~B.~J.; Pitts,~T.; Lathiotakis,~N.; Hodgson,~M. J.~P.;
  Gidopoulos,~N. Improving the exchange and correlation potential in density
  functional approximations through constraints. \emph{Faraday Discuss.}
  \textbf{2020}, \relax
\mciteBstWouldAddEndPunctfalse
\mciteSetBstMidEndSepPunct{\mcitedefaultmidpunct}
{}{\mcitedefaultseppunct}\relax
\EndOfBibitem
\bibitem[Hirata \latin{et~al.}(2001)Hirata, Ivanov, Grabowski, Bartlett, Burke,
  and Talman]{hirata2001can}
Hirata,~S.; Ivanov,~S.; Grabowski,~I.; Bartlett,~R.~J.; Burke,~K.;
  Talman,~J.~D. Can optimized effective potentials be determined uniquely?
  \emph{The Journal of Chemical Physics} \textbf{2001}, \emph{115},
  1635--1649\relax
\mciteBstWouldAddEndPuncttrue
\mciteSetBstMidEndSepPunct{\mcitedefaultmidpunct}
{\mcitedefaultendpunct}{\mcitedefaultseppunct}\relax
\EndOfBibitem
\bibitem[Staroverov \latin{et~al.}(2006)Staroverov, Scuseria, and
  Davidson]{staroverov2006optimized}
Staroverov,~V.~N.; Scuseria,~G.~E.; Davidson,~E.~R. Optimized effective
  potentials yielding Hartree--Fock energies and densities. \emph{J. Chem.
  Phys.} \textbf{2006}, \emph{124}, 141103\relax
\mciteBstWouldAddEndPuncttrue
\mciteSetBstMidEndSepPunct{\mcitedefaultmidpunct}
{\mcitedefaultendpunct}{\mcitedefaultseppunct}\relax
\EndOfBibitem
\bibitem[Glushkov \latin{et~al.}(2009)Glushkov, Fesenko, and
  Polatoglou]{glushkov2009finite}
Glushkov,~V.~N.; Fesenko,~S.~I.; Polatoglou,~H.~M. On finite basis set
  implementation of the exchange-only optimized effective potential method.
  \emph{Theor. Chem. Acc.} \textbf{2009}, \emph{124}, 365--376\relax
\mciteBstWouldAddEndPuncttrue
\mciteSetBstMidEndSepPunct{\mcitedefaultmidpunct}
{\mcitedefaultendpunct}{\mcitedefaultseppunct}\relax
\EndOfBibitem
\bibitem[Heaton-Burgess \latin{et~al.}(2007)Heaton-Burgess, Bulat, and
  Yang]{PhysRevLett.98.256401}
Heaton-Burgess,~T.; Bulat,~F.~A.; Yang,~W. Optimized Effective Potentials in
  Finite Basis Sets. \emph{Phys. Rev. Lett.} \textbf{2007}, \emph{98},
  256401\relax
\mciteBstWouldAddEndPuncttrue
\mciteSetBstMidEndSepPunct{\mcitedefaultmidpunct}
{\mcitedefaultendpunct}{\mcitedefaultseppunct}\relax
\EndOfBibitem
\bibitem[He{\ss}elmann \latin{et~al.}(2007)He{\ss}elmann, G{\"o}tz, Della~Sala,
  and G{\"o}rling]{hesselmann2007numerically}
He{\ss}elmann,~A.; G{\"o}tz,~A.~W.; Della~Sala,~F.; G{\"o}rling,~A. Numerically
  stable optimized effective potential method with balanced Gaussian basis
  sets. \emph{The Journal of chemical physics} \textbf{2007}, \emph{127},
  054102\relax
\mciteBstWouldAddEndPuncttrue
\mciteSetBstMidEndSepPunct{\mcitedefaultmidpunct}
{\mcitedefaultendpunct}{\mcitedefaultseppunct}\relax
\EndOfBibitem
\bibitem[Bredas \latin{et~al.}(2009)Bredas, Norton, Cornil, and
  Coropceanu]{Bredas09}
Bredas,~J.-L.; Norton,~J.~E.; Cornil,~J.; Coropceanu,~V. {Molecular
  Understanding of Organic Solar Cells: The Challenges}. \emph{Acc. Chem. Res.}
  \textbf{2009}, \emph{42}, 1691--1699\relax
\mciteBstWouldAddEndPuncttrue
\mciteSetBstMidEndSepPunct{\mcitedefaultmidpunct}
{\mcitedefaultendpunct}{\mcitedefaultseppunct}\relax
\EndOfBibitem
\bibitem[Venkataraman \latin{et~al.}(2010)Venkataraman, Yurt, Venkatraman, and
  Gavvalapalli]{venkataraman2010role}
Venkataraman,~D.; Yurt,~S.; Venkatraman,~B.~H.; Gavvalapalli,~N. Role of
  molecular architecture in organic photovoltaic cells. \emph{J. Phys. Chem.
  Lett.} \textbf{2010}, \emph{1}, 947--958\relax
\mciteBstWouldAddEndPuncttrue
\mciteSetBstMidEndSepPunct{\mcitedefaultmidpunct}
{\mcitedefaultendpunct}{\mcitedefaultseppunct}\relax
\EndOfBibitem
\bibitem[Deibel \latin{et~al.}(2010)Deibel, Strobel, and
  Dyakonov]{deibel2010role}
Deibel,~C.; Strobel,~T.; Dyakonov,~V. Role of the charge transfer state in
  organic donor--acceptor solar cells. \emph{Adv. Mater.} \textbf{2010},
  \emph{22}, 4097--4111\relax
\mciteBstWouldAddEndPuncttrue
\mciteSetBstMidEndSepPunct{\mcitedefaultmidpunct}
{\mcitedefaultendpunct}{\mcitedefaultseppunct}\relax
\EndOfBibitem
\bibitem[Liu \latin{et~al.}(2014)Liu, Shen, He, Luo, and Li]{liu2014strategy}
Liu,~X.; Shen,~W.; He,~R.; Luo,~Y.; Li,~M. Strategy to modulate the
  electron-rich units in donor--acceptor copolymers for improvements of organic
  photovoltaics. \emph{J. Phys. Chem. C} \textbf{2014}, \emph{118},
  17266--17278\relax
\mciteBstWouldAddEndPuncttrue
\mciteSetBstMidEndSepPunct{\mcitedefaultmidpunct}
{\mcitedefaultendpunct}{\mcitedefaultseppunct}\relax
\EndOfBibitem
\bibitem[Chen \latin{et~al.}(2018)Chen, Liu, Li, Zhao, Lu, Huang, and
  Xu]{chen2018density}
Chen,~J.; Liu,~Q.; Li,~H.; Zhao,~Z.; Lu,~Z.; Huang,~Y.; Xu,~D. {Density
  functional theory investigations of DAD'structural molecules as donor
  materials in organic solar cell}. \emph{Front. Chem.} \textbf{2018},
  \emph{6}, 200\relax
\mciteBstWouldAddEndPuncttrue
\mciteSetBstMidEndSepPunct{\mcitedefaultmidpunct}
{\mcitedefaultendpunct}{\mcitedefaultseppunct}\relax
\EndOfBibitem
\bibitem[Trang \latin{et~al.}(2020)Trang, Dung, Cuong, Hai, Escudero, Nguyen,
  and Nguyen]{trang2020theoretical}
Trang,~N.~V.; Dung,~T.~N.; Cuong,~N.~T.; Hai,~L. T.~H.; Escudero,~D.;
  Nguyen,~M.~T.; Nguyen,~H. M.~T. {Theoretical Study of a Class of Organic
  D-$\pi$-A Dyes for Polymer Solar Cells: Influence of Various $\pi$-Spacers}.
  \emph{Crystals} \textbf{2020}, \emph{10}, 163\relax
\mciteBstWouldAddEndPuncttrue
\mciteSetBstMidEndSepPunct{\mcitedefaultmidpunct}
{\mcitedefaultendpunct}{\mcitedefaultseppunct}\relax
\EndOfBibitem
\bibitem[Nitzan and Ratner(2003)Nitzan, and Ratner]{nitzan2003electron}
Nitzan,~A.; Ratner,~M.~A. Electron transport in molecular wire junctions.
  \emph{Science} \textbf{2003}, \emph{300}, 1384--1389\relax
\mciteBstWouldAddEndPuncttrue
\mciteSetBstMidEndSepPunct{\mcitedefaultmidpunct}
{\mcitedefaultendpunct}{\mcitedefaultseppunct}\relax
\EndOfBibitem
\bibitem[Evers \latin{et~al.}(2004)Evers, Weigend, and
  Koentopp]{PhysRevB.69.235411}
Evers,~F.; Weigend,~F.; Koentopp,~M. Conductance of molecular wires and
  transport calculations based on density-functional theory. \emph{Phys. Rev.
  B} \textbf{2004}, \emph{69}, 235411\relax
\mciteBstWouldAddEndPuncttrue
\mciteSetBstMidEndSepPunct{\mcitedefaultmidpunct}
{\mcitedefaultendpunct}{\mcitedefaultseppunct}\relax
\EndOfBibitem
\bibitem[Stefanucci and Kurth(2015)Stefanucci, and Kurth]{stefanucci2015steady}
Stefanucci,~G.; Kurth,~S. Steady-state density functional theory for finite
  bias conductances. \emph{Nano letters} \textbf{2015}, \emph{15},
  8020--8025\relax
\mciteBstWouldAddEndPuncttrue
\mciteSetBstMidEndSepPunct{\mcitedefaultmidpunct}
{\mcitedefaultendpunct}{\mcitedefaultseppunct}\relax
\EndOfBibitem
\bibitem[Kurth and Stefanucci(2017)Kurth, and Stefanucci]{kurth2017transport}
Kurth,~S.; Stefanucci,~G. Transport through correlated systems with density
  functional theory. \emph{Journal of Physics: Condensed Matter} \textbf{2017},
  \emph{29}, 413002\relax
\mciteBstWouldAddEndPuncttrue
\mciteSetBstMidEndSepPunct{\mcitedefaultmidpunct}
{\mcitedefaultendpunct}{\mcitedefaultseppunct}\relax
\EndOfBibitem
\bibitem[Zelovich \latin{et~al.}(2017)Zelovich, Hansen, Liu, Neaton, Kronik,
  and Hod]{Zelovich17}
Zelovich,~T.; Hansen,~T.; Liu,~Z.-F.; Neaton,~J.~B.; Kronik,~L.; Hod,~O.
  Parameter-free driven Liouville-von Neumann approach for time-dependent
  electronic transport simulations in open quantum systems. \emph{J. Chem.
  Phys.} \textbf{2017}, \emph{146}, 092331\relax
\mciteBstWouldAddEndPuncttrue
\mciteSetBstMidEndSepPunct{\mcitedefaultmidpunct}
{\mcitedefaultendpunct}{\mcitedefaultseppunct}\relax
\EndOfBibitem
\bibitem[Bhandari \latin{et~al.}(2018)Bhandari, Cheung, Geva, Kronik, and
  Dunietz]{Dunietz18}
Bhandari,~S.; Cheung,~M.~S.; Geva,~E.; Kronik,~L.; Dunietz,~B.~D. Fundamental
  Gaps of Condensed-Phase Organic Semiconductors from Single-Molecule
  Calculations using Polarization-Consistent Optimally Tuned Screened
  Range-Separated Hybrid Functionals. \emph{J. Chem. Theory Comput.}
  \textbf{2018}, \emph{14}, 6287--6294\relax
\mciteBstWouldAddEndPuncttrue
\mciteSetBstMidEndSepPunct{\mcitedefaultmidpunct}
{\mcitedefaultendpunct}{\mcitedefaultseppunct}\relax
\EndOfBibitem
\bibitem[Begam \latin{et~al.}(2020)Begam, Bhandari, Maiti, and
  Dunietz]{Dunietz20}
Begam,~K.; Bhandari,~S.; Maiti,~B.; Dunietz,~B.~D. Screened Range-Separated
  Hybrid Functional with Polarizable Continuum Model Overcomes Challenges in
  Describing Triplet Excitations in the Condensed Phase Using TDDFT. \emph{J.
  Chem. Theory Comput.} \textbf{2020}, \emph{16}, 3287--3293\relax
\mciteBstWouldAddEndPuncttrue
\mciteSetBstMidEndSepPunct{\mcitedefaultmidpunct}
{\mcitedefaultendpunct}{\mcitedefaultseppunct}\relax
\EndOfBibitem
\bibitem[Note1()]{Note1}
The criterion for a large separation $d$ is such that in the region between the
  atoms the L- and R-densities have reached the regime of exponential decay;
  see below.\relax
\mciteBstWouldAddEndPunctfalse
\mciteSetBstMidEndSepPunct{\mcitedefaultmidpunct}
{}{\mcitedefaultseppunct}\relax
\EndOfBibitem
\bibitem[Prodan and Kohn(2005)Prodan, and Kohn]{ProdanKohn05}
Prodan,~E.; Kohn,~W. {Nearsightedness of electronic matter.} \emph{PNAS}
  \textbf{2005}, \emph{102}, 11635--11638\relax
\mciteBstWouldAddEndPuncttrue
\mciteSetBstMidEndSepPunct{\mcitedefaultmidpunct}
{\mcitedefaultendpunct}{\mcitedefaultseppunct}\relax
\EndOfBibitem
\bibitem[Note2()]{Note2}
It is also possible, both in the exact case and within common xc
  approximations, that such hopping of an electron will not solve the problem:
  in the system $\protect \mathrm {L}^+ \protect \cdots \protect \mathrm {R}^-$
  the lu of $\protect \mathrm {L}^+$ will lie below the ho of $\protect \mathrm
  {R}^-$, which will require the electron to jump back. In common
  approximations, such as the LSDA, this results in a spuriously fractional
  number of electrons on each of the atoms (see Refs.~\cite
  {PPLB82,KraislerKronik15} and references therein), violating the principle of
  integer preference~\cite {Perdew90}, and being the manifestation of the
  \protect \emph {delocalization error}~\cite {MoriS08}\relax
\mciteBstWouldAddEndPuncttrue
\mciteSetBstMidEndSepPunct{\mcitedefaultmidpunct}
{\mcitedefaultendpunct}{\mcitedefaultseppunct}\relax
\EndOfBibitem
\bibitem[Note3()]{Note3}
For generality, we introduced two constants here, $v'$ and $v''$, to allow both
  atomic potentials to be vertically shifted. In the case depicted in Fig.~\ref
  {fig:graphs_illustration_LR} it is actually convenient to set $v'$ to 0, thus
  far from both atoms $v^\protect \mathrm {KS}_\protect \mathrm {L \protect
  \cdots R}(\protect \bm {r})$ approaches 0.\relax
\mciteBstWouldAddEndPunctfalse
\mciteSetBstMidEndSepPunct{\mcitedefaultmidpunct}
{}{\mcitedefaultseppunct}\relax
\EndOfBibitem
\bibitem[Note4()]{Note4}
Hartree atomic units are used throughout.\relax
\mciteBstWouldAddEndPunctfalse
\mciteSetBstMidEndSepPunct{\mcitedefaultmidpunct}
{}{\mcitedefaultseppunct}\relax
\EndOfBibitem
\bibitem[Katriel and Davidson(1980)Katriel, and Davidson]{KatrielDavidson80}
Katriel,~J.; Davidson,~E.~R. Asymptotic behavior of atomic and molecular wave
  functions. \emph{PNAS} \textbf{1980}, \emph{77}, 4403--4406\relax
\mciteBstWouldAddEndPuncttrue
\mciteSetBstMidEndSepPunct{\mcitedefaultmidpunct}
{\mcitedefaultendpunct}{\mcitedefaultseppunct}\relax
\EndOfBibitem
\bibitem[Hoffmann-Ostenhof and Hoffmann-Ostenhof(1977)Hoffmann-Ostenhof, and
  Hoffmann-Ostenhof]{PhysRevA.16.1782}
Hoffmann-Ostenhof,~M.; Hoffmann-Ostenhof,~T. "Schr\"odinger inequalities" and
  asymptotic behavior of the electron density of atoms and molecules.
  \emph{Phys. Rev. A} \textbf{1977}, \emph{16}, 1782--1785\relax
\mciteBstWouldAddEndPuncttrue
\mciteSetBstMidEndSepPunct{\mcitedefaultmidpunct}
{\mcitedefaultendpunct}{\mcitedefaultseppunct}\relax
\EndOfBibitem
\bibitem[Gori-Giorgi and Baerends(2018)Gori-Giorgi, and Baerends]{GoriGiorgi18}
Gori-Giorgi,~P.; Baerends,~E.~J. {Asymptotic nodal planes in the electron
  density and the potential in the effective equation for the square root of
  the density}. \emph{Eur. Phys. J. B} \textbf{2018}, \emph{91}, 160\relax
\mciteBstWouldAddEndPuncttrue
\mciteSetBstMidEndSepPunct{\mcitedefaultmidpunct}
{\mcitedefaultendpunct}{\mcitedefaultseppunct}\relax
\EndOfBibitem
\bibitem[Kraisler()]{Kraisler_IJC20}
Kraisler,~E. Asymptotic Behavior of the Exchange-Correlation Energy Density and
  the Kohn-Sham Potential in Density Functional Theory: Exact Results and
  Strategy for Approximations. \emph{Isr. J. Chem.} doi:
  10.1002/ijch.201900103\relax
\mciteBstWouldAddEndPuncttrue
\mciteSetBstMidEndSepPunct{\mcitedefaultmidpunct}
{\mcitedefaultendpunct}{\mcitedefaultseppunct}\relax
\EndOfBibitem
\bibitem[Note5()]{Note5}
We note in passing that the criterion for the separation $d$ to be considered
  large follows from the exponential decay rate analysis we just performed: $d$
  has to be larger than the decay lengths of both Atoms L and R, i.e., $d \gg
  1/\protect \sqrt {I_\protect \mathrm {L}}$ and $d \gg 1/\protect \sqrt
  {I_\protect \mathrm {R}}$.\relax
\mciteBstWouldAddEndPunctfalse
\mciteSetBstMidEndSepPunct{\mcitedefaultmidpunct}
{}{\mcitedefaultseppunct}\relax
\EndOfBibitem
\bibitem[Wetherell \latin{et~al.}(2019)Wetherell, Hodgson, Talirz, and
  Godby]{PhysRevB.99.045129}
Wetherell,~J.; Hodgson,~M. J.~P.; Talirz,~L.; Godby,~R.~W. Advantageous
  nearsightedness of many-body perturbation theory contrasted with Kohn-Sham
  density functional theory. \emph{Phys. Rev. B} \textbf{2019}, \emph{99},
  045129\relax
\mciteBstWouldAddEndPuncttrue
\mciteSetBstMidEndSepPunct{\mcitedefaultmidpunct}
{\mcitedefaultendpunct}{\mcitedefaultseppunct}\relax
\EndOfBibitem
\bibitem[Note6()]{Note6}
In the bonded case both the bonding and the anti-bonding molecular orbitals
  delocalize over both atoms, and in the infinite limit the bonding orbital can
  equally be described by two half-filled orbitals of the same energy, one
  localized on L, and one on R.\relax
\mciteBstWouldAddEndPunctfalse
\mciteSetBstMidEndSepPunct{\mcitedefaultmidpunct}
{}{\mcitedefaultseppunct}\relax
\EndOfBibitem
\bibitem[Harbola(1998)]{Harbola98}
Harbola,~M.~K. Differential virial theorem for the fractional electron number:
  Derivative discontinuity of the Kohn-Sham exchange-correlation potential.
  \emph{Phys. Rev. A} \textbf{1998}, \emph{57}, 4253\relax
\mciteBstWouldAddEndPuncttrue
\mciteSetBstMidEndSepPunct{\mcitedefaultmidpunct}
{\mcitedefaultendpunct}{\mcitedefaultseppunct}\relax
\EndOfBibitem
\bibitem[Yang \latin{et~al.}(2012)Yang, Cohen, and Mori-S\'{a}nchez]{Yang12}
Yang,~W.; Cohen,~A.~J.; Mori-S\'{a}nchez,~P. {Derivative discontinuity, bandgap
  and lowest unoccupied molecular orbital in density functional theory.}
  \emph{J. Chem. Phys.} \textbf{2012}, \emph{136}, 204111\relax
\mciteBstWouldAddEndPuncttrue
\mciteSetBstMidEndSepPunct{\mcitedefaultmidpunct}
{\mcitedefaultendpunct}{\mcitedefaultseppunct}\relax
\EndOfBibitem
\bibitem[Cohen and Wasserman(2007)Cohen, and Wasserman]{cohen2007foundations}
Cohen,~M.~H.; Wasserman,~A. On the foundations of chemical reactivity theory.
  \emph{J. Phys. Chem. A} \textbf{2007}, \emph{111}, 2229--2242\relax
\mciteBstWouldAddEndPuncttrue
\mciteSetBstMidEndSepPunct{\mcitedefaultmidpunct}
{\mcitedefaultendpunct}{\mcitedefaultseppunct}\relax
\EndOfBibitem
\bibitem[Oueis and Wasserman(2018)Oueis, and Wasserman]{oueis2018exact}
Oueis,~Y.; Wasserman,~A. Exact partition potential for model systems of
  interacting electrons in 1-D. \emph{The European Physical Journal B}
  \textbf{2018}, \emph{91}, 247\relax
\mciteBstWouldAddEndPuncttrue
\mciteSetBstMidEndSepPunct{\mcitedefaultmidpunct}
{\mcitedefaultendpunct}{\mcitedefaultseppunct}\relax
\EndOfBibitem
\bibitem[Elliott \latin{et~al.}(2010)Elliott, Burke, Cohen, and
  Wasserman]{elliott2010partition}
Elliott,~P.; Burke,~K.; Cohen,~M.~H.; Wasserman,~A. Partition
  density-functional theory. \emph{Phys. Rev. A} \textbf{2010}, \emph{82},
  024501\relax
\mciteBstWouldAddEndPuncttrue
\mciteSetBstMidEndSepPunct{\mcitedefaultmidpunct}
{\mcitedefaultendpunct}{\mcitedefaultseppunct}\relax
\EndOfBibitem
\bibitem[G{\'o}mez \latin{et~al.}(2019)G{\'o}mez, Oueis, Restrepo, and
  Wasserman]{gomez2019partition}
G{\'o}mez,~S.; Oueis,~Y.; Restrepo,~A.; Wasserman,~A. Partition potential for
  hydrogen bonding in formic acid dimers. \emph{Int. J. Quantum Chem.}
  \textbf{2019}, \emph{119}, e25814\relax
\mciteBstWouldAddEndPuncttrue
\mciteSetBstMidEndSepPunct{\mcitedefaultmidpunct}
{\mcitedefaultendpunct}{\mcitedefaultseppunct}\relax
\EndOfBibitem
\bibitem[Nafziger \latin{et~al.}(2011)Nafziger, Wu, and
  Wasserman]{nafziger2011molecular}
Nafziger,~J.; Wu,~Q.; Wasserman,~A. Molecular binding energies from partition
  density functional theory. \emph{J. Chem. Phys.} \textbf{2011}, \emph{135},
  234101\relax
\mciteBstWouldAddEndPuncttrue
\mciteSetBstMidEndSepPunct{\mcitedefaultmidpunct}
{\mcitedefaultendpunct}{\mcitedefaultseppunct}\relax
\EndOfBibitem
\bibitem[Nafziger and Wasserman(2015)Nafziger, and
  Wasserman]{nafziger2015fragment}
Nafziger,~J.; Wasserman,~A. Fragment-based treatment of delocalization and
  static correlation errors in density-functional theory. \emph{J. Chem. Phys.}
  \textbf{2015}, \emph{143}, 234105\relax
\mciteBstWouldAddEndPuncttrue
\mciteSetBstMidEndSepPunct{\mcitedefaultmidpunct}
{\mcitedefaultendpunct}{\mcitedefaultseppunct}\relax
\EndOfBibitem
\bibitem[Jiang \latin{et~al.}(2018)Jiang, Nafziger, and
  Wasserman]{jiang2018constructing}
Jiang,~K.; Nafziger,~J.; Wasserman,~A. Constructing a non-additive
  non-interacting kinetic energy functional approximation for covalent bonds
  from exact conditions. \emph{J. Chem. Phys.} \textbf{2018}, \emph{149},
  164112\relax
\mciteBstWouldAddEndPuncttrue
\mciteSetBstMidEndSepPunct{\mcitedefaultmidpunct}
{\mcitedefaultendpunct}{\mcitedefaultseppunct}\relax
\EndOfBibitem
\bibitem[Cohen \latin{et~al.}(2009)Cohen, Wasserman, Car, and
  Burke]{cohen2009charge}
Cohen,~M.~H.; Wasserman,~A.; Car,~R.; Burke,~K. Charge transfer in partition
  theory. \emph{J. Phys. Chem. A} \textbf{2009}, \emph{113}, 2183--2192\relax
\mciteBstWouldAddEndPuncttrue
\mciteSetBstMidEndSepPunct{\mcitedefaultmidpunct}
{\mcitedefaultendpunct}{\mcitedefaultseppunct}\relax
\EndOfBibitem
\bibitem[Nafziger and Wasserman(2014)Nafziger, and
  Wasserman]{nafziger2014density}
Nafziger,~J.; Wasserman,~A. Density-based partitioning methods for ground-state
  molecular calculations. \emph{J. Phys. Chem. A} \textbf{2014}, \emph{118},
  7623--7639\relax
\mciteBstWouldAddEndPuncttrue
\mciteSetBstMidEndSepPunct{\mcitedefaultmidpunct}
{\mcitedefaultendpunct}{\mcitedefaultseppunct}\relax
\EndOfBibitem
\bibitem[G{\'o}mez \latin{et~al.}(2017)G{\'o}mez, Nafziger, Restrepo, and
  Wasserman]{gomez2017partition}
G{\'o}mez,~S.; Nafziger,~J.; Restrepo,~A.; Wasserman,~A. Partition-DFT on the
  water dimer. \emph{J. Chem. Phys.} \textbf{2017}, \emph{146}, 074106\relax
\mciteBstWouldAddEndPuncttrue
\mciteSetBstMidEndSepPunct{\mcitedefaultmidpunct}
{\mcitedefaultendpunct}{\mcitedefaultseppunct}\relax
\EndOfBibitem
\bibitem[Ch{\'a}vez and Wasserman(2020)Ch{\'a}vez, and
  Wasserman]{chavez2020towards}
Ch{\'a}vez,~V.~H.; Wasserman,~A. Towards a density functional theory of
  molecular fragments. What is the shape of atoms in molecules? \emph{Revista
  de la Academia Colombiana de Ciencias Exactas, F{\'\i}sicas y Naturales}
  \textbf{2020}, \emph{44}, 269--279\relax
\mciteBstWouldAddEndPuncttrue
\mciteSetBstMidEndSepPunct{\mcitedefaultmidpunct}
{\mcitedefaultendpunct}{\mcitedefaultseppunct}\relax
\EndOfBibitem
\bibitem[Landau and Lifshitz(1991)Landau, and Lifshitz]{LL3}
Landau,~L.~D.; Lifshitz,~E.~M. \emph{Quantum Mechanics (Non-Relativistic
  Theory)}, 3rd ed.; Course of Theoretical Physics; Pergamon, 1991;
  Vol.~3\relax
\mciteBstWouldAddEndPuncttrue
\mciteSetBstMidEndSepPunct{\mcitedefaultmidpunct}
{\mcitedefaultendpunct}{\mcitedefaultseppunct}\relax
\EndOfBibitem
\bibitem[Lieb(1983)]{Lieb}
Lieb,~E.~H. Density Functionals for Coulomb Systems. \emph{Int. J. Quantum
  Chem.} \textbf{1983}, \emph{24}, 243--277\relax
\mciteBstWouldAddEndPuncttrue
\mciteSetBstMidEndSepPunct{\mcitedefaultmidpunct}
{\mcitedefaultendpunct}{\mcitedefaultseppunct}\relax
\EndOfBibitem
\bibitem[van Leeuwen()]{RvL_PhD}
van Leeuwen,~R. {K}ohn-{S}ham Potentials in Density Functional Theory. Ph.D.\
  thesis, Vrije Universiteit, Amsterdam, The Netherlands (1994)\relax
\mciteBstWouldAddEndPuncttrue
\mciteSetBstMidEndSepPunct{\mcitedefaultmidpunct}
{\mcitedefaultendpunct}{\mcitedefaultseppunct}\relax
\EndOfBibitem
\bibitem[van Leeuwen(2003)]{RvL_adv}
van Leeuwen,~R. Density functional approach to the many-body problem: key
  concepts and exact functionals. \emph{Adv. Quantum Chem.} \textbf{2003},
  \emph{43}, 24\relax
\mciteBstWouldAddEndPuncttrue
\mciteSetBstMidEndSepPunct{\mcitedefaultmidpunct}
{\mcitedefaultendpunct}{\mcitedefaultseppunct}\relax
\EndOfBibitem
\bibitem[Janak(1978)]{Janak78}
Janak,~J.~F. Proof that $\partial{}E\partial{}n_i=\epsilon{}$ in
  density-functional theory. \emph{Phys. Rev. B} \textbf{1978}, \emph{18},
  7165--7168\relax
\mciteBstWouldAddEndPuncttrue
\mciteSetBstMidEndSepPunct{\mcitedefaultmidpunct}
{\mcitedefaultendpunct}{\mcitedefaultseppunct}\relax
\EndOfBibitem
\bibitem[Yang \latin{et~al.}(2000)Yang, Zhang, and Ayers]{Yang00}
Yang,~W.; Zhang,~Y.; Ayers,~P.~W. {Degenerate Ground States and a Fractional
  Number of Electrons in Density and Reduced Density Matrix Functional Theory}.
  \emph{Phys. Rev. Lett.} \textbf{2000}, \emph{84}, 5172--5175\relax
\mciteBstWouldAddEndPuncttrue
\mciteSetBstMidEndSepPunct{\mcitedefaultmidpunct}
{\mcitedefaultendpunct}{\mcitedefaultseppunct}\relax
\EndOfBibitem
\bibitem[Senjean and Fromager()Senjean, and Fromager]{doi:10.1002/qua.26190}
Senjean,~B.; Fromager,~E. N-centered ensemble density-functional theory for
  open systems. \emph{Int. J. Quantum Chem.} \emph{n/a}, e26190\relax
\mciteBstWouldAddEndPuncttrue
\mciteSetBstMidEndSepPunct{\mcitedefaultmidpunct}
{\mcitedefaultendpunct}{\mcitedefaultseppunct}\relax
\EndOfBibitem
\bibitem[Sai \latin{et~al.}(2011)Sai, Barbara, and Leung]{Sai11}
Sai,~N.; Barbara,~P.~F.; Leung,~K. \emph{Phys. Rev. Lett.} \textbf{2011},
  \emph{106}, 226403\relax
\mciteBstWouldAddEndPuncttrue
\mciteSetBstMidEndSepPunct{\mcitedefaultmidpunct}
{\mcitedefaultendpunct}{\mcitedefaultseppunct}\relax
\EndOfBibitem
\bibitem[Zheng \latin{et~al.}(2011)Zheng, Cohen, Mori-S\'{a}nchez, Hu, and
  Yang]{Zheng11}
Zheng,~X.; Cohen,~A.~J.; Mori-S\'{a}nchez,~P.; Hu,~X.; Yang,~W. {Improving Band
  Gap Prediction in Density Functional Theory from Molecules to Solids}.
  \emph{Phys. Rev. Lett.} \textbf{2011}, \emph{107}, 026403\relax
\mciteBstWouldAddEndPuncttrue
\mciteSetBstMidEndSepPunct{\mcitedefaultmidpunct}
{\mcitedefaultendpunct}{\mcitedefaultseppunct}\relax
\EndOfBibitem
\bibitem[Refaely-Abramson \latin{et~al.}(2011)Refaely-Abramson, Baer, and
  Kronik]{Refaely11}
Refaely-Abramson,~S.; Baer,~R.; Kronik,~L. {Fundamental and excitation gaps in
  molecules of relevance for organic photovoltaics from an optimally tuned
  range-separated hybrid functional}. \emph{Phys. Rev. B} \textbf{2011},
  \emph{84}, 075144\relax
\mciteBstWouldAddEndPuncttrue
\mciteSetBstMidEndSepPunct{\mcitedefaultmidpunct}
{\mcitedefaultendpunct}{\mcitedefaultseppunct}\relax
\EndOfBibitem
\bibitem[Yang \latin{et~al.}(2013)Yang, Mori-S{\'{a}}nchez, and Cohen]{Yang13}
Yang,~W.; Mori-S{\'{a}}nchez,~P.; Cohen,~A.~J. {Extension of many-body theory
  and approximate density functionals to fractional charges and fractional
  spins}. \emph{J. Chem. Phys.} \textbf{2013}, \emph{139}, 104114\relax
\mciteBstWouldAddEndPuncttrue
\mciteSetBstMidEndSepPunct{\mcitedefaultmidpunct}
{\mcitedefaultendpunct}{\mcitedefaultseppunct}\relax
\EndOfBibitem
\bibitem[Atalla \latin{et~al.}(2013)Atalla, Yoon, Caruso, Rinke, and
  Scheffler]{Atalla13}
Atalla,~V.; Yoon,~M.; Caruso,~F.; Rinke,~P.; Scheffler,~M. Hybrid density
  functional theory meets quasiparticle calculations: A consistent electronic
  structure approach. \emph{Phys. Rev. B} \textbf{2013}, \emph{88},
  165122\relax
\mciteBstWouldAddEndPuncttrue
\mciteSetBstMidEndSepPunct{\mcitedefaultmidpunct}
{\mcitedefaultendpunct}{\mcitedefaultseppunct}\relax
\EndOfBibitem
\bibitem[Armiento and K\"ummel(2013)Armiento, and K\"ummel]{ArmientoKummel13}
Armiento,~R.; K\"ummel,~S. Orbital Localization, Charge Transfer, and Band Gaps
  in Semilocal Density-Functional Theory. \emph{Phys. Rev. Lett.}
  \textbf{2013}, \emph{111}, 036402\relax
\mciteBstWouldAddEndPuncttrue
\mciteSetBstMidEndSepPunct{\mcitedefaultmidpunct}
{\mcitedefaultendpunct}{\mcitedefaultseppunct}\relax
\EndOfBibitem
\bibitem[Refaely-Abramson \latin{et~al.}(2013)Refaely-Abramson, Sharifzadeh,
  Jain, Baer, Neaton, and Kronik]{Refaely13}
Refaely-Abramson,~S.; Sharifzadeh,~S.; Jain,~M.; Baer,~R.; Neaton,~J.~B.;
  Kronik,~L. {Gap renormalization of molecular crystals from density-functional
  theory}. \emph{Phys. Rev. B} \textbf{2013}, \emph{88}, 081204\relax
\mciteBstWouldAddEndPuncttrue
\mciteSetBstMidEndSepPunct{\mcitedefaultmidpunct}
{\mcitedefaultendpunct}{\mcitedefaultseppunct}\relax
\EndOfBibitem
\bibitem[Dabo \latin{et~al.}(2014)Dabo, Ferretti, and Marzari]{Dabo14}
Dabo,~I.; Ferretti,~A.; Marzari,~N. \emph{Top. Curr. Chem.} \textbf{2014},
  \emph{347}, 193\relax
\mciteBstWouldAddEndPuncttrue
\mciteSetBstMidEndSepPunct{\mcitedefaultmidpunct}
{\mcitedefaultendpunct}{\mcitedefaultseppunct}\relax
\EndOfBibitem
\bibitem[Borghi \latin{et~al.}(2014)Borghi, Ferretti, Nguyen, Dabo, and
  Marzari]{Borghi14}
Borghi,~G.; Ferretti,~A.; Nguyen,~N.~L.; Dabo,~I.; Marzari,~N. \emph{Phys. Rev.
  B} \textbf{2014}, \emph{90}, 075135\relax
\mciteBstWouldAddEndPuncttrue
\mciteSetBstMidEndSepPunct{\mcitedefaultmidpunct}
{\mcitedefaultendpunct}{\mcitedefaultseppunct}\relax
\EndOfBibitem
\bibitem[Borghi \latin{et~al.}(2015)Borghi, Park, Nguyen, Ferretti, and
  Marzari]{Borghi15}
Borghi,~G.; Park,~C.-H.; Nguyen,~N.~L.; Ferretti,~A.; Marzari,~N. Variational
  minimization of orbital-density-dependent functionals. \emph{Phys. Rev. B}
  \textbf{2015}, \emph{91}, 155112\relax
\mciteBstWouldAddEndPuncttrue
\mciteSetBstMidEndSepPunct{\mcitedefaultmidpunct}
{\mcitedefaultendpunct}{\mcitedefaultseppunct}\relax
\EndOfBibitem
\bibitem[Refaely-Abramson \latin{et~al.}(2015)Refaely-Abramson, Jain,
  Sharifzadeh, Neaton, and Kronik]{Refaely15}
Refaely-Abramson,~S.; Jain,~M.; Sharifzadeh,~S.; Neaton,~J.~B.; Kronik,~L.
  {Solid-state optical absorption from optimally-tuned time-dependent
  range-separated hybrid density functional theory}. \emph{Phys. Rev. B}
  \textbf{2015}, \emph{92}, 081204\relax
\mciteBstWouldAddEndPuncttrue
\mciteSetBstMidEndSepPunct{\mcitedefaultmidpunct}
{\mcitedefaultendpunct}{\mcitedefaultseppunct}\relax
\EndOfBibitem
\bibitem[Kraisler and Kronik(2015)Kraisler, and Kronik]{KraislerKronik15}
Kraisler,~E.; Kronik,~L. Elimination of the asymptotic fractional dissociation
  problem in Kohn-Sham density functional theory using the
  ensemble-generalization approach. \emph{Phys. Rev. A} \textbf{2015},
  \emph{91}, 032504\relax
\mciteBstWouldAddEndPuncttrue
\mciteSetBstMidEndSepPunct{\mcitedefaultmidpunct}
{\mcitedefaultendpunct}{\mcitedefaultseppunct}\relax
\EndOfBibitem
\bibitem[Kraisler \latin{et~al.}(2015)Kraisler, Schmidt, K\"{u}mmel, and
  Kronik]{KraislerSchmidt15}
Kraisler,~E.; Schmidt,~T.; K\"{u}mmel,~S.; Kronik,~L. Effect of ensemble
  generalization on the highest-occupied Kohn-Sham eigenvalue. \emph{J. Chem.
  Phys.} \textbf{2015}, \emph{143}, 104105\relax
\mciteBstWouldAddEndPuncttrue
\mciteSetBstMidEndSepPunct{\mcitedefaultmidpunct}
{\mcitedefaultendpunct}{\mcitedefaultseppunct}\relax
\EndOfBibitem
\bibitem[Nguyen \latin{et~al.}(2015)Nguyen, Borghi, Ferretti, Dabo, and
  Marzari]{Nguyen15}
Nguyen,~N.~L.; Borghi,~G.; Ferretti,~A.; Dabo,~I.; Marzari,~N. First-Principles
  Photoemission Spectroscopy and Orbital Tomography in Molecules from
  Koopmans-Compliant Functionals. \emph{Phys. Rev. Lett.} \textbf{2015},
  \emph{114}, 166405\relax
\mciteBstWouldAddEndPuncttrue
\mciteSetBstMidEndSepPunct{\mcitedefaultmidpunct}
{\mcitedefaultendpunct}{\mcitedefaultseppunct}\relax
\EndOfBibitem
\bibitem[Li \latin{et~al.}(2015)Li, Zheng, Cohen, Mori-S{\'{a}}nchez, and
  Yang]{LiZhengYang15}
Li,~C.; Zheng,~X.; Cohen,~A.~J.; Mori-S{\'{a}}nchez,~P.; Yang,~W. {Local
  Scaling Correction for Reducing Delocalization Error in Density Functional
  Approximations}. \emph{Phys. Rev. Lett.} \textbf{2015}, \emph{114},
  053001\relax
\mciteBstWouldAddEndPuncttrue
\mciteSetBstMidEndSepPunct{\mcitedefaultmidpunct}
{\mcitedefaultendpunct}{\mcitedefaultseppunct}\relax
\EndOfBibitem
\bibitem[Yang \latin{et~al.}(2016)Yang, Peng, Sun, and Perdew]{Yang16}
Yang,~Z.-h.; Peng,~H.; Sun,~J.; Perdew,~J.~P. More realistic band gaps from
  meta-genera\-lized gradient approximations: Only in a generalized Kohn-Sham
  scheme. \emph{Phys. Rev. B} \textbf{2016}, \emph{93}, 205205\relax
\mciteBstWouldAddEndPuncttrue
\mciteSetBstMidEndSepPunct{\mcitedefaultmidpunct}
{\mcitedefaultendpunct}{\mcitedefaultseppunct}\relax
\EndOfBibitem
\bibitem[Atalla \latin{et~al.}(2016)Atalla, Zhang, Hofmann, Ren, Rinke, and
  Scheffler]{Atalla16}
Atalla,~V.; Zhang,~I.~Y.; Hofmann,~O.~T.; Ren,~X.; Rinke,~P.; Scheffler,~M.
  Enforcing the linear behavior of the total energy with hybrid functionals:
  Implications for charge transfer, interaction energies, and the random-phase
  approximation. \emph{Phys. Rev. B} \textbf{2016}, \emph{94}, 035140\relax
\mciteBstWouldAddEndPuncttrue
\mciteSetBstMidEndSepPunct{\mcitedefaultmidpunct}
{\mcitedefaultendpunct}{\mcitedefaultseppunct}\relax
\EndOfBibitem
\bibitem[Tran \latin{et~al.}(2016)Tran, Stezl, and Blaha]{Tran16}
Tran,~F.; Stezl,~J.; Blaha,~P. Rungs 1 to 4 of {DFT} {J}acob's ladder:
  Extensive test on the lattice constant, bulk modulus, and cohesive energy of
  solids. \emph{J. Chem. Phys.} \textbf{2016}, \emph{144}, 204120\relax
\mciteBstWouldAddEndPuncttrue
\mciteSetBstMidEndSepPunct{\mcitedefaultmidpunct}
{\mcitedefaultendpunct}{\mcitedefaultseppunct}\relax
\EndOfBibitem
\bibitem[Nguyen \latin{et~al.}(2018)Nguyen, Colonna, Ferretti, and
  Marzari]{NguyenColonna18}
Nguyen,~N.~L.; Colonna,~N.; Ferretti,~A.; Marzari,~N. Koopmans-Compliant
  Spectral Functionals for Extended Systems. \emph{Phys. Rev. X} \textbf{2018},
  \emph{8}, 021051\relax
\mciteBstWouldAddEndPuncttrue
\mciteSetBstMidEndSepPunct{\mcitedefaultmidpunct}
{\mcitedefaultendpunct}{\mcitedefaultseppunct}\relax
\EndOfBibitem
\bibitem[Senjean and Fromager(2018)Senjean, and Fromager]{SenjeanFromager18}
Senjean,~B.; Fromager,~E. Unified formulation of fundamental and optical gap
  problems in density-functional theory for ensembles. \emph{Phys. Rev. A}
  \textbf{2018}, \emph{98}, 022513\relax
\mciteBstWouldAddEndPuncttrue
\mciteSetBstMidEndSepPunct{\mcitedefaultmidpunct}
{\mcitedefaultendpunct}{\mcitedefaultseppunct}\relax
\EndOfBibitem
\bibitem[Kronik and K\"ummel(2018)Kronik, and K\"ummel]{KronikKummel18_review}
Kronik,~L.; K\"ummel,~S. Dielectric Screening Meets Optimally Tuned Density
  Functionals. \emph{Adv. Mater.} \textbf{2018}, \emph{30}, 1706560\relax
\mciteBstWouldAddEndPuncttrue
\mciteSetBstMidEndSepPunct{\mcitedefaultmidpunct}
{\mcitedefaultendpunct}{\mcitedefaultseppunct}\relax
\EndOfBibitem
\bibitem[Gould \latin{et~al.}(2019)Gould, Pittalis, Toulouse, Kraisler, and
  Kronik]{Gould19}
Gould,~T.; Pittalis,~S.; Toulouse,~J.; Kraisler,~E.; Kronik,~L. {A}symptotic
  Behavior of the {H}artree-exchange and Correlation Potentials in ensemble
  density functional theory. \emph{Phys. Chem. Chem. Phys.} \textbf{2019},
  \emph{21}, 19805\relax
\mciteBstWouldAddEndPuncttrue
\mciteSetBstMidEndSepPunct{\mcitedefaultmidpunct}
{\mcitedefaultendpunct}{\mcitedefaultseppunct}\relax
\EndOfBibitem
\bibitem[Deur and Fromager(2019)Deur, and Fromager]{DeurFromager19}
Deur,~K.; Fromager,~E. Ground and excited energy levels can be extracted
  exactly from a single ensemble density-functional theory calculation.
  \emph{J. Chem. Phys.} \textbf{2019}, \emph{150}, 094106\relax
\mciteBstWouldAddEndPuncttrue
\mciteSetBstMidEndSepPunct{\mcitedefaultmidpunct}
{\mcitedefaultendpunct}{\mcitedefaultseppunct}\relax
\EndOfBibitem
\bibitem[Wing \latin{et~al.}(2019)Wing, Haber, Noff, Barker, Egger,
  Ramasubramaniam, Louie, Neaton, and Kronik]{WingKronik19}
Wing,~D.; Haber,~J.~B.; Noff,~R.; Barker,~B.; Egger,~D.~A.;
  Ramasubramaniam,~A.; Louie,~S.~G.; Neaton,~J.~B.; Kronik,~L. Comparing
  time-dependent density functional theory with many-body perturbation theory
  for semiconductors: Screened range-separated hybrids and the $GW$ plus
  Bethe-Salpeter approach. \emph{Phys. Rev. Materials} \textbf{2019}, \emph{3},
  064603\relax
\mciteBstWouldAddEndPuncttrue
\mciteSetBstMidEndSepPunct{\mcitedefaultmidpunct}
{\mcitedefaultendpunct}{\mcitedefaultseppunct}\relax
\EndOfBibitem
\bibitem[Dreuw \latin{et~al.}(2003)Dreuw, Weisman, and
  Head-Gordon]{dreuw2003long}
Dreuw,~A.; Weisman,~J.~L.; Head-Gordon,~M. Long-range charge-transfer excited
  states in time-dependent density functional theory require non-local
  exchange. \emph{J. Chem. Phys.} \textbf{2003}, \emph{119}, 2943--2946\relax
\mciteBstWouldAddEndPuncttrue
\mciteSetBstMidEndSepPunct{\mcitedefaultmidpunct}
{\mcitedefaultendpunct}{\mcitedefaultseppunct}\relax
\EndOfBibitem
\bibitem[Gritsenko and Baerends(2004)Gritsenko, and
  Baerends]{gritsenko2004asymptotic}
Gritsenko,~O.; Baerends,~E.~J. Asymptotic correction of the
  exchange--correlation kernel of time-dependent density functional theory for
  long-range charge-transfer excitations. \emph{J. Chem. Phys.} \textbf{2004},
  \emph{121}, 655--660\relax
\mciteBstWouldAddEndPuncttrue
\mciteSetBstMidEndSepPunct{\mcitedefaultmidpunct}
{\mcitedefaultendpunct}{\mcitedefaultseppunct}\relax
\EndOfBibitem
\bibitem[Gross \latin{et~al.}(1988)Gross, Oliveira, and Kohn]{GOK1}
Gross,~E.~K.~U.; Oliveira,~L.~N.; Kohn,~W. {Rayleigh-Ritz variational principle
  for ensembles of fractionally occupied states}. \emph{Phys. Rev. A}
  \textbf{1988}, \emph{37}, 2805\relax
\mciteBstWouldAddEndPuncttrue
\mciteSetBstMidEndSepPunct{\mcitedefaultmidpunct}
{\mcitedefaultendpunct}{\mcitedefaultseppunct}\relax
\EndOfBibitem
\bibitem[Gross \latin{et~al.}(1988)Gross, Oliveira, and Kohn]{GOK2}
Gross,~E.~K.~U.; Oliveira,~L.~N.; Kohn,~W. {Density-functional theory for
  ensembles of fractionall occupied states. I. Basic formalism}. \emph{Phys.
  Rev. A} \textbf{1988}, \emph{37}, 2809\relax
\mciteBstWouldAddEndPuncttrue
\mciteSetBstMidEndSepPunct{\mcitedefaultmidpunct}
{\mcitedefaultendpunct}{\mcitedefaultseppunct}\relax
\EndOfBibitem
\bibitem[Oliveira \latin{et~al.}(1988)Oliveira, Gross, and Kohn]{GOK3}
Oliveira,~L.~N.; Gross,~E.~K.~U.; Kohn,~W. {Density-funcional theory for
  ensembles of fractionally occupied states. II. Application to the He atom}.
  \emph{Phys. Rev. A} \textbf{1988}, \emph{37}, 2821\relax
\mciteBstWouldAddEndPuncttrue
\mciteSetBstMidEndSepPunct{\mcitedefaultmidpunct}
{\mcitedefaultendpunct}{\mcitedefaultseppunct}\relax
\EndOfBibitem
\bibitem[Deur \latin{et~al.}(2017)Deur, Mazouin, and
  Fromager]{PhysRevB.95.035120}
Deur,~K.; Mazouin,~L.; Fromager,~E. Exact ensemble density functional theory
  for excited states in a model system: Investigating the weight dependence of
  the correlation energy. \emph{Phys. Rev. B} \textbf{2017}, \emph{95},
  035120\relax
\mciteBstWouldAddEndPuncttrue
\mciteSetBstMidEndSepPunct{\mcitedefaultmidpunct}
{\mcitedefaultendpunct}{\mcitedefaultseppunct}\relax
\EndOfBibitem
\bibitem[Loos and Fromager(2020)Loos, and Fromager]{loos2020weight}
Loos,~P.-F.; Fromager,~E. A weight-dependent local correlation
  density-functional approximation for ensembles. \emph{arXiv preprint
  arXiv:2003.05553} \textbf{2020}, \relax
\mciteBstWouldAddEndPunctfalse
\mciteSetBstMidEndSepPunct{\mcitedefaultmidpunct}
{}{\mcitedefaultseppunct}\relax
\EndOfBibitem
\bibitem[Fromager(2020)]{fromager2020individual}
Fromager,~E. Individual correlations in ensemble density-functional theory:
  State-driven/density-driven decomposition without additional Kohn-Sham
  systems. \emph{arXiv preprint arXiv:2001.08605} \textbf{2020}, \relax
\mciteBstWouldAddEndPunctfalse
\mciteSetBstMidEndSepPunct{\mcitedefaultmidpunct}
{}{\mcitedefaultseppunct}\relax
\EndOfBibitem
\bibitem[Veldman \latin{et~al.}(2009)Veldman, Meskers, and
  Janssen]{veldman2009energy}
Veldman,~D.; Meskers,~S.~C.; Janssen,~R.~A. The energy of charge-transfer
  states in electron donor--acceptor blends: insight into the energy losses in
  organic solar cells. \emph{Adv. Funct. Mater.} \textbf{2009}, \emph{19},
  1939--1948\relax
\mciteBstWouldAddEndPuncttrue
\mciteSetBstMidEndSepPunct{\mcitedefaultmidpunct}
{\mcitedefaultendpunct}{\mcitedefaultseppunct}\relax
\EndOfBibitem
\bibitem[Hodgson \latin{et~al.}(2013)Hodgson, Ramsden, Chapman, Lillystone, and
  Godby]{Hodgson13}
Hodgson,~M. J.~P.; Ramsden,~J.~D.; Chapman,~J. B.~J.; Lillystone,~P.;
  Godby,~R.~W. Exact time-dependent density-functional potentials for strongly
  correlated tunneling electrons. \emph{Phys. Rev. B} \textbf{2013}, \emph{88},
  241102\relax
\mciteBstWouldAddEndPuncttrue
\mciteSetBstMidEndSepPunct{\mcitedefaultmidpunct}
{\mcitedefaultendpunct}{\mcitedefaultseppunct}\relax
\EndOfBibitem
\bibitem[Entwistle and Godby(2019)Entwistle, and Godby]{PhysRevB.99.161102}
Entwistle,~M.~T.; Godby,~R.~W. Exact exchange-correlation kernels for optical
  spectra of model systems. \emph{Phys. Rev. B} \textbf{2019}, \emph{99},
  161102\relax
\mciteBstWouldAddEndPuncttrue
\mciteSetBstMidEndSepPunct{\mcitedefaultmidpunct}
{\mcitedefaultendpunct}{\mcitedefaultseppunct}\relax
\EndOfBibitem
\bibitem[Kraisler \latin{et~al.}(2010)Kraisler, Makov, and
  Kelson]{KraislerMakovKelson10}
Kraisler,~E.; Makov,~G.; Kelson,~I. Ensemble $v$-representable ab initio
  density-functional calculation of energy and spin in atoms: A test of
  exchange-correlation approximations. \emph{Phys. Rev. A} \textbf{2010},
  \emph{82}, 042516\relax
\mciteBstWouldAddEndPuncttrue
\mciteSetBstMidEndSepPunct{\mcitedefaultmidpunct}
{\mcitedefaultendpunct}{\mcitedefaultseppunct}\relax
\EndOfBibitem
\bibitem[Gritsenko and Baerends(1996)Gritsenko, and Baerends]{Gritsenko96}
Gritsenko,~O.~V.; Baerends,~E.~J. {Effect of molecular dissociation on the
  exchange-correlation Kohn-Sham potential}. \emph{Phys. Rev. A} \textbf{1996},
  \emph{54}, 1957--1972\relax
\mciteBstWouldAddEndPuncttrue
\mciteSetBstMidEndSepPunct{\mcitedefaultmidpunct}
{\mcitedefaultendpunct}{\mcitedefaultseppunct}\relax
\EndOfBibitem
\bibitem[Zhang and Yang(2000)Zhang, and Yang]{zhang2000perspective}
Zhang,~Y.; Yang,~W. \emph{Theor. Chem. Acc.}; Springer, 2000; pp 346--348\relax
\mciteBstWouldAddEndPuncttrue
\mciteSetBstMidEndSepPunct{\mcitedefaultmidpunct}
{\mcitedefaultendpunct}{\mcitedefaultseppunct}\relax
\EndOfBibitem
\bibitem[Qian and Sahni(2000)Qian, and Sahni]{PhysRevB.62.16364}
Qian,~Z.; Sahni,~V. Origin of the derivative discontinuity in density
  functional theory. \emph{Phys. Rev. B} \textbf{2000}, \emph{62},
  16364--16369\relax
\mciteBstWouldAddEndPuncttrue
\mciteSetBstMidEndSepPunct{\mcitedefaultmidpunct}
{\mcitedefaultendpunct}{\mcitedefaultseppunct}\relax
\EndOfBibitem
\bibitem[Note7()]{Note7}
The spin of the electron affects the gap in a quantitate but not qualitative
  way~\cite {capelle2010spin}.\relax
\mciteBstWouldAddEndPunctfalse
\mciteSetBstMidEndSepPunct{\mcitedefaultmidpunct}
{}{\mcitedefaultseppunct}\relax
\EndOfBibitem
\bibitem[Gould and Toulouse(2014)Gould, and Toulouse]{GouldToulouse14}
Gould,~T.; Toulouse,~J. {Kohn-Sham potentials in exact density-functional
  theory at noninteger electron numbers}. \emph{Phys. Rev. A} \textbf{2014},
  \emph{90}, 050502 (R)\relax
\mciteBstWouldAddEndPuncttrue
\mciteSetBstMidEndSepPunct{\mcitedefaultmidpunct}
{\mcitedefaultendpunct}{\mcitedefaultseppunct}\relax
\EndOfBibitem
\bibitem[Tempel \latin{et~al.}(2009)Tempel, Martinez, and Maitra]{Tempel09}
Tempel,~D.; Martinez,~T.; Maitra,~N. {Revisiting molecular dissociation in
  density functional theory: A simple model}. \emph{J. Chem. Theory and
  Comput.} \textbf{2009}, \emph{5}, 770--780\relax
\mciteBstWouldAddEndPuncttrue
\mciteSetBstMidEndSepPunct{\mcitedefaultmidpunct}
{\mcitedefaultendpunct}{\mcitedefaultseppunct}\relax
\EndOfBibitem
\bibitem[Helbig \latin{et~al.}(2009)Helbig, Tokatly, and
  Rubio]{HelbigTokatlyRubio09}
Helbig,~N.; Tokatly,~I.~V.; Rubio,~A. {Exact Kohn-Sham potential of strongly
  correlated finite systems}. \emph{J. Chem. Phys.} \textbf{2009}, \emph{131},
  224105\relax
\mciteBstWouldAddEndPuncttrue
\mciteSetBstMidEndSepPunct{\mcitedefaultmidpunct}
{\mcitedefaultendpunct}{\mcitedefaultseppunct}\relax
\EndOfBibitem
\bibitem[Aschebrock \latin{et~al.}(2017)Aschebrock, Armiento, and
  K\"ummel]{Aschebrock17b}
Aschebrock,~T.; Armiento,~R.; K\"ummel,~S. Challenges for semilocal density
  functionals with asymptotically nonvanishing potentials. \emph{Phys. Rev. B}
  \textbf{2017}, \emph{96}, 075140\relax
\mciteBstWouldAddEndPuncttrue
\mciteSetBstMidEndSepPunct{\mcitedefaultmidpunct}
{\mcitedefaultendpunct}{\mcitedefaultseppunct}\relax
\EndOfBibitem
\bibitem[Note8()]{Note8}
For a stretched diatomic molecule the transfer of an infinitesimal amount of
  charge cannot yield a change in the height of the interatomic step because,
  as shown in Sec.~\ref {sec:relationship.Delta} and \ref {sec:excited_atom},
  when the additional or excited charge is infinitesimal the xc potential can
  only change by an overall constant in the vicinity of the atoms, i.e., the
  only change to the potential is at the periphery of the system.\relax
\mciteBstWouldAddEndPunctfalse
\mciteSetBstMidEndSepPunct{\mcitedefaultmidpunct}
{}{\mcitedefaultseppunct}\relax
\EndOfBibitem
\bibitem[Gould \latin{et~al.}(2018)Gould, Kronik, and
  Pittalis]{gould2018charge}
Gould,~T.; Kronik,~L.; Pittalis,~S. Charge transfer excitations from exact and
  approximate ensemble Kohn-Sham theory. \emph{J. Chem. Phys.} \textbf{2018},
  \emph{148}, 174101\relax
\mciteBstWouldAddEndPuncttrue
\mciteSetBstMidEndSepPunct{\mcitedefaultmidpunct}
{\mcitedefaultendpunct}{\mcitedefaultseppunct}\relax
\EndOfBibitem
\bibitem[Note9()]{Note9}
Note our system consists of same-spin electrons and hence each electron
  occupies a distinct KS orbital.\relax
\mciteBstWouldAddEndPunctfalse
\mciteSetBstMidEndSepPunct{\mcitedefaultmidpunct}
{}{\mcitedefaultseppunct}\relax
\EndOfBibitem
\bibitem[Fuks and Maitra(2014)Fuks, and Maitra]{Fuks14}
Fuks,~J.~I.; Maitra,~N.~T. {Charge transfer in time-dependent
  density-functional theory : Insights from the asymmetric Hubbard dimer}.
  \emph{Phys. Rev. A} \textbf{2014}, \emph{89}, 062502\relax
\mciteBstWouldAddEndPuncttrue
\mciteSetBstMidEndSepPunct{\mcitedefaultmidpunct}
{\mcitedefaultendpunct}{\mcitedefaultseppunct}\relax
\EndOfBibitem
\bibitem[Maitra(2017)]{Maitra_2017}
Maitra,~N.~T. Charge transfer in time-dependent density functional theory.
  \emph{J. Phys. Condens. Matter} \textbf{2017}, \emph{29}, 423001\relax
\mciteBstWouldAddEndPuncttrue
\mciteSetBstMidEndSepPunct{\mcitedefaultmidpunct}
{\mcitedefaultendpunct}{\mcitedefaultseppunct}\relax
\EndOfBibitem
\bibitem[Pribram-Jones \latin{et~al.}(2014)Pribram-Jones, Yang, Trail, Burke,
  Needs, and Ullrich]{pribram2014excitations}
Pribram-Jones,~A.; Yang,~Z.-h.; Trail,~J.~R.; Burke,~K.; Needs,~R.~J.;
  Ullrich,~C.~A. Excitations and benchmark ensemble density functional theory
  for two electrons. \emph{The Journal of chemical physics} \textbf{2014},
  \emph{140}, 18A541\relax
\mciteBstWouldAddEndPuncttrue
\mciteSetBstMidEndSepPunct{\mcitedefaultmidpunct}
{\mcitedefaultendpunct}{\mcitedefaultseppunct}\relax
\EndOfBibitem
\bibitem[Yang \latin{et~al.}(2014)Yang, Trail, Pribram-Jones, Burke, Needs, and
  Ullrich]{YangZH14}
Yang,~Z.~H.; Trail,~J.~R.; Pribram-Jones,~A.; Burke,~K.; Needs,~R.~J.;
  Ullrich,~C.~A. {Exact and approximate Kohn-Sham potentials in ensemble
  density-functional theory}. \emph{Phys. Rev. A} \textbf{2014}, \emph{90},
  042501\relax
\mciteBstWouldAddEndPuncttrue
\mciteSetBstMidEndSepPunct{\mcitedefaultmidpunct}
{\mcitedefaultendpunct}{\mcitedefaultseppunct}\relax
\EndOfBibitem
\bibitem[Senjean and Fromager(2018)Senjean, and Fromager]{senjean2018unified}
Senjean,~B.; Fromager,~E. Unified formulation of fundamental and optical gap
  problems in density-functional theory for ensembles. \emph{Phys. Rev. A}
  \textbf{2018}, \emph{98}, 022513\relax
\mciteBstWouldAddEndPuncttrue
\mciteSetBstMidEndSepPunct{\mcitedefaultmidpunct}
{\mcitedefaultendpunct}{\mcitedefaultseppunct}\relax
\EndOfBibitem
\bibitem[G\"orling(1996)]{PhysRevA.54.3912}
G\"orling,~A. Density-functional theory for excited states. \emph{Phys. Rev. A}
  \textbf{1996}, \emph{54}, 3912--3915\relax
\mciteBstWouldAddEndPuncttrue
\mciteSetBstMidEndSepPunct{\mcitedefaultmidpunct}
{\mcitedefaultendpunct}{\mcitedefaultseppunct}\relax
\EndOfBibitem
\bibitem[Umrigar \latin{et~al.}(1998)Umrigar, Savin, and
  Gonze]{umrigar1998unoccupied}
Umrigar,~C.~J.; Savin,~A.; Gonze,~X. \emph{Electronic Density Functional
  Theory}; Springer, 1998; pp 167--176\relax
\mciteBstWouldAddEndPuncttrue
\mciteSetBstMidEndSepPunct{\mcitedefaultmidpunct}
{\mcitedefaultendpunct}{\mcitedefaultseppunct}\relax
\EndOfBibitem
\bibitem[Hellgren and Gross(2012)Hellgren, and Gross]{Hellgren12_PRA}
Hellgren,~M.; Gross,~E.~K.~U. Discontinuities of the exchange-correlation
  kernel and charge-transfer excitations in time-dependent density-functional
  theory. \emph{Phys. Rev. A} \textbf{2012}, \emph{85}, 022514\relax
\mciteBstWouldAddEndPuncttrue
\mciteSetBstMidEndSepPunct{\mcitedefaultmidpunct}
{\mcitedefaultendpunct}{\mcitedefaultseppunct}\relax
\EndOfBibitem
\bibitem[Hellgren and Gross(2013)Hellgren, and
  Gross]{hellgren2013discontinuous}
Hellgren,~M.; Gross,~E.~K.~U. Discontinuous functional for linear-response
  time-dependent density-functional theory: The exact-exchange kernel and
  approximate forms. \emph{Phys. Rev. A} \textbf{2013}, \emph{88}, 052507\relax
\mciteBstWouldAddEndPuncttrue
\mciteSetBstMidEndSepPunct{\mcitedefaultmidpunct}
{\mcitedefaultendpunct}{\mcitedefaultseppunct}\relax
\EndOfBibitem
\bibitem[Cavo \latin{et~al.}(2020)Cavo, Berger, and
  Romaniello]{PhysRevB.101.115109}
Cavo,~S.; Berger,~J.~A.; Romaniello,~P. Accurate optical spectra of solids from
  pure time-dependent density functional theory. \emph{Phys. Rev. B}
  \textbf{2020}, \emph{101}, 115109\relax
\mciteBstWouldAddEndPuncttrue
\mciteSetBstMidEndSepPunct{\mcitedefaultmidpunct}
{\mcitedefaultendpunct}{\mcitedefaultseppunct}\relax
\EndOfBibitem
\bibitem[Perdew(1985)]{NATO85_Perdew_p284-286}
Perdew,~J.~P. In \emph{Density Functional Methods in Physics}; Dreizler,~R.~M.,
  da~Provid\^{e}ncia,~J., Eds.; NATO ASI Series; Plenum Press, 1985; Vol. 123;
  pp 284--286\relax
\mciteBstWouldAddEndPuncttrue
\mciteSetBstMidEndSepPunct{\mcitedefaultmidpunct}
{\mcitedefaultendpunct}{\mcitedefaultseppunct}\relax
\EndOfBibitem
\bibitem[Note10()]{Note10}
Obviously, now the KS potential does not approach zero as $|\protect \bm {r}|
  \rightarrow \infty $. While this should happen for the exact KS potential,
  this does not happen for the LDA (or invLDA).\relax
\mciteBstWouldAddEndPunctfalse
\mciteSetBstMidEndSepPunct{\mcitedefaultmidpunct}
{}{\mcitedefaultseppunct}\relax
\EndOfBibitem
\bibitem[Perdew(1990)]{Perdew90}
Perdew,~J.~P. {Size-consistency, self-interaction correction, and derivative
  discontinuity in density functional theory}. \emph{Adv. Quantum Chem.}
  \textbf{1990}, \emph{21}, 113\relax
\mciteBstWouldAddEndPuncttrue
\mciteSetBstMidEndSepPunct{\mcitedefaultmidpunct}
{\mcitedefaultendpunct}{\mcitedefaultseppunct}\relax
\EndOfBibitem
\bibitem[Mori-S\'{a}nchez \latin{et~al.}(2008)Mori-S\'{a}nchez, Cohen, and
  Yang]{MoriS08}
Mori-S\'{a}nchez,~P.; Cohen,~A.~J.; Yang,~W. {Localization and Delocalization
  Errors in Density Functional Theory and Implications for Band-Gap
  Prediction}. \emph{Phys. Rev. Lett.} \textbf{2008}, \emph{100}, 146401\relax
\mciteBstWouldAddEndPuncttrue
\mciteSetBstMidEndSepPunct{\mcitedefaultmidpunct}
{\mcitedefaultendpunct}{\mcitedefaultseppunct}\relax
\EndOfBibitem
\bibitem[Capelle \latin{et~al.}(2010)Capelle, Vignale, and
  Ullrich]{capelle2010spin}
Capelle,~K.; Vignale,~G.; Ullrich,~C.~A. Spin gaps and spin-flip energies in
  density-functional theory. \emph{Physical Review B} \textbf{2010}, \emph{81},
  125114\relax
\mciteBstWouldAddEndPuncttrue
\mciteSetBstMidEndSepPunct{\mcitedefaultmidpunct}
{\mcitedefaultendpunct}{\mcitedefaultseppunct}\relax
\EndOfBibitem
\end{mcitethebibliography}

\end{document}